# Multilevel non-linear interrupted time series analysis

RJ Waken, PhD[1, 2]
Fengxian Wang, PhD[2, 3]
Sarah A. Eisenstein, PhD[2, 4]
Tim McBride, PhD[2, 4]
Kim Johnson, PhD[2, 4, 5]
Karen Joynt-Maddox, MD, MPH[2, 3, 4]

1. Center for Biostatistics, Institute for Informatics, Data Science, and Biostatistics, Washington University in St. Louis School of Medicine, St. Louis, Missouri
2. Center for Advancing Health Services, Policy & Economics Research, Washington University in St. Louis School of Medicine, St. Louis, Missouri
3. Cardiovascular Division, Department of Medicine, Washington University in St. Louis School of Medicine, St. Louis, Missouri
4. School of Public Health, Washington University in St. Louis, St. Louis, Missouri
5. Brown School of Social Work, Washington University in St. Louis, St. Louis, Missouri

Corresponding author: RJ Waken, rjwaken@wustl.edu

## Abstract

**Background.** Recent advances in interrupted time series analysis permit characterization of a non-linear interruption effect through use of generalized additive models, but subgroup specific effects are typically estimated through separate or sequential models.
**Methods.** We propose to characterize interruption effects in a multilevel structure, incorporating meaningful variability in existing time-varying trends and causal effects across subpopulations of interest, encouraging parsimony and partial pooling while allowing poststratification.
**Results.** We compare the proposed model to a single-level model in a simulation study. Both approaches declare an overall effect in 2.3% of replications with no effect (nominal 5%). Our method reduces subgroup effect RMSE by up to 20%, detects true subgroup differences more often (73-79% compared to 40-43% at the highest signal-to-noise ratio [SNR]), and only exhibits higher RMSE under subgroup effect structure misspecification at the highest SNR. We then present three applications for our multilevel model while highlighting our ability to characterize overall, cell level, and subgroup specific effects: 1) the effect of the introduction of the prostate-specific antigen (PSA) test on prostate cancer diagnosis rates, 2) the change in stroke or transient ischemic attack (TIA) hospitalization rates in Medicare beneficiaries early in the COVID-19 pandemic, and 3) the effect of Medicaid expansion in Missouri on the proportion of inpatient hospitalizations discharged with Medicaid as a primary payer. We show in 1) that the typical increase in prostate cancer diagnoses was greatest in the 45-54 age group (poststratified IRR = 2.11, sample IRR = 2.27), in 2) March 1st through May 24th 2020 saw more pronounced decreases in Stroke/TIA hospitalizations than the rest of 2020, and 3) Medicaid expansion increased the share of Medicaid-paid hospital discharges most in men aged 19-26 or 27-44 compared to other age-sex subgroups.
**Conclusions.** We present a fully Bayesian framework that recovers overall and subgroup-specific intervention effects from a single fit with uncertainty, supports poststratification to a target population, and offers applied researchers a principled alternative to sequential subgroup analyses for investigating intervention effects in time series data.

## Background

Quasi-experimental designs allow researchers to uncover causal relationships between treatments, interventions, or interruptions and outcomes of interest. When analyzing data where an event or treatment affects all individuals in a population or in a significant share of the subpopulations, interrupted time series designs give a framework for describing the causal relationship between treatment and outcomes while accounting for secular and natural variability in outcomes[1]. Implementation strategies are varied, and additionally, they may incorporate little to account for natural time varying patterns[2], despite readily available methods that account for serial dependency[3]. This consideration is crucial to effect characterization since, in the potential outcomes framework for causal inference[4], we must accurately characterize both the mean effect of and variability about the counterfactual to recover the causal effect. Interrupted time series modeling approaches that fail to address serial dependency or time-varying patterns in a principled way may have conclusions with higher type I error rates[5].

Researchers are often interested in describing causal effects within subgroups as well as overall in an affected population. Many statistical practitioners take a sequential, model building like approach to describing inferences; first, researchers fit an "overall" pre-post model, describing the typical effect across a population without addressing within subgroup differences, which may suffer from Simpson's paradox[6]. This is then followed with either a stratified by subgroup analysis, which loses information that may be borrowed across subgroups, or an overall pre-post effect as well as subgroup by pre-post interaction terms to describe how each subgroup experiences the intervention effect differently than the reference group, which often inflates type I error rates[7]. Multilevel models can help us to characterize overall effects, subgroup differences, and cell level effects (defined as the combination of subgroups) in a single model specification while encouraging parsimony via shrinkage to the overall intervention mean[8]. Further, characterizing intervention effects using a multilevel regression model allows us to address sample representativeness issues via multilevel regression with poststratification if using demographic variables available in the sample and in the overall population of interest[9]. Recent computational and software advances make multilevel model implementation feasible in a time series setting, and also allow for generalized additive models to be structured alongside stochastic latent time-varying model specifications[10].

All statistical analyses are complicated by the tendency for natural processes to eschew convenient, linear patterns. To address these issues in interrupted time series designs, authors in health services research often specify segmented regression components designated based on knowledge of interruption timing or output of visualization tools to address non-linear post intervention effects[11–13]. Others incorporate dynamic linear modeling structures[14, 15]. More recent approaches incorporate non-linear trends via generalized additive models, or GAMs, and specify smoothed curves to describe the post-intervention effect[5]. This approach allows for a changing, non-linear intervention effect in the post period. These models parallel some approaches for similar problems encountered in ecology, and can additionally accommodate latent time series in addition to semiparametric trend modeling[10].

We propose an interrupted time series model that facilitates causal inference across multiple subgroups with potentially different interruption effects and counterfactuals that accounts for sequential, serial variability with a latent time series process, a non-linear interruption effect characterized with a GAM, using a multilevel modeling structure. This proposal is a multilevel extension of the model proposed in Cho[5]. Our paper progresses as follows: in the Methods section, we will describe each of a) latent time series and secular trend specification approach, b) intervention effect characterization via GAM, c) prior specification, and d) the final model. In the Results section, we will compare our model specification to the single-level model proposed in Cho[5] with a simulation study where we characterize type I error rates as well as intervention effect coverage, root mean squared error (RMSE), and bias for overall, subgroup specific, and cell level effects and subgroup effect difference detection. We will then showcase our method with three applications: prostate cancer diagnosis rates changing with the introduction of the prostate specific antigen test, changes in hospitalization rates for stroke and transient ischemic attack early in the COVID-19 pandemic amongst Medicare beneficiaries, and changes in proportion of inpatient discharges with Medicaid as a primary payer after Medicaid expansion in Missouri. In the Discussion and Conclusions, we will conclude with a brief discussion, limitations, and a recap of key findings.

# Methods

## Notation, assumptions, conceptual model, and estimand

Let $y_{st}$ be our outcome in series $s \in 1, \dots, S$ at time $t \in 1, \dots T$, $T_1$ be the interruption time where $1 < T_1 < T$. Each series index $s$ represents membership in one level $k_{j(s)}$ for each population grouping factor $j = 1, \dots, J$ such as age group or race.

Consider the following model:

$$E[y_{st}] = f^{-1}\big(\mu_{st} + g(t, s)\big),$$

$f^{-1}(\cdot)$ represents an inverse link function as typically specified in a generalized linear model, $\mu_{st}$ represents the realization of secular components and a latent time series process for series $s$ at time $t$, and $g(t, s)$ is the output of a function of time and grouping structures that describes an interruption effect of interest, where, for any time $t_0 < T_1$ we define $g(t_0, s) = 0$.

We make the following assumptions:

1) The intervention time point $T_1$ is known and ascertained exactly
2) The treatment state is absorbing; once a series $s$ is treated at time $T_1$, it remains in the treated state for all $t \geq T_1$
3) There is no treatment anticipation
4) The composition of the treated at risk sample does not change over the course of the study period in a way that affects the outcome process described in the above model

5) The functional form of the counterfactual is correctly specified, and faithfully represents the trajectory of the outcome process sans interruption
6) No concurrent or subsequent shocks occur in the period $t \geq T_1$ that affect the latent mean process $\mu_{st}$

Let $\tilde{y}_{st}$ be the series of potential outcomes under no interruption whose expectation is defined as
$$E[\tilde{y}_{st}] = f^{-1}(\mu_{st}).$$

Under the assumptions above, we define our interruption effect for series $s$ at time $t$ on the link function scale as
$$\begin{aligned} f(E[y_{st}]) - f(E[\tilde{y}_{st}]) &= \mu_{st} + g(t,s) - \mu_{st} \\ &= g(t,s). \end{aligned}$$

Let $w_{st}$ be weights, $G$ represent a subset of series of interest, and let $Q$ represent a subset of post interruption time points of interest. We may report our estimand, the average treatment on the treated effect ($ATT$) for $G$ during times $Q$ directly on the outcome scale using
$$\bar{\Delta}_{GQ} = \frac{\sum_{t\in Q}\sum_{s\in G} w_{st}(E[y_{st}] - E[\tilde{y}_{st}])}{\sum_{t\in Q}\sum_{s\in G} w_{st}}.$$

Alternatively, we can summarize information with a weighted average for $s$ with
$$\bar{g}_{GQ} = \frac{\sum_{t\in Q}\sum_{s\in G} w_{st} g(t,s)}{\sum_{t\in Q}\sum_{s\in G} w_{st}}.$$

$\bar{g}_G = \sum_{t\geq T_1}\sum_{s\in G} w_{st} g(t,s)$, and like in the application examples that follow, report our $ATT(G)$ as a rate ratio or odds ratio with $\exp(\bar{g}_G)$ for generalized linear models with a log or logistic link function. In situations where our sample frame has subgroup characteristics that differ from a generalizable population of interest, we may also poststratify our estimates into a population frame by using population weights (as opposed to sample weights) for $w_{st}$ using multilevel regression with poststratification[9]. These quantities are produced per posterior draw such that we obtain posterior distributions for $\bar{\Delta}_{GQ}$ or $\bar{g}_{GQ}$.

## Considerations for specific model components

We present the below two items in a simpler, more approachable format to facilitate an intuitive read of the methods proposed; for a more in depth guide, including prior parameterization, likelihood characterization, and computational specifics, see the Supplemental File. For the below example of our model, suppose we have one grouping factor with levels $s = 1, \ldots, S$ so that each series corresponds to one subgroup level $s$.

### *Intervention effect specification*

We specify a generalized additive model[16] (GAM), specifically thin plate splines, to characterize the intervention effect. For all $t \geq T_1$, let $\tau = t - T_1$. We specify an overall intervention effect $\xi(t)$ as
$$\xi(t) = \omega_0 + \boldsymbol{b}(\tau)^{\mathrm{T}}\boldsymbol{\alpha},$$
where $\boldsymbol{b}(\tau)$ is our spline basis, and $\boldsymbol{\alpha} = [\alpha_1 \quad \cdots \quad \alpha_M]$ are spline coefficients with a penalized prior, and in this formulation, $\omega_0$ is the average post-period interruption effect. We also characterize deviations from the overall mean effect with a multilevel structure, where for subgroup $s \in 1, \ldots, S$, with $\boldsymbol{\alpha}_s = [\alpha_{1s} \quad \cdots \quad \alpha_{Ms}]$

$$\xi_s(t) = \omega_s + \boldsymbol{b}(\tau)^{\mathrm{T}} \boldsymbol{\alpha}_s,$$

subject to $\sum \omega_s = 0$ and $\sum_{s=1}^{S} \alpha_{ms} = 0$ for each $m$. Our full intervention effect characterization is then $g(t, s) = \xi(t) + \xi_s(t)$.

### *Latent time-varying mean specification*

Our latent time-varying mean component $\mu_{st}$ operates as the mean function during the pre-period, and counterfactual in the post-period; it accounts for secular components whose coefficients do not vary over time and serial dependency in our observation streams through a latent time series structure which allows the process to evolve throughout the study period. For the remainder of this paper, we will assume that a latent random walk structure is sufficient to describe the time varying structure and dependency amongst observations over time in our serial process[15], but it is possible to incorporate other time-varying structures if appropriate. To build this component across our subgroup levels $s = 1, \ldots, S$, we specify

$$\mu_{st} = \boldsymbol{x}_{st}^{\mathrm{T}} \boldsymbol{\beta} + \theta_s + \gamma_t + \gamma_{st},$$

where $\boldsymbol{x}_{st}^{\mathrm{T}} \boldsymbol{\beta}$ is a secular mean component, $\theta_s$ is a random intercept subject to $\sum \theta_s = 0$, $\gamma_t$ is the overall time varying mean component where $\sigma_\gamma$ is the standard deviation associated with the overall random walk process, $\gamma_t \sim N(\gamma_{t-1}, \sigma_\gamma)$ and $\gamma_1 = 0$, and $\gamma_{st}$ are subgroup specific deviations from the time varying component, where $\sigma_{s\gamma}$ is the standard deviation associated with the random walk process in subgroup $s$, with $\gamma_{st} \sim N(\gamma_{s,t-1}, \sigma_{s\gamma})$, $\gamma_{s1} = 0$, and $\sum_{s=1}^{S} \gamma_{st} = 0$.

Unlike other approaches such as Bayesian Structural Time Series[14] (BSTS), these components are informed throughout the entirety of the study period, both pre and post interruption, such that the underlying latent process does not cease at time $T_1$. This is appropriate if assumptions 5 and 6 above are reasonable. In situations where the post period inappropriately shifts this latent trend, we would expect pre-period posterior predictive coverage to suffer; users should take care to evaluate whether posterior predictive coverage in the pre-period achieved nominal rates to determine if assumptions 5 and 6 are reasonable.

Critically, in order to appropriately characterize the intervention effect under the potential outcomes framework, the intervention effect structure must be either a) equivalent to or b) nested in the latent multilevel random effects and time varying evolution structure; intuitively, this is because each intervention effect needs a representative counterfactual – if this is not the case, our counterfactual would fail to act as a potential outcome for all subgroups evaluated.

## Modeling and Implementation Details

We use Stan[17] Version 2.32.2 for all analyses. All code is hosted in the public repository at https://doi.org/10.5281/zenodo.24178292.

All inferences in our application section are based on 95% credible intervals (CIs) based on quantiles from Markov-Chain Monte Carlo (MCMC) draws. While the Medicare and Missouri Hospital Association data used in this study are not publicly available due to data use agreement (DUA) restrictions, we provide a small cell censored DUA compliant version of the SEER data with code in the linked repository. We tuned all application models such that we have no divergences, all rank-normalized Rhats were less than 1.01, and bulk effective sample size was ≥ 400 for all parameters. We additionally inspected posterior predictive coverage in both the pre- and post-intervention periods to investigate model fit, and in all applications, Ljung-Box[18] tests for residual autocorrelation on Dunn-Smyth[19] residuals to assess how well our time-varying mean structure accounted for serial dependency.

For notational convenience and clarity, we describe the above random walks with a "centered" parameterization; in practice, we sample all random walks and group level components in our multilevel specification with a non-centered parameterization to avoid problematic posterior geometry[20].

# Results

## Simulation study

We conduct extensive simulation studies to evaluate the performance of our proposed method against a single level comparator. We evaluate each method on cell level, aggregated subgroup level, and overall level results across all time points in the interruption period, and characterize performance with the following metrics:

- Bias, defined as the $mean(estimate - truth)$
- Root mean squared error (RMSE), defined as $\sqrt{mean((estimate - truth)^2)}$
- Coverage, which is the proportion of 95% credible intervals (CI) that contain the true parameter value
- "Declared" effect, which is the proportion of 95% CIs that exclude 0
    - When there is no true effect, this metric describes type I error rates
    - This metric is also used to describe each method's ability to detect true (or false) subgroup differences

We report estimates of each of the above metrics per inferential estimate of interest per simulation replication, then report averages and Monte Carlo standard errors[21].

In each simulation iteration we also describe sampler diagnostics, such as Rhat, effective sample size, number of divergences, treedepth hits, and estimated Bayesian fraction of missing information (E-BFMI)[22], as well as important diagnostics for our method, such as pre-period latent mean trend coverage.

### *Simulation study design*

In each simulation iteration, we simulate 60 time points for two crossed subgrouping factors F1 and F2 with an interruption set at time point 41; this yields a simulation scenario with 9 total series and 20 post-intervention points (time points 41 through 60). We simulate noise independently in each of the 540 observations using randomly generated standard normal errors.

Our latent time varying structure consists of

- An overall random walk term $\gamma_t \sim N(\gamma_{t-1}, 0.30)$, where $\gamma_1 = 0$, and
- Series specific random walk terms within each level of the factors F1 and F2, where we independently generate $\gamma_{jkt} \sim N(\gamma_{jk,t-1}, 0.15)$, then center the within factor level such that $\sum_k \gamma_{jkt} = 0$ for a given $t$ and define $\gamma_{jk1} = 0$.

Our intervention effect structure is factor dependent.

- Factor F1 sets the scale of the intervention effect, with $s = (1.0, 0.3, -0.6)$, which yields an overall mean of $\bar{s} = 0.233$, and
- Factor F2 sets the shape of the intervention effect as a function of $\tau = t - 41$, with
  - The first series representing an immediate shock, where $h_1(\tau) = 1$ throughout
  - The second series representing a linear trend, where $h_2(\tau) = \tau/19$
  - The third series representing a logistic curve with an inflection at the post-period midpoint ($t \approx 50.5$); this term runs from 0.05 to 0.95, and is rescaled to a maximum of 1
  - $\bar{h}(\tau)$ is the average of the three shapes

We investigate one scenario with no intervention effect, and three different scenarios at three signal to noise ($SNR$) ratios $SNR \in 1, 2, 4$, which is defined at peak magnitude such that $m = SNR \times$ noise $SD$: i) Null, where $g(t, s) = 0$, ii) Additive, where $g(t, s) = m \times \left(s_{F1} \times \bar{h}(\tau) + \bar{s} \times (h_{F2}(\tau) - \bar{h}(\tau))\right)$, iii) Multiplicative, where $g(t, s) = m \times s_{F1} \times h_{F2}(\tau)$, and iv) F2 homogeneous, $g(t, s) = m \times s_{F1} \times \bar{h}(\tau)$ where all F2 contrasts are defined as exactly zero so we can investigate false subgroup differences.

We repeat our simulation experiment 300 times for each scenario and $SNR$ combination, and characterize

$$\bar{g}_Q = \frac{\sum_{t \geq 41} \sum_{s \in G} w_{st} g(t, s)}{\sum_{t \geq 41} \sum_{s \in G} w_{st}}$$

as our estimate of interest; in these simulations, we set all weights $w_{st}$ to be equal, and because we are fitting a model with an identity link and a normal error structure, the estimate above yields our $ATT \, \bar{g}_Q$ directly.

### *Simulation methods*

We fit our MLITS method as specified in the Supplemental File once per simulation replication; intuitively, this extension requires us to add one additional grouping factor to the example model proposed in methods above. This creates an additive multilevel structure, which is consistent with one of the simulation scenarios investigated, and inconsistent with the multiplicative scenario proposed. All quantities of interest can be ascertained from the same model. In practice (and in one of the applications below), this structure can be specified in a way that is better suited to the problem at hand. Our single level comparator is also fully described in the Supplemental File, but intuitively, is parameterized like the models above with no factor specific components. This model is fit 16 times per simulated dataset: once for the overall effect, one time for each factor level (each level of F1 aggregated across F2, each level of F2 aggregated across F1), and once for each of the individual 9 series. All models have a normal likelihood, and greater detail regarding prior parameterization, input control parameters, and convergence diagnostics used are included in the Supplemental File. There are convergence issues that arose, rarely for MLITS, but frequently in the single-level model for aggregated data fits (in overall, F1 and F2 subgroups); while we use the full results in the plots presented here, we recreate the same results plots while only using comparisons where all relevant fits converged, which do not materially change the results we report below.

### *Simulation results*

We present our simulation results for post-period effects in a plotted summary in Figure 1, where all results are plotted with error bars representing $\pm 1.96 \times$ Monte Carlo SE. Under the Null scenario, both models declare an overall effect in 2.3% of cases, below the nominal rate of 5%. The MLITS model declares false subgroup differences in 3.1% of F1 and 4.4% of F2 contrasts in the Null scenario and 3.1-3.9% of F2 contrasts in the F2-homogeneous scenario; contrast false positive rates are essentially zero for the single-level model.

Cell-level RMSE is superior in the MLITS model in the Additive and F2-homogeneous scenarios, showing that pooling information under an appropriately specified intervention structure is beneficial at the cell level. Cell coverage for MLITS is at or above nominal in all of these scenarios. The single-cell model has higher cell-level RMSE and lower CI coverage at the highest $SNR$ across outside of the Multiplicative scenario, where the MLITS model is purposefully misspecified.

The overall effect estimates are difficult for either model to detect, and neither model stands out because the true overall effects largely cancel out after averaging over positive and negative effects; similarly, detecting the small F1 subgroup effect (scale 0.3) is difficult for both models. The MLITS model outperforms the single-level model in detecting true F1 subgroup effects at scale 1.0 (at $SNR = 4$, 0.85 vs 0.65, Multiplicative: 0.82 vs. 0.63, F2-homogeneous: 0.86 vs 0.65), and -0.6 (at $SNR = 4$, Additive: 0.47 vs 0.30, Multiplicative – some overlap in Monte Carlo SE intervals: 0.33 vs. 0.27, F2-homogeneous: 0.48 vs 0.30) for the other two subgroup levels at $SNR$ 2 and 4 as well as true between F1 subgroup differences in all scenarios and $SNR$ levels (at $SNR = 4$, Additive: 0.79 vs 0.40, Multiplicative: 0.73 vs. 0.42, F2-homogeneous: 0.79 vs 0.43) even in the Multiplicative scenario where the MLITS model is misspecified.

We additionally evaluated bias for the overall, within subgroup levels, and cell specific estimates from each of our models. We found that the MLITS model tends to have absolute bias lower than or statistically indistinguishable from the single-level model for overall and subgroup-level summaries, and is only consistently bested for cell-level effects by the single-level model only in the most extreme cells of the Multiplicative scenario (scale 1.0 and −0.6 with the shock shape) where the MLITS model is misspecified. For specifics and greater detail, see Supplemental File Figures S8 and S9.

The above simulation study concerns a comparison between the MLITS model and single-level model, both fit with a normal likelihood to data generated with a normally distributed noise process. We additionally present the following in the Supplemental File:

- Results from BSTS models fit using the CausalImpact[23] package to the full simulation above, refit at each aggregation level (similar to the single-level model), with both the default and a data-scaled prior specification
- Full coverage and bias results for the full simulation and BSTS models at each post-interruption time point
- Using the MLITS model only, we generate

- Results in above simulation study design with a Negative Binomial data generating process and model likelihood in place of a normal error structure
- Results from changes to the level shift, random walk, and smoothing parameter prior structure, with each change implemented one at a time, in the Additive scenario.
- Results from a shorter simulation study at $SNR = 0, 2$ based on a latent autoregressive time-varying process.

## Applications

All applications fit the model described above, but have different grouping structure configurations. We introduce simplified notation below that can help describe each of the application models fit; we use

$$\mu_{st} = \boldsymbol{x}_{st}^{\mathrm{T}}\boldsymbol{\beta} + \sum \theta_{f[s]}^{F} + \gamma_t + \sum \gamma_{f[s],t}^{F},$$
$$g(t,s) = \xi(t) + \sum \xi_{f[s]}^{F}(t), \qquad t \geq T_1,$$
$$E[y_{st}] = f^{-1}\big(\mu_{st} + g(t,s)\big)$$

where the superscript $F$ indicates a grouping factor that has levels $f[s]$, and the summands above sum across all sub-group level memberships that the observation belongs to.

## SEER Prostate Cancer Diagnosis Incidence

The Prostate Specific Antigen (PSA) test for prostate cancer, a non-invasive alternative to biopsy for a quick, easily accessible test for prostate cancer diagnosis (DX) based on presence of a biomarker that became widely commercially available in 1986[24]. Although the biomarker levels for positive results changed in 1998, this test is still widely used today, and in spite of critiques that this test leads to overtreatment[25], it is estimated that one prostate cancer death is avoided for every 769 men screened[26]. We intend to investigate both a) did the introduction of the PSA test increase the number of prostate cancer diagnoses in the US, b) was this effect different by subgroup, and c) how do our findings from these non-representative registries translate to expected changes in prostate cancer DX rates in the whole US?

Using Surveillance, Epidemiology and End Results (SEER) 8 data for men aged 45 and older from 1975 to 1998, we characterize yearly prostate cancer counts and as well as total men at risk in age brackets of 45 – 54, 55 – 64, 65 – 74, and 75+, as well as in the SEER provided race groups of Black, Other, and White. The "8" in the data name indicates that these data are the product of 8 different regional registries[27]. We use U.S. decennial census and intercensal data from the U.S. Census bureau to create poststratification weights to describe our results in men by age group, race, and overall to be more representative of the at-risk U.S. population as a whole[28–30]. SEER race categorizations include individuals of Hispanic ethnicity in all race categories in the time periods considered here, which was also practice in the Census data utilized.

We fit the above model using a multilevel random intercept and slope (over time), random walk, and GAM intervention effect structure with race nested in age group with a Negative Binomial likelihood (specified with mean and overdispersion term), and use an offset term describing number of individuals at risk. For offset term $O_{st}$, overdispersion term $\phi$, and random slopes $\zeta_{a[s]}^{\text{Age}}$ and $\zeta_{r[s]}^{\text{Race(Age)}}$, we specify

$$
\begin{aligned}
&\mu_{st} = \beta_0 + \beta_1 t + \theta^{\text{Age}}_{a[s]} + \theta^{\text{Race(Age)}}_{r[s]} + \left(\zeta^{\text{Age}}_{a[s]} + \zeta^{\text{Race(Age)}}_{r[s]}\right) t_{\text{dec}} + \gamma_t + \gamma^{\text{Age}}_{a[s],t} + \gamma^{\text{Race(Age)}}_{r[s],t},\\
&g(t,s) = \xi(t) + \xi^{\text{Age}}_{a[s]}(t) + \xi^{\text{Race(Age)}}_{r[s]}(t), \qquad t \geq T_1,\\
&y_{st} \sim \text{NB}\big(\exp\big(O_{st} + \mu_{st} + g(t,s)\big), \phi\big).
\end{aligned}
$$

We begin estimating our intervention effect at the commercial introduction of the PSA test in 1986 (corresponding to time $t = 12$). We focus on 95% CIs for our intervention effect based on both our sample as well as poststratified results reflecting the at-risk U.S. population.

Figure 2 gives the Incidence Rate Ratio (IRR) for the change from 1986 - 1998 in DX Rate attributable to the interruption overall, by age group, and by race, as well as all within grouping factor pairwise comparisons, both in the sample, as well as poststratified onto the at risk male US population makeup by age group and race. We can see that poststratification tends to attenuate our DX rate increase predictions; this is most apparent in the 55-64 age group, Black race group, and overall estimates. We also see that younger males saw a larger increase in prostate cancer DX rates after commercial approval and introduction of the PSA test. Figure 3 gives poststratified estimates of prostate cancer DX incidence both modeled to the observed process as well as the counterfactual overall, by race, and by age group, where the bands represent 95% CIs. We show that the latent time-varying process used as a counterfactual evolves and expands linearly as expected under a random walk structure, even while updating posterior parameters information throughout the post-interruption period.

We present a prior sensitivity analysis detailing how prior choices on our random walk innovation standard deviations (both increasing and decreasing standard deviations), interruption shifts, and smoothing terms affect inferences; we do not see appreciable changes in visual summaries, but do observe that decreasing random walk standard deviations results in model convergence issues.

## Drop in stroke and transient ischemic attack hospitalization episodes during the COVID-19 pandemic in Medicare

The COVID-19 pandemic resulted in a drop in utilization due to both health care system avoidance and administrative directives[31]. This drop included emergent and non-emergent care in hospital settings[32].

We access Medicare Fee-for-service Part A claims and Medicare Encounter Medicare Advantage data that describes hospital utilization in the Virtual Research Data Center (VRDC) environment. The VRDC is a secure system that gives researchers with privileged access the capability to see, analyze, and interpret patterns in a variety of data sets provided at the Medicare enrolled beneficiary level using a variety of software options. Output from this system is reviewed to ensure beneficiary privacy, any datum describing between 1 and 10 Medicare beneficiaries must be censored before a download is allowed, and all data or results must be presented in a text based tabular format.

We characterize stroke counts and days at risk of hospitalization on a weekly basis from January 7th, 2018, through December 31st, 2020 and investigate the change in stroke hospitalizations on or after March 1st, 2020 during the COVID pandemic for all beneficiaries in Medicare actively enrolled and in a county with a valid Federal Information Processing Standards (FIPS) code. We describe rurality based on commuting based statistical area definitions on a per FIPS code basis: each FIPS is classified as metropolitan, micropolitan, or rural, which is defined as the absence of membership in a Core Based Statistical Area[33]. We define hospitalization episodes based on the rules outlined in Supplemental File Figure S1, and we classify as hospitalization episode as a stroke hospitalization event if any of the diagnosis codes in included in Supplemental File Table S1. For offset term $O_{st}$, overdispersion term $\phi$, and starting day of week as indexed to the beginning of the year $d_t$, we fit the model

$$\mu_{st} = \beta_0 + \beta_1 \cos(2\pi d_t/365) + \beta_2 \sin(2\pi d_t/365) + \theta_{r[s]}^{\text{Rurality}} + \gamma_t + \gamma_{r[s],t}^{\text{Rurality}}$$
$$g(t,s) = \xi(t) + \xi_{r[s]}^{\text{Rurality}}(t), \qquad t \geq T_1,$$
$$y_{st} \sim \text{NB}(\exp\{\log(\text{person-days}_{st}) + \mu_{st} + g(t,s)\},\ \phi).$$

In Table 1, we give estimated incidence rate ratios attributable to the COVID interruption over the first 12 weeks, estimated as

$$\text{IRR}_{s\in R} = \frac{\sum_{t=113}^{124} \exp(O_{st} + \mu_{st} + g(t,s))}{\sum_{t=113}^{124} \exp(O_{st} + \mu_{st})}$$

for $g(s,t) = \xi(t) + \xi_{r[s]}^{Rurality}(t)$, as well as similar IRRs throughout the end of the study period, and additionally present pairwise comparisons between groups by rurality. The post-interruption rates are depressed across the full post-period, more so in the first 12 weeks than in the last. Rates are depressed in the post-period across all levels of rurality, and we do not have strong evidence to suggest that these IRRs differ by beneficiary rurality.

## Changes in proportion of inpatient discharges covered by Medicaid in Missouri after Medicaid Expansion

Missouri voted to expand Medicaid eligibility to individuals aged 19 to 64 based on income thresholds in the 2020 election, and after some pauses in implementation, the expanded eligibility requirements became law on July 1st, 2021 (enrollment officially started in October, but eligible beneficiaries were covered a posteriori back to July). Before expansion, adults aged 19–64 who were not pregnant and had no qualifying disability were eligible for Medicaid only as primary caretakers with extremely low income; after expansion, these adults became eligible with incomes up to 133% (138% after the standard 5% disregard) of the federal poverty level[34]. Pregnant women were already eligible up to 196% of the federal poverty level[34], so expansion changed did not affect eligibility for that group of individuals. The combination of state-by-state implementation of laws dictating eligibility requirements and other states expanding eligibility requirements in the time period around Missouri's expansion law makes finding a suitable control group or using differences in differences methods difficult. Additionally, the Department of Health and Human Services declared a Public Health Emergency on January 31st, 2020, after which the federal Families First Coronavirus Response Act was passed which financially incentivized states from stopping Medicaid eligibility redetermination as of March 18th, 2020[35, 36]; these occurred during the study period, and while it did not permit new enrollment pathways, it did increase Missouri Medicaid enrollment counts[37] although the effect on inpatient hospitalization payer mix at discharge in Missouri from this was less clear[38].

We use Missouri Hospital Association data that describes all inpatient hospitalization claims in Missouri, and keep all inpatient claims with a discharge date in 2016 – 2022, from individuals aged 19 – 64, and characterize hospitalization by combination of age group (19 – 26, 27 – 44, and 45 – 64) and whether the data identifies individuals with a sex code as male or female, as well as by patient rurality. We count the total number of hospitalizations per month both a) overall, and b) that are paid for with Medicaid as a primary payer, defining the latter as a success. We fit a logistic regression model describing the number of inpatient hospital discharges monthly with Medicaid as a primary payer over time with the above multilevel latent RW time series structure with proposed GAM intervention effect characterization overall, by age-sex group combination, and by rurality. For total hospitalization discharges in a series and month $n_{st}$, hospitalization discharges with Medicaid as a primary payer $y_{st}$ and term indicating that $t$ is during the period where Medicaid redetermination was disincentivized (on or after March 18th, 2020) as PHE, we fit

$$\mu_{st} = \beta_0 + \beta_1 \text{PHE} + \theta_s^{\text{S}} + \zeta_s^{\text{S}} \text{PHE} + \gamma_t + \gamma_{st},$$
$$g(t,s) = \xi(t) + \xi_{a[s]}^{\text{Age-sex}}(t) + \xi_{r[s]}^{\text{Rurality}}(t), \qquad t \geq T_1,$$
$$y_{st} \sim \text{Binomial}\left(n_{st}, \text{logit}^{-1}\left(\mu_{st} + g(t,s)\right)\right).$$

In this case, our interruption effect structure is nested in the more complex time-varying trends structure; each of the series has a random intercept, is allowed to evolve through a joint and series specific random walk, and has series specific PHE random slopes, and we characterize interruption effects overall, by rurality, and by age-sex group.

Figure 4 gives the predicted effect of Medicaid expansion on the difference in proportions of discharges by age-sex group integrated across levels of rurality after the start of Medicaid expansion; we present results a) over time (top panel) as well as b) pairwise differences between groups. We see distinctive differences by age-sex group, which was expected given that pregnant women, who often give birth in inpatient settings, had a less restrictive pathway toward eligibility than other groups prior to Medicaid expansion. Males in the youngest two age categories saw the largest increases in Medicaid as payer at discharge.

# Discussion

In this paper, we have proposed a novel application of multilevel GAMs with an underlying stochastic multilevel time varying process to describe causal effects in an interrupted time series design. This builds on existing work that has previously described how GAMs can be utilized to describe causal effects, and allows us to describe different effects by subgroups in a single cohesive model. The multilevel implementation also supports poststratification, which can assist researchers in easily communicating results to stakeholders in a way that is representative of the targeted inferential population.

Most commonly, researchers investigating overall and subgroup specific trends fit models to an overall trend, then subgroup specific component parts; this modeling framework has several benefits, but does not easily permit a multilevel approach that incorporates subgroup specific trends and information without iteratively refitting models on smaller subsets of data. Our modeling framework allows applied researchers to investigate overall and subgroup specific trends in a single model fit, which provides three key benefits: 1) it shares information across population subgroups that aids in fitting the time varying trend components and the intervention effect components, 2) it allows researchers to make comparisons across population subgroups for effects of interest, which may uncover subgroup specific intervention effects, and 3) in the event that analyses are based on a sample with subgroup cell counts that are unlike the population of interest, it allows researchers to utilize poststratification to ascertain the expected intervention effect in the target population.

There are however some limitations to this work. Our recommended procedures are somewhat computationally intensive given our full Bayesian implementation, which we feel is beneficial given both the time varying and multilevel modeling components. New advances in software for deterministic Bayesian inference, such as programs built for variational inference or the Pathfinder algorithm[39] could serve to speed up model fitting, but are not explored here due to availability restrictions on some of the platforms where our secure data are housed and analyzed. GAMs are excellent for estimating effects over observed nuisance factors, such as time, but do not allow researchers to forecast outside of the observed nuisance factor bounds in the data. Time is a nuisance factor in our analysis considered above; we are interested in making inferences at specific observed time points in a retrospective fashion, but do not care to project our inferences forward beyond the bounds of the observed data. Researchers interested in the above strategy with forecasting may be able to accomplish this through the use of dynamic linear models as the mean function describing intervention effects[10, 14].

Our counterfactual framework relies on appropriate specification of both a) the time-varying structure, and b) the underlying latent multilevel structure. If the latent time-varying trend is misspecified, we may see convergence issues[40], or may see undercoverage in the pre-period and lower true effect detection rates in the post period as we show in the Supplmental Files (Figure S13). When random effects structures are misspecified, prior research has found that while fixed effects estimates remain consistent with little bias, predictions conditional on the random effects structure suffer and are less reliable[41]; this point is crucial here because our proposed counterfactual relies on proper specification of the multilevel structure. In that regard, misspecification of the multilevel structure issue mirrors violations of the parallel trends assumption (and how progressively more extreme violations affect inferences) in the differences-in-differences literature[42]. Our approach to assessing potential violations involves posterior predictive checks for the pre-period as well as prior sensitivity analyses that tighten or relax the within subgroup evolutionary processes. Prior sensitivity analyses are especially crucial for multilevel model specifications where some (or all) factors have few subgroups, because the factor level standard deviations are almost solely informed by the prior on those standard deviation terms[43].

We did not explore concepts like differential treatment timing or multiple treatments per unit. This is possible, however, based on prior work in this field[5] which is easily extendable to incorporate a multilevel structure. Further, if units subject to differential treatment times could be attributed membership in a distinct subpopulation, we could treat treatment initiation timing as a subgrouping as well. This is conceptually similar to recent advances in the differences in differences literature[44, 45].

# Conclusions

We have developed a multilevel non-linear interrupted time series model that combines a latent random-walk counterfactual with a generalized-additive intervention effect under parsimony-encouraging priors. The model yields subgroup-level inferences from a single fit, supports poststratification to a target population, and outperforms an independently-refit single-level comparator in scenarios with subgroup heterogeneity. In three applications spanning cancer surveillance, pandemic-era utilization, and Medicaid policy, the method recovered interpretable subgroup-specific intervention effects with full uncertainty propagation. The approach is implemented in Stan and is released as reproducible R code alongside this manuscript, lowering the bar for applied health-services researchers to adopt principled multilevel ITS as a default analytic tool in observational policy evaluation.

## List of abbreviations

CI: credible interval
FIPS: Federal Information Processing Standards
GAM: generalized additive model
IRR: incidence rate ratio
ITS: interrupted time series
MCMC: Markov-chain Monte Carlo
PSA: prostate-specific antigen
SEER: Surveillance, Epidemiology and End Results
RW: random walk
VRDC: Virtual Research Data Center
DUA: data use agreement
MLITS: Multilevel Non-linear ITS

## Declarations

### Consent for publication

Not applicable.

### Competing interests

The authors declare that they have no competing interests.

### Availability of data and materials

All data used in this study, save the US Census Bureau Data cited, are available by request from the data brokering body but cannot be shared with other interested researchers as per the DUAs put in place to obtain data access. R and Stan code implementing the proposed model, simulation code, a censored and disclosable version of the SEER cancer surveillance data, and analysis scripts are available in the public repository at https://doi.org/10.5281/zenodo.24178292.

### Funding

This work was supported by the Missouri Foundation for Health under Grant 21-0035-ME-21 and National Institutes of Health National Institute for Nursing Research under Grant U01NR020555.

### Authors' contributions

RJW conceived the methodological proposal, implemented the model in Stan, performed the simulation study and three applications, and drafted the manuscript. KJ contributed to the analytic plan for the SEER application and to data preparation. FW contributed to the analytic plan for the Medicare application and to data preparation. SAE contributed to the analytic plan and writing for the Missouri Medicaid expansion application. KJM, KJ, and TM provided clinical and policy context in elements of the three applications. All authors reviewed the manuscript and approved the final submitted version.

## Acknowledgements

Not applicable.

## AI Disclosure Statement

During revision of this work, the authors used Anthropic's Claude Opus 5.5 for programming tasks, code cleanup, code editing for clarity, language editing assistance, and material organization. After using this tool, the authors reviewed and edited any content affected as needed and take full responsibility for the work presented.

## Ethics approval and consent to participate

This study was approved by the Human Research Protection Office at Washington University School of Medicine, and the requirement for informed consent was waived due to the de-identified nature of the data.

# References

1. Turner SL, Karahalios A, Forbes AB, Taljaard M, Grimshaw JM, McKenzie JE. Comparison of six statistical methods for interrupted time series studies: empirical evaluation of 190 published series. BMC Med Res Methodol. 2021;21:134. https://doi.org/10.1186/s12874-021-01306-w.

2. Zhang Y, Ren Y, Huang Y, Yao M, Jia Y, Wang Y, et al. Design and statistical analysis reporting among interrupted time series studies in drug utilization research: a cross-sectional survey. BMC Med Res Methodol. 2024;24:62. https://doi.org/10.1186/s12874-024-02184-8.

3. Pang J, Chen H, Valente MJ. Evaluation of statistical methods in R for estimating intervention effects using segmented linear regression in the AB interrupted time series design. BMC Med Res Methodol. 2026;26:42. https://doi.org/10.1186/s12874-026-02771-x.

4. Rubin DB. Causal inference using potential outcomes: design, modeling, decisions. J Am Stat Assoc. 2005;100:322–31. https://doi.org/10.1198/016214504000001880.

5. Cho S-J. Modelling change processes in multivariate interrupted time series data using a multivariate dynamic additive model: an application to heart rate and blood pressure self-monitoring in heart failure with drug changes. J R Stat Soc Ser C Appl Stat. 2024;73:123–42. https://doi.org/10.1093/jrsssc/qlad088.

6. Pearl J. Comment: understanding Simpson's paradox. Am Stat. 2014;68:8–13. https://doi.org/10.1080/00031305.2014.876829.

7. Burke JF, Sussman JB, Kent DM, Hayward RA. Three simple rules to ensure reasonably credible subgroup analyses. BMJ. 2015;351:h5651. https://doi.org/10.1136/bmj.h5651.

8. Gelman A, Hill J. Data Analysis Using Regression and Multilevel/Hierarchical Models. Cambridge University Press; 2007.

9. Wang W, Rothschild D, Goel S, Gelman A. Forecasting elections with non-representative polls. Int J Forecast. 2015;31:980–91. https://doi.org/10.1016/j.ijforecast.2014.06.001.

10. Clark NJ, Wells K. Dynamic generalised additive models (DGAMs) for forecasting discrete ecological time series. Methods Ecol Evol. 2023;14:771–84. https://doi.org/10.1111/2041-210X.13974.

11. Tran PT, Antonelli PJ, Hincapie-Castillo JM, Winterstein AG. Association of US Food and Drug Administration removal of indications for use of oral quinolones with prescribing trends. JAMA Intern Med. 2021;181:808–16. https://doi.org/10.1001/jamainternmed.2021.1154.

12. Lopez Bernal J, Cummins S, Gasparrini A. Interrupted time series regression for the evaluation of public health interventions: a tutorial. Int J Epidemiol. 2017;46:348–55. https://doi.org/10.1093/ije/dyw098.

13. Wagner AK, Soumerai SB, Zhang F, Ross-Degnan D. Segmented regression analysis of interrupted time series studies in medication use research. J Clin Pharm Ther. 2002;27:299–309. https://doi.org/10.1046/j.1365-2710.2002.00430.x.

14. Brodersen KH, Gallusser F, Koehler J, Remy N, Scott SL. Inferring causal impact using Bayesian structural time-series models. Ann Appl Stat. 2015;9:247–74. https://doi.org/10.1214/14-AOAS788.

15. Gianacas C, Liu B, Kirk M, Di Tanna GL, Belcher J, Blogg S, et al. Bayesian structural time series, an alternative to interrupted time series in the right circumstances. J Clin Epidemiol. 2023;163:102–10. https://doi.org/10.1016/j.jclinepi.2023.10.003.

16. Wood SN. Generalized Additive Models: An Introduction with R. 2nd edition. Chapman and Hall/CRC; 2017. https://doi.org/10.1201/9781315370279.

17. Carpenter B, Gelman A, Hoffman MD, Lee D, Goodrich B, Betancourt M, et al. Stan: a probabilistic programming language. J Stat Softw. 2017;76:1–32. https://doi.org/10.18637/jss.v076.i01.

18. Ljung GM, Box GEP. On a measure of lack of fit in time series models. Biometrika. 1978;65:297–303. https://doi.org/10.1093/biomet/65.2.297.

19. Dunn PK, Smyth GK. Randomized quantile residuals. J Comput Graph Stat. 1996;5:236–44. https://doi.org/10.1080/10618600.1996.10474708.

20. Papaspiliopoulos O, Roberts GO, Sköld M. A General Framework for the Parametrization of Hierarchical Models. Statist Sci. 2007;22:59–73. https://doi.org/10.1214/088342307000000014.

21. Morris TP, White IR, Crowther MJ. Using simulation studies to evaluate statistical methods. Statistics in Medicine. 2019;38:2074–102. https://doi.org/10.1002/sim.8086.

22. Vehtari A, Gelman A, Simpson D, Carpenter B, Bürkner PC. Rank-normalization, folding, and localization: an improved R-hat for assessing convergence of MCMC. Bayesian Anal. 2021;16:667–718. https://doi.org/10.1214/20-BA1221.

23. Brodersen KH, Hauser A. CausalImpact: inferring causal effects using Bayesian structural time-series models. R package version 1.4.1. 2022.

24. Stamey TA, Yang N, Hay AR, McNeal JE, Freiha FS, Redwine E. Prostate-specific antigen as a serum marker for adenocarcinoma of the prostate. N Engl J Med. 1987;317:909–16. https://doi.org/10.1056/NEJM198710083171501.

25. Shao YH, Albertsen PC, Shih W, Roberts CB, Lu-Yao GL. The impact of PSA testing frequency on prostate cancer incidence and treatment in older men. Prostate Cancer Prostatic Dis. 2011;14:332–9. https://doi.org/10.1038/pcan.2011.29.

26. US Preventive Services Task Force. Screening for prostate cancer: US Preventive Services Task Force recommendation statement. JAMA. 2018;319:1901–13. https://doi.org/10.1001/jama.2018.3710.

27. National Cancer Institute. Characteristics of the SEER population compared with the total United States population. https://seer.cancer.gov/registries/characteristics.html. Accessed 5 Aug 2025.

28. US Census Bureau. State intercensal datasets: 1970-1980. https://www.census.gov/data/datasets/time-series/demo/popest/1970s-state.html.

29. US Census Bureau. County intercensal datasets: 1980-1990. https://www.census.gov/data/datasets/time-series/demo/popest/1980s-county.html.

30. US Census Bureau. State and county intercensal datasets: 1990-2000. https://www.census.gov/data/datasets/time-series/demo/popest/intercensal-1990-2000-state-and-county-characteristics.html.

31. Johnson KJ, O'Connell CP, Waken RJ, Barnes JM. Impact of COVID-19 pandemic on breast cancer screening in a large midwestern United States academic medical center. PLoS One. 2024;19:e0303280. https://doi.org/10.1371/journal.pone.0303280.

32. Yu J, Hammond G, Waken RJ, Fox D, Joynt Maddox KE. Changes in non-COVID-19 emergency department visits by acuity and insurance status during the COVID-19 pandemic. Health Aff (Millwood). 2021;40:896–903. https://doi.org/10.1377/hlthaff.2020.02464.

33. National Bureau of Economic Research. Census core-based statistical area (CBSA) to Federal Information Processing Series (FIPS) county crosswalk. https://www.nber.org/research/data/census-core-based-statistical-area-cbsa-federal-information-processing-series-fips-county-crosswalk.

34. Missouri Department of Social Services. Benefit program income limits. https://dss.mo.gov/benefit-program-income-limits.

35. Azar A. Determination that a Public Health Emergency Exists Nationwide as the Result of the 2019 Novel Coronavirus | ASPR. 2020. https://www.aspr.gov/about/law-policy-strategy/emergency-declarations/declarations-public-health-emergency/2020-01-nationwide. Accessed 16 Nov 2026.

36. Rep. Lowey NM [D-N-17. Text - H.R.6201 - 116th Congress (2019-2020): Families First Coronavirus Response Act. 2020. https://www.congress.gov/bill/116th-congress/house-bill/6201/text. Accessed 21 Nov 2026.

37. Center for Advancing Health Services P& ER (CAHSPER) Washington University in St Louis. Missouri Medicaid enrollment dashboard. https://cahsper.wustl.edu/impact/medicaid-policy-analysis/missouri-medicaid-enrollment-dashboard/.

38. Center for Advancing Health Services P& ER (CAHSPER) Washington University in St Louis. Missouri hospital payor mix dashboards: inpatient. https://cahsper.wustl.edu/impact/medicaid-policy-analysis/missouri-hospital-payor-mix_dashboards/missouri-hospital-payor-mix-dashboards-inpatient/.

39. Zhang L, Carpenter B, Gelman A, Vehtari A. Pathfinder: parallel quasi-Newton variational inference. J Mach Learn Res. 2022;23:1–49.

40. Pitt MK, Shephard N. Analytic Convergence Rates and Parameterization Issues for the Gibbs Sampler Applied to State Space Models. Journal Time Series Analysis. 1999;20:63–85. https://doi.org/10.1111/1467-9892.00126.

41. Verbeke G, Lesaffre E. The effect of misspecifying the random-effects distribution in linear mixed models for longitudinal data. Computational Statistics & Data Analysis. 1997;23:541–56. https://doi.org/10.1016/S0167-9473(96)00047-3.

42. Rambachan A, Roth J. A More Credible Approach to Parallel Trends. Review of Economic Studies. 2023;90:2555–91. https://doi.org/10.1093/restud/rdad018.

43. Gelman A. Prior distributions for variance parameters in hierarchical models. Bayesian Anal. 2006;1:515–34. https://doi.org/10.1214/06-BA117A.

44. Sun L, Abraham S. Estimating dynamic treatment effects in event studies with heterogeneous treatment effects. J Econom. 2021;225:175–99. https://doi.org/10.1016/j.jeconom.2020.09.006.

45. Callaway B, Sant'Anna PHC. Difference-in-differences with multiple time periods. J Econom. 2021;225:200–30. https://doi.org/10.1016/j.jeconom.2020.12.001.

# Tables and Figures

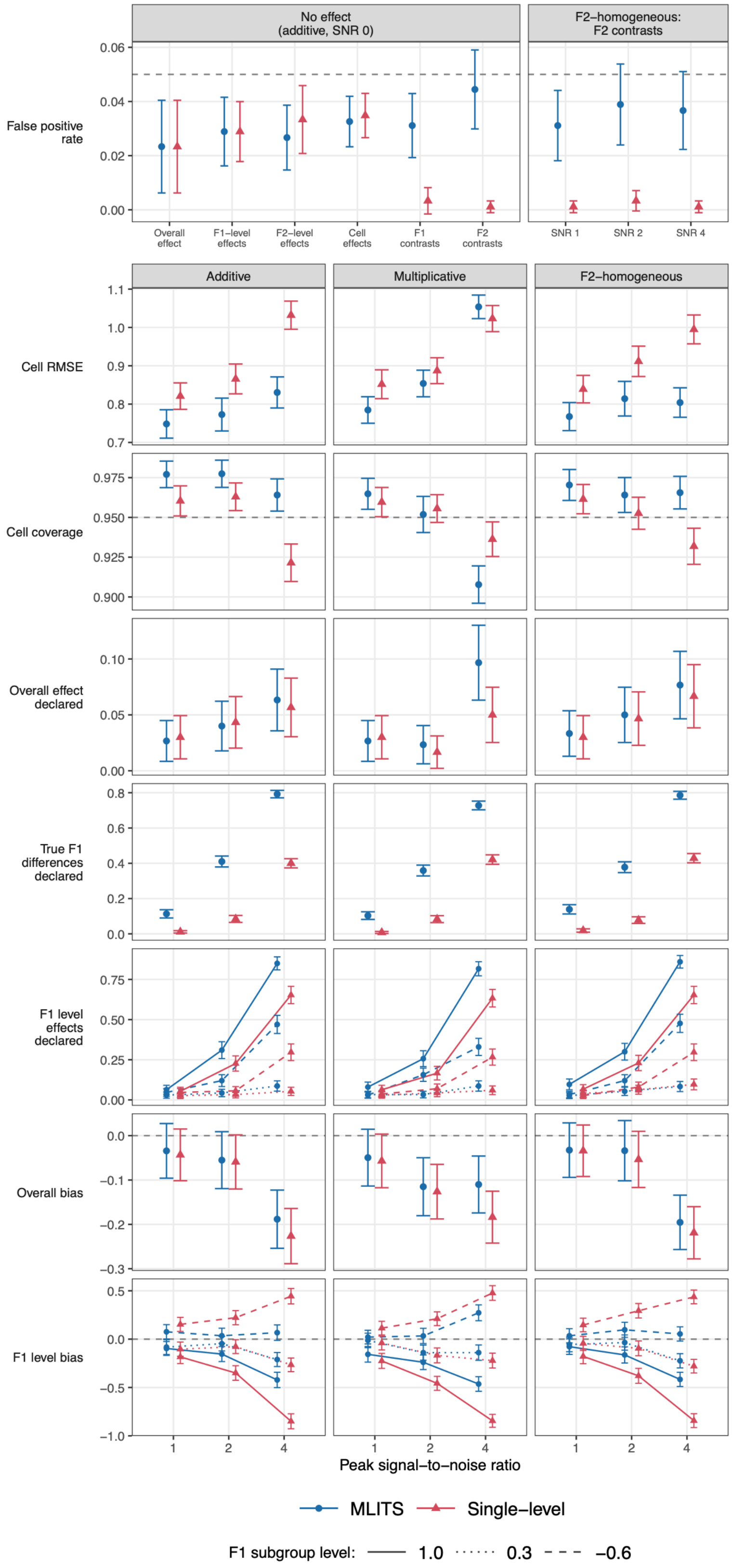

*Figure 1: Simulation study results*

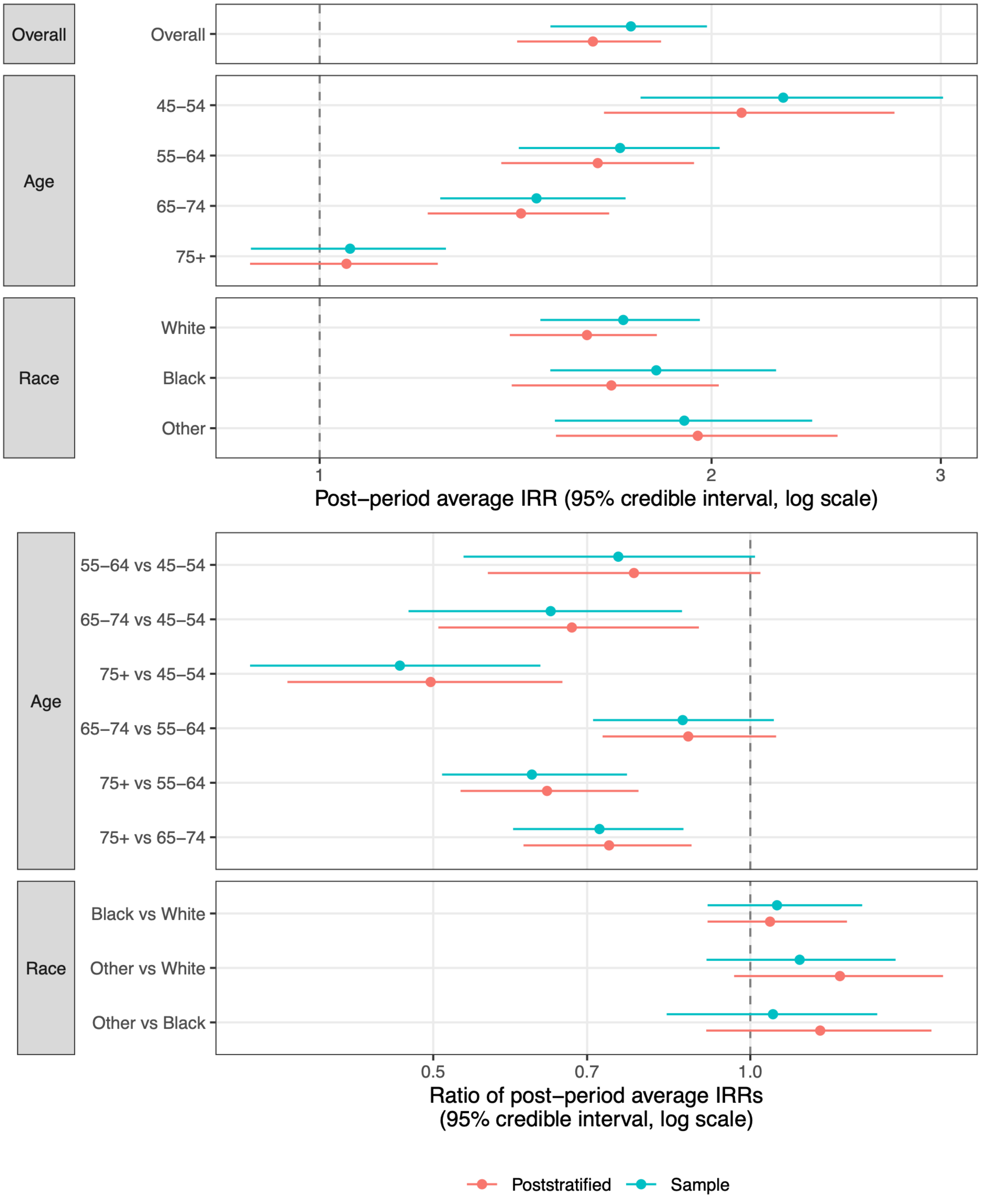

*Figure 2: Full post-period prostate cancer DX IRRs pre-to-post introduction of the PSA test, overall and by subgroup, presented for the observed sample and poststratified into the US male population ≥45 years of age.*

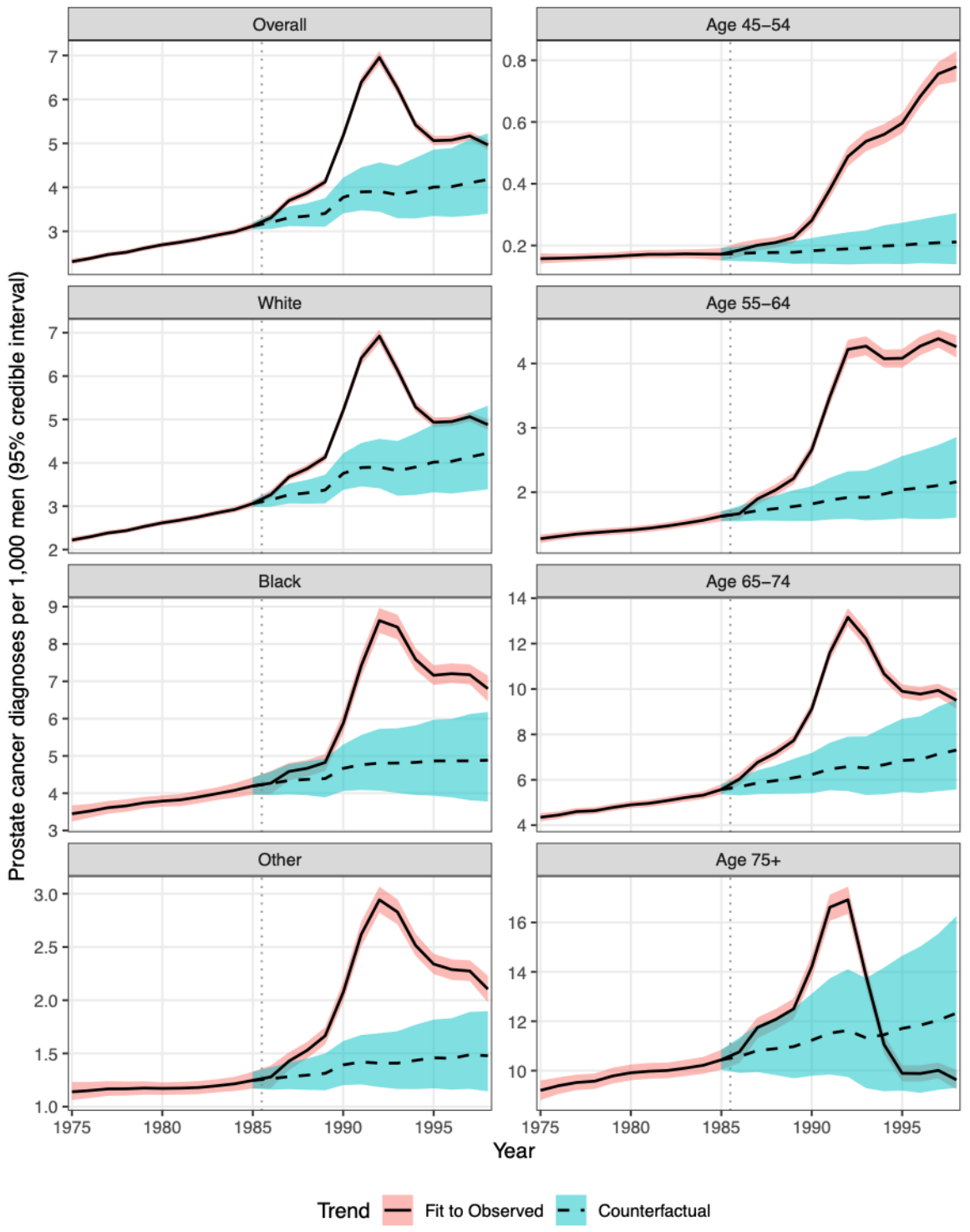

*Figure 3: Poststratified US male 45+ prostate cancer rate DX incidence predicted by model (red) and counterfactual if the PSA test had not been introduced (blue), overall, by race, and by age group.*

| Period | Contrast | IRR (95% CI) |
|---|---|---|
| Weeks 1-12 | Overall | 0.782, (0.757, 0.810) |
| | Urban | 0.780, (0.753, 0.809) |
| | Micro | 0.784, (0.755, 0.814) |
| | Rural | 0.805, (0.776, 0.838) |
| | Micro vs Urban | 1.005, (0.973, 1.038) |
| | Rural vs Urban | 1.033, (1.000, 1.067) |
| | Rural vs Micro | 1.028, (0.992, 1.064) |
| Full post-period | Overall | 0.890, (0.855, 0.929) |
| | Urban | 0.889, (0.854, 0.929) |
| | Micro | 0.890, (0.856, 0.931) |
| | Rural | 0.895, (0.860, 0.936) |
| | Micro vs Urban | 1.002, (0.981, 1.023) |
| | Rural vs Urban | 1.007, (0.985, 1.030) |
| | Rural vs Micro | 1.005, (0.983, 1.028) |
| Weeks 1-12 vs Weeks 13-44 | Overall | 0.840, (0.807, 0.876) |
| | Urban | 0.837, (0.802, 0.875) |
| | Micro | 0.841, (0.806, 0.880) |
| | Rural | 0.866, (0.828, 0.905) |

*Table 1: Post-pandemic onset (start 3/1/2020) vs pre-period IRRs for stroke hospitalization rates in early (first 12 weeks) and full post-period, given overall, by subgroup, and as pairwise IRRs of between subgroup changes..*

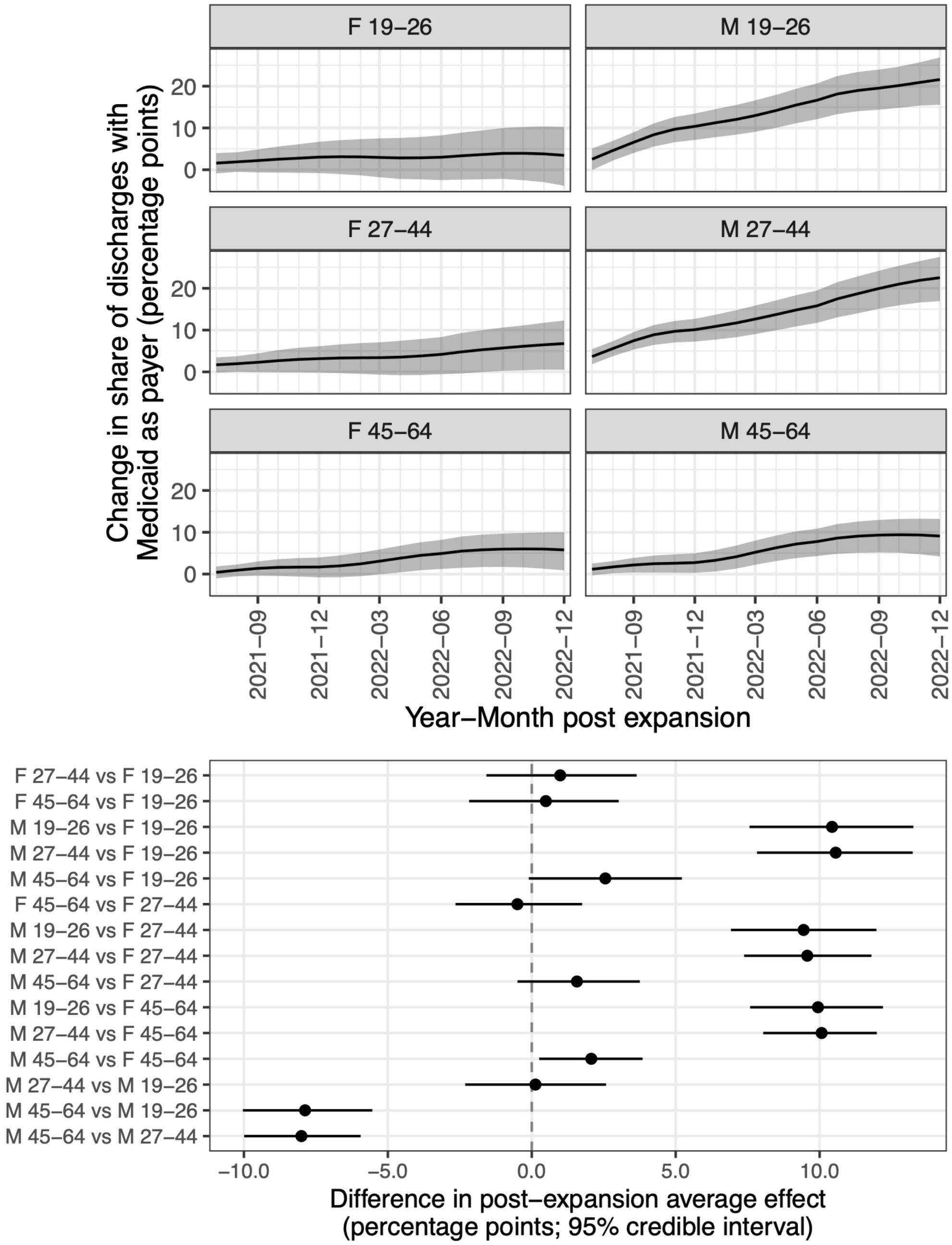

*Figure 3: Age-sex subgroup changes (in percentage points) of inpatient discharges with Medicaid as a primary payer, both presented over time within subgroup and as pairwise differences in changes averaged over the post-period.*

# Supplemental File

## A1. Notation and general model

Let $y_{st}$ be the outcome for series $s = 1, \dots, S$ at time $t = 1, \dots, T$, with the interruption at time $T_1$, and let $t \geq T_1$, $\tau_t = t - T_1$ be since interruption ($\tau = 0$ at onset). Each series has membership in exactly one level per grouping factor $j = 1, \dots, J$; factor $j$ has $K_j$ levels and $k_j(s)$ is the level of factor $j$ that series $s$ belongs to.

For a link function $f$,

$$f\big(E(y_{st})\big) = \eta_{st} = \mu_{st} + g(t, s),$$

where $\mu_{st}$ is the linear predictor without interruption and $g(t, s)$ is the interruption effect, where

$$\mu_{st} = \boldsymbol{x}_{st}^{\mathrm{T}}\boldsymbol{\beta} + \sum_{j=1}^{J} \theta_{j,k_j(s)} + \gamma_t + \sum_{j=1}^{J} \gamma_{j,k_j(s),t}\,,$$

$$\xi_0(\tau_t) = \omega_0 + \boldsymbol{b}(\tau_t)^{\mathrm{T}}\boldsymbol{\alpha}, \qquad \xi_{j,k_j(s)}(\tau_t) = \omega_{j,k_j(s)} + \boldsymbol{b}(\tau_t)^{\mathrm{T}}\boldsymbol{\alpha}_{j,k_j(s)}$$

$$g(t, s) = 1_{(t \geq T_1)}\left(\xi_0(\tau_t) + \sum_{j=1}^{J} \xi_{j,k_j(s)}\,(\tau_t)\right).$$

Here $1_{(t \geq T_1)}$ is the indicator function that takes the value 0 if $t < T_1$ and 1 if $t \geq T_1$, $\boldsymbol{\beta}$ holds time-invariant coefficients (an intercept, and in some applications a linear time term, seasonal terms or other covariates); $\theta_{jk}$ are random intercepts, $\gamma_t$ and $\gamma_{jkt}$ form the latent time-varying structure, $\omega_0$ and $\omega_{jk}$ are level shifts at the interruption, and $\xi_0$ and $\xi_{jk}$ are penalized splines in time since the interruption. Each spline averages zero over the post-period (A3), so the level shifts are average post-period effects on the link scale: $\omega_0 + \sum_j \omega_{j,k_j(s)}$ is the average effect for series $s$, and the splines describe how the effect varies around that average. We fit this model across all time points, where in the pre-period $\mu_{st}$ is the linear predictor, and in the post-post period, $\mu_{st}$ serves as counterfactual and $g(t, s)$ serves as the interruption attributable deviation from $\mu_{st}$. Where appropriate, such as situations where we fit a count data model with a log link function to series with different time at risk of outcome $y_{st}$, we also incorporate an offset term $O_{st}$ into $\mu_{st}$; we can also accommodate time invariant random slopes into the random effects structure in $\mu_{st}$, which we notate as $\zeta_{j,k_j(s)}$.

The main text and A7 write subgroup terms with the grouping factor as a superscript and the level as a bracketed subscript for ease of interpretation such that $\theta_{j,k_j(s)} = \theta_{f[s]}^{\mathrm{F}_j}$, and likewise for $\gamma$, $\omega$ and $\xi$. There, $\xi(t) = \omega_0 + \xi_0(\tau_t)$ and $\xi_{f[s]}^{\mathrm{F}}(t) = \omega_{f[s]}^{\mathrm{F}} + \xi_{f[s]}^{\mathrm{F}}(\tau_t)$ include the level shifts.

## A2. Latent time-varying structure: random walks

We use first-order random walks anchored at zero at $t = 1$ throughout this paper to describe existing trends over time such that

$$\gamma_1 = 0, \qquad \gamma_t = \gamma_{t-1} + \sigma_\gamma \, \varepsilon_t, \qquad \varepsilon_t \sim N(0,1),$$

for the overall trend, and subgroup specific random walks

$$\gamma_{jk1} = 0, \qquad \gamma_{jkt} = \gamma_{jk,t-1} + \sigma_{jk} \, \varepsilon_{jkt} - \frac{1}{K_j} \sum_{k'=1}^{K_j} \sigma_{jk'} \, \varepsilon_{jk't}, \qquad \varepsilon_{jkt} \sim N(0,1).$$

Subgroup innovations are centered across the levels of each factor at every time point such that $\sum_k \gamma_{jkt} = 0$ for all $t$ for identifiability.

In practice, users should carefully consider their choice of time-varying trend structure; other structures may better represent the processes users intend to model, and in some cases, may "tame" the post-interruption trajectories such that ascertainment of interruption effects is more efficient.

## A3. Interruption effect variability over time: penalized thin-plate splines

We characterize deviation from the overall interruption effect $\omega_0$ with $\xi(\tau) = \boldsymbol{b}(\tau)^\top \boldsymbol{\alpha}$, where $\boldsymbol{b}(\tau)$ is a thin-plate regression spline basis in time. In practice, we build this with `mgcv::jagam()[1]`, which is the construction `mvgam[2]`. The basis has two penalty matrices, $\boldsymbol{S}_w$ (wiggliness) and $\boldsymbol{S}_n$ (the null space of $\boldsymbol{S}_w$), with smoothing parameters $\lambda_w$ and $\lambda_n$. We set a penalized-spline prior on the coefficients[3] with penalties entering the precision matrix in a multivariate normal structure with

$$\boldsymbol{\alpha} \sim N(\boldsymbol{0}, \ (\lambda_w \boldsymbol{S}_w + \lambda_n \boldsymbol{S}_n)^{-1}).$$

This is a proper prior because $\boldsymbol{S}_n$ penalizes the null space of $\boldsymbol{S}_w$, meaning that the sum $\lambda_w \boldsymbol{S}_w + \lambda_n \boldsymbol{S}_n$ is full rank for $\lambda_w, \lambda_n > 0$.

**Rotation.** We facilitate more efficient sampling by rotating the above spline basis. Let $\boldsymbol{U}$ hold the eigenvectors of $\boldsymbol{S}_w$. $\boldsymbol{U}$ diagonalizes both penalties, $\boldsymbol{U}^\top \boldsymbol{S}_w \boldsymbol{U} = \mathrm{diag}(\boldsymbol{d}_w)$ and $\boldsymbol{U}^\top \boldsymbol{S}_n \boldsymbol{U} = \mathrm{diag}(\boldsymbol{d}_n)$ For $\boldsymbol{\alpha} = \boldsymbol{U}\tilde{\boldsymbol{\alpha}}$,

$$\tilde{\alpha}_k = \frac{z_k}{\sqrt{\lambda_w d_{w,k} + \lambda_n d_{n,k}}}, \qquad z_k \sim N(0,1),$$

which facilitates use of the same prior. This is the mixed-model reparameterization of a penalized smooth used by `mgcv::smooth2random[4]`. This increases sampling efficiency and in practice, where necessary, we report the unrotated spline coefficients after back-transformation.

**Subgroup smooths.** We repeat the above process for each level $k$ of factor $j$, where each of these smooths have their own estimated coefficients and spline coefficients $\left(\boldsymbol{\alpha}_{jk}, \lambda_{w,jk}, \lambda_{n,jk}\right)$. The subgroup coefficient vectors are centered across levels coefficient by coefficient (A4), so $\sum_k \xi_{jk}(\tau) = 0$ for every $\tau$.

| Analysis | Basis | $K$ | Post-period length |
|---|---|---|---|
| Simulation | `s(t_since_int)` (default $k = 10$) | 9 | 20 time points |
| SEER | `s(time, k = 7)` | 6 | 13 years |
| Medicaid | `s(time, k = 7)` | 6 | 18 months |
| Stroke | thin-plate basis built outside the VRDC | 10 | 44 weeks |

# A4. Sum-to-zero constraints

We facilitate identifiability in our models that estimate overall and subgroup within factor effects, such as the subgroup random intercepts, random-walk innovations (at each $t$), level shifts and spline coefficients by specifying sum to zero constraints across subgroup levels within each factor. We achieve this by centering: a term with $K$ levels is specified as $a_k = \sigma_{a_k}(z_k - \bar{z})$ where $z_k \sim N(0,1)$ and $\bar{z} = \sum_{k=1}^{K} z_k / K$, and $\sigma_{a_k}$ is the standard deviation for $a_k$. Centering reduces the marginal variance by a factor $(K-1)/K$, so terms with a fixed prior SD (the level shifts) are multiplied by $\sqrt{K/(K-1)}$ to keep their stated marginal SD; terms with an estimated SD are not rescaled. For a factor nested within another, centering is within each level of the outer factor.

In our work, we do not use the above in the Medicare stroke model; this model uses soft sum-to-zero constraints instead, where for any subgroup term term $a_k$, $\sum_k a_k \sim N(0, \psi)$ (parameterized using scale = SD) with $\psi = 0.03$ (A7).

# A5. Prior structure

Notation: $N^+(m, s)$ (parameterized using scale = SD) is a normal distribution truncated to positive values; $t_\nu(m, s)$ a Student-t with $\nu$ degrees of freedom, location $m$ and scale $s$.

**Level shifts: data-scaled.** We give overall and subgroup prior SD $c \cdot s_{\text{pre}}$ with $c = 2.5$, where $s_{\text{pre}}$ is the pooled within-subgroup (or overall) SD of the pre-interruption outcome on the scale of the linear predictor: $y$ (Gaussian), $\log(y + 0.5) - o$ (counts) and the empirical logit $\log\{(y + 0.5)/(n - y + 0.5)\}$ (binomial).

**Variance components.** Half-normal priors on all SDs: random intercepts $N^+(0,1)$; overall random-walk SD $\sigma_\gamma \sim N^+(0,0.5)$; subgroup random-walk SDs $N^+(0,0.25)$; SEER random time slopes (per decade) $N^+(0,0.5)$. With two to four levels per factor, the data carry little information about the subgroup SDs, so these priors matter[5]; we intentionally specified subgroup random walk innovation SDs smaller to shrink to the overall random walk trajectory. This choice is evaluated in sensitivity analyses (A7.4, A10.8).

**Smoothing parameters.** $\lambda_w, \lambda_n \sim N^+(5,10)$ (simulation, SEER) or $N^+(5,3)$ (Medicaid, stroke). The prior is on the scale of the penalty matrices returned by `jagam` which rescales each penalty matrix to a size comparable to $\boldsymbol{X}^\top \boldsymbol{X}$ for its basis, so $\lambda$ is relative to that scaling rather than to the raw basis[1].

**Fixed effects.** Intercepts $t_3(m_0, s_0)$ with a location set from the data where noted; other coefficients as listed per model.

**Sensitivity.** Simulation: shift SD 1, 10 and data-scaled; overall walk SD scale 1.5; subgroup walk SD scales 0.1 and 1; smoothing $N^+(1,5)$ (A10.8, Figure S13). SEER: subgroup walk SD scales 0.1 and 1, smoothing $N^+(1,5)$, shift prior four times wider (A7.4, Figure S2).

# A6. Simulation models

## A6.1 Gaussian outcome (main simulation)

Two crossed factors with three levels each ($J = 2, K_1 = K_2 = 3, S = 9$ cells), $T = 60, T_1 = 41$:

$$y_{st} \sim N(\eta_{st}, \sigma_s), \qquad \eta_{st} = \beta_0 + \theta_{1,k_1(s)} + \theta_{2,k_2(s)} + \gamma_t + \gamma_{1,k_1(s),t} + \gamma_{2,k_2(s),t} + g(t,s),$$

with $g(t,s)$ as in A1 (overall plus both factors). Priors: $\beta_0 \sim t_3(0,2.5)$; $\sigma_s \sim t_5^+(0,1)$ (one observation SD per series); random-intercept SDs $N^+(0,1)$; walks and smooths as in A2, A3 and A5.

## A6.2 Single-level comparator

We fit a model to each "aggregated" series at the level of inference desired (each cell, each level of each factor, and the overall series), with no subgroup terms such that $\mu_{st} = \beta_0 + \gamma_t$ and $g(t,s) = g(t)$[6], so

$$y_t \sim N(\beta_0 + \gamma_t + g(t), \sigma), \qquad g(t) = \omega_0 + \xi_0(\tau_t) \text{ for } t \geq T_1.$$

Priors are scaled to the series' own pre-period data, with $s_\Delta$ the SD of its pre-period first differences: $\sigma \sim t_5^+\left(0, s_\Delta/\sqrt{2}\right)$, $\sigma_\gamma \sim N^+(0, s_\Delta/2)$, $\omega_0 \sim N\left(0{,}2.5 \cdot \mathrm{SD}\left(y_{\mathrm{pre}}\right)\right)$, $\beta_0 \sim t_3(0,2.5)$, smoothing as in A5. The random walk is non-centered.

## A6.3 Count outcome (simulation extension)

Same design, with a negative binomial likelihood on the log scale with mean and overdispersion parameterization such that

$$y_{st} \sim \mathrm{NB}(e^{\eta_{st}}, \phi_s), \qquad \mathrm{Var}(y_{st}) = e^{\eta_{st}} + e^{2\eta_{st}}/\phi_s.$$

Differences from A6.1:

- we only fit this for the multilevel model

- one intercept per series, $\beta_{0s} \sim N(\log\bar{y}_{s,\text{pre}}, 1)$, instead of additive random intercepts (cell baselines are not additive in the factors);
- series-specific dispersion partially pooled toward the Poisson: $\phi_s = 1/(\kappa_s^2 + 10^{-4})$, $\kappa_s = \tau_\kappa\,\tilde{\kappa}_s, \tilde{\kappa}_s \sim N^+(0,1), \tau_\kappa \sim N^+(0,0.5)$ (the $10^{-4}$ caps $\phi$ at $10^4$, effectively Poisson);
- the walk innovations and spline coefficients are multiplied by a data-based noise unit $u = (\text{mean pre-period count})^{-1/2}$, so the walk and smoothing priors have the same meaning as in the Gaussian model with noise SD 1;
- shift prior SD 2.5 × the pooled within-series SD of pre-period $\log(y + 0.5)$.

# A7. Application models

Notation follows the main text: a superscript names the grouping factor a term belongs to and a bracketed subscript maps a series to its level of that factor ($a[s]$ is the age group of series $s$, $r[s]$ its race or rurality level) [ref: Gao et al. 2021]. Terms without a superscript are overall. $\xi(t)$ and $\xi^{\text{F}}_{f[s]}(t)$ include the level shifts $\omega_0$ and $\omega^{\text{F}}_{f[s]}$ (A1, A3), $g(t,s) = 0$ for $t < T_1$, and the rotated parameterization used for sampling (A3) is omitted, as it does not change the model.

## A7.1 SEER: PSA testing and prostate cancer incidence (negative binomial)

Annual counts of incident prostate cancer, 1975-1998 ($T = 24$; $T_1 = 12$, 1986), for $S = 12$ series: age group $a \in$ {45-54, 55-64, 65-74, 75+} and race $r \in$ {White, Black, Other} within age group. Race is nested in age (there is no race main effect); "Race(Age)" terms are centered within each age group.

$$y_{st} \sim \text{NB}\big(\exp\big(\log(\text{pop}_{st}/1000) + \mu_{st} + g(t,s)\big),\ \phi\big)$$

$$\mu_{st} = \beta_0 + \beta_1 t + \theta^{\text{Age}}_{a[s]} + \theta^{\text{Race(Age)}}_{r[s]} + \left(\zeta^{\text{Age}}_{a[s]} + \zeta^{\text{Race(Age)}}_{r[s]}\right) t_{\text{dec}} + \gamma_t + \gamma^{\text{Age}}_{a[s],t} + \gamma^{\text{Race(Age)}}_{r[s],t}$$

$$g(t,s) = 1_{(t \geq T_1)}\left(\xi(\tau_t) + \xi^{\text{Age}}_{a[s]}(\tau_t) + \xi^{\text{Race(Age)}}_{r[s]}(\tau_t)\right),$$

where $t_{\text{dec}}$ is time in decades centered at the middle of the series, $\theta$ are random intercepts and $\zeta$ random time slopes.

| Parameter | Prior |
|---|---|
| $\beta_0$; $\beta_1$ (per year) | $t_3(m_0, 0.5)$, $m_0$ the mean empirical log rate at $t = 1$; $t_3(0,0.5)$ |
| $\theta^{\text{Age}}_a$; $\theta^{\text{Race(Age)}}_r$ | $N(0, \sigma^{\text{Age}}_\theta)$; $N\big(0, \sigma^{\text{Race(Age)}}_\theta\big)$, each SD $N^+(0,1)$ |

| Parameter | Prior |
|---|---|
| $\zeta_a^{\text{Age}}$; $\zeta_r^{\text{Race(Age)}}$ | $N\left(0,\sigma_\zeta^{\text{Age}}\right)$; $N\left(0,\sigma_\zeta^{\text{Race(Age)}}\right)$, each SD $N^+(0,0.5)$ |
| $\sigma_\gamma$ | $N^+(0,0.5)$ |
| $\sigma_{\gamma,a}^{\text{Age}}$ (one per age group); $\sigma_\gamma^{\text{Race(Age)}}$ (shared) | $N^+(0,0.25)$ |
| $\omega_0$, $\omega_a^{\text{Age}}$, $\omega_r^{\text{Race(Age)}}$ | $N(0,2.5\, s_{\text{pre}})$ |
| $\lambda_w, \lambda_n$: overall; one pair per age group; one shared pair for race within age | $N^+(5,10)$ |
| $\phi$ | $\phi = \sigma_d^{-1/2}$, $\sigma_d \sim N^+(0,3)$ |

Repository version: with the publicly shared data, counts of 1-4 are suppressed (7 of 288 cells) and enter as interval-censored, $\log\Pr(1 \le y_{st} \le 4) = \log\sum_{k=1}^{4} \mathrm{NB}\,(k \mid e^{\eta_{st}}, \phi)$. We fit the version presented in the paper with full data.

## A7.2 Medicare stroke/TIA hospitalizations during COVID-19 (negative binomial)

Weekly counts of stroke/TIA hospitalization episodes from the week of January 7, 2018 through 2020 ($T = 156$; $T_1 = 113$, the week of March 1, 2020; 44 post-period weeks), for $S = 3$ series by rurality $r \in$ {metropolitan, micropolitan, rural}, so $r[s] = s$.

$$y_{st} \sim \mathrm{NB}\big(\exp\big(\log(\text{person-days}_{st}) + \mu_{st} + g(t,s)\big),\ \phi\big)$$

$$\mu_{st} = \beta_0 + \beta_1\cos(2\pi d_t/365) + \beta_2\sin(2\pi d_t/365) + \theta_{r[s]}^{\text{Rurality}} + \gamma_t + \gamma_{r[s],t}^{\text{Rurality}}$$

$$g(t,s) = 1_{(t\ge T_1)}\left(\xi(\tau_t) + \xi_{r[s]}^{\text{Rurality}}(\tau_t)\right),$$

where $d_t$ is the day of the year. This application uses an unrotated spline basis with $\boldsymbol{\alpha} \sim N(\mathbf{0}, (\lambda_w \boldsymbol{S}_w + \lambda_n \boldsymbol{S}_n)^{-1})$ sampled directly, and soft sum-to-zero constraints, $\sum_k a_k \sim N(0, 0.03^2)$, on $\theta$, $\gamma^{\text{Rurality}}$ at each $t$, $\omega^{\text{Rurality}}$ and each spline coefficient (A4). The rotated spline basis was added upon revision to facilitate more efficient sampling, and other applications were changed to match, however, given the strict guidelines for data uploads and downloads in the CMS VRDC we elected to retain the same parameterization for this model.

| Parameter | Prior |
|---|---|
| $\beta_0$ | $t_3(m_0, 2.5)$ |
| $\beta_1, \beta_2$ | $t_3(0,2.5)$ |
| $\theta_r^{\text{Rurality}}$ | $N(0,\sigma_\theta^{\text{Rurality}})$, $\sigma_\theta^{\text{Rurality}} \sim N^+(0,1)$ |
| $\sigma_\gamma$; $\sigma_{\gamma,r}^{\text{Rurality}}$ - one per level | $N^+(0,0.5)$; $N^+(0,0.25)$ |
| $\omega_0$; $\omega_r^{\text{Rurality}}$ | $N(0,2)$; $N(0,1)$ |

| Parameter | Prior |
| --- | --- |
| $\lambda_w, \lambda_n$: overall; one pair per rurality level | $N^+(5,3)$ |
| $\phi$ | Gamma(2,0.1) (shape, rate) |

Posterior predictive draws use the gamma-Poisson representation of the NB2 [$\lambda \sim \mathrm{Gamma}(\phi, \phi)$, $y \sim \mathrm{Poisson}(e^{\eta}\lambda)$], which avoids numerical overflow in the NB random number generator in Stan[6]; this does not affect the fit or the effect estimates.

**Hospitalization episodes.** Claims are payment records rather than clinical episodes: a single hospitalization can appear as several claims, for example an emergency department claim followed by a transfer to an inpatient setting (inpatients cannot be transferred to an outpatient setting). We grouped claims into episodes with the algorithm in Figure S1, which assigns each episode an identifier (CE_ID) so that transfers, interim claims and within-hospital transfers are not counted twice. An episode is a stroke/TIA hospitalization if any diagnosis position on any of its claims carries one of the codes in Table S1.

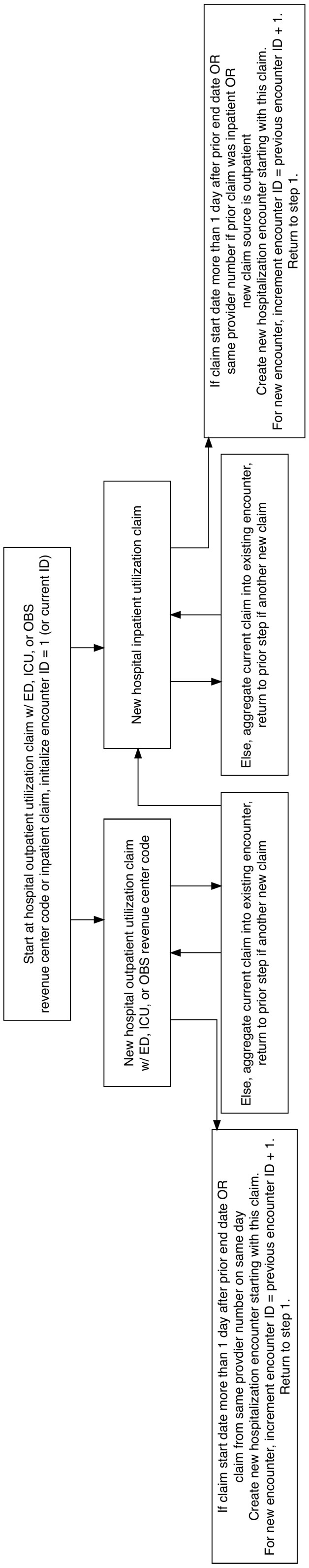

*Figure S1. Algorithm for grouping claims into hospitalization episodes (rotated; read with the page turned clockwise).*

*Table S1. ICD-10 diagnosis codes identifying stroke or transient ischemic attack; one of these codes in any position qualifies a hospitalization episode.*

| Diagnosis | ICD-10 codes |
|---|---|
| Ischemic stroke | I6300, I63011, I63012, I63013, I63019, I6302, I63031, I63032, I63033, I63039, I6309, I6310, I63111, I63112, I63113, I63119, I6312, I63131, I63132, I63133, I63139, I6319, I6320, I63211, I63212, I63213, I63219, I6322, I63231, I63232, I63233, I63239, I6329, I6330, I63311, I63312, I63313, I63319, I63321, I63322, I63323, I63329, I63331, I63332, I63333, I63339, I63341, I63342, I63343, I63349, I6339, I6340, I63411, I63412, I63413, I63419, I63421, I63422, I63423, I63429, I63431, I63432, I63433, I63439, I63441, I63442, I63443, I63449, I6349, I6350, I63511, I63512, I63513, I63519, I63521, I63522, I63523, I63529, I63531, I63532, I63533, I63539, I63541, I63542, I63543, I63549, I6359, I636, I638, I6381, I6389, I639, I6789 |
| Transient ischemic attack | G450, G451, G452, G453, G458, G459 |

## A7.3 Missouri Medicaid expansion (binomial)

Monthly inpatient discharges among adults aged 19-64, January 2016-December 2022 ($T = 84$; $T_1 = 67$, July 2021; 18 post-period months), for $S = 12$ series: age-sex group $a \in$ {female, male} × {19-26, 27-44, 45-64} and rurality $r \in$ {rural, urban}. Of $n_{st}$ discharges, $y_{st}$ had Medicaid as the primary payer. PHE indicates the COVID-19 public health emergency (from February 2020).

$$y_{st} \sim \text{Binomial}\left(n_{st},\ \text{logit}^{-1}\big(\mu_{st} + g(t,s)\big)\right)$$

$$\mu_{st} = \beta_0 + \beta_1 \text{PHE} + \theta_s^{\text{S}} + \zeta_s^{\text{S}} \text{PHE} + \gamma_t + \gamma_{st}$$

$$g(t,s) = 1_{(t \geq T_1)} \left( \xi(\tau_t) + \xi_{a[s]}^{\text{Age-sex}}(\tau_t) + \xi_{r[s]}^{\text{Rurality}}(\tau_t) \right).$$

The latent structure has one overall walk $\gamma_t$ and one walk per series $\gamma_{st}$ with a common SD $\sigma_{s\gamma}$, centered across the 12 series ($\sum_s \gamma_{st} = 0$). The series specific deviations ($\theta$ and $\zeta$ terms) let the intercept and the PHE level shift vary by series and are centered across groups; rurality has two levels and enters through fixed effects. The effect structure (age-sex and rurality) is nested in the series-level latent structure.

| Parameter | Prior |
|---|---|
| $\beta_0$ | $N(\mu_0, 1)$, where $\mu_0$ is the pre PHE log odds |
| $\beta_1, \beta_2, \beta_3$ | $N(0{,}0.5)$ |
| $\theta_s^{\text{S}}$ | normal with SDs $N^+(0{,}0.26)$; |
| $\zeta_s^{\text{S}}$ | normal with SDs $N^+(0{,}0.26)$; |

| Parameter | Prior |
|---|---|
| $\sigma_\gamma$; $\sigma_{s\gamma}$ (shared) | $N^+(0{,}0.26)$; $N^+(0{,}0.13)$ |
| $\omega_0, \omega_a^{\text{Age-sex}}, \omega_r^{\text{Rurality}}$ | $N(0, s_{\text{pre}})$, $s_{\text{pre}} = 0.258$ (empirical logit) |
| $\lambda_w, \lambda_n$: overall; one pair per age-sex group; one pair per rurality level | $N^+(10{,}5)$ |

## A7.4 Supplementary application results

**SEER prior sensitivity.** We refit the SEER model with one prior changed at a time: subgroup random-walk SD scale 0.1 and 1 (main model 0.25), smoothing $N^+(1{,}5)$ (main model $N^+(5{,}10)$) and a shift prior four times wider. Post-period IRRs, sample and poststratified, overall and for every subgroup with 95% credible intervals (CI), changed by at most 0.01 for the random-walk and smoothing variants and 0.03 for the wider shift prior (Figure S2). No variant had divergent transitions. Maximum $\hat{R}$ was 1.009-1.014, except 1.021 (minimum bulk ESS 314) with the tight subgroup random walk, and the wider shift prior reached the maximum treedepth in 1,898 of 2,000 iterations, suggesting less efficient sampling.

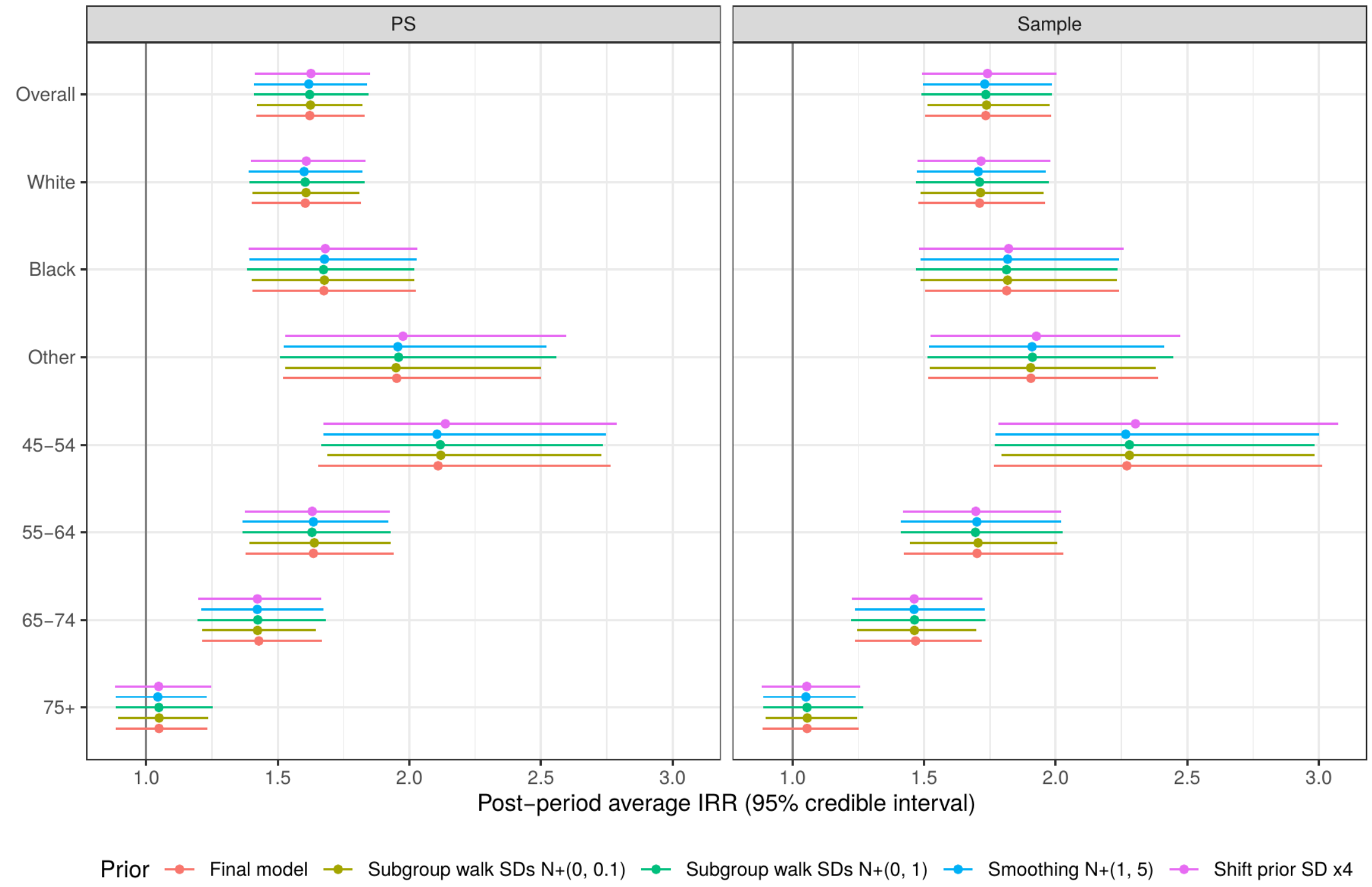

*Figure S2. SEER post-period IRRs (95% credible intervals) under the main model and four prior changes, poststratified (left) and sample (right).*

**Medicaid by rurality.** Figure S3 shows the expansion effect over time by rurality. Averaged over the post-period, the increase in the share of discharges with Medicaid as payer was 1.9 percentage points smaller in rural than in urban areas (95% CI 0.6 to 3.2; ratio of odds ratios 0.92, 0.86 to 0.99).

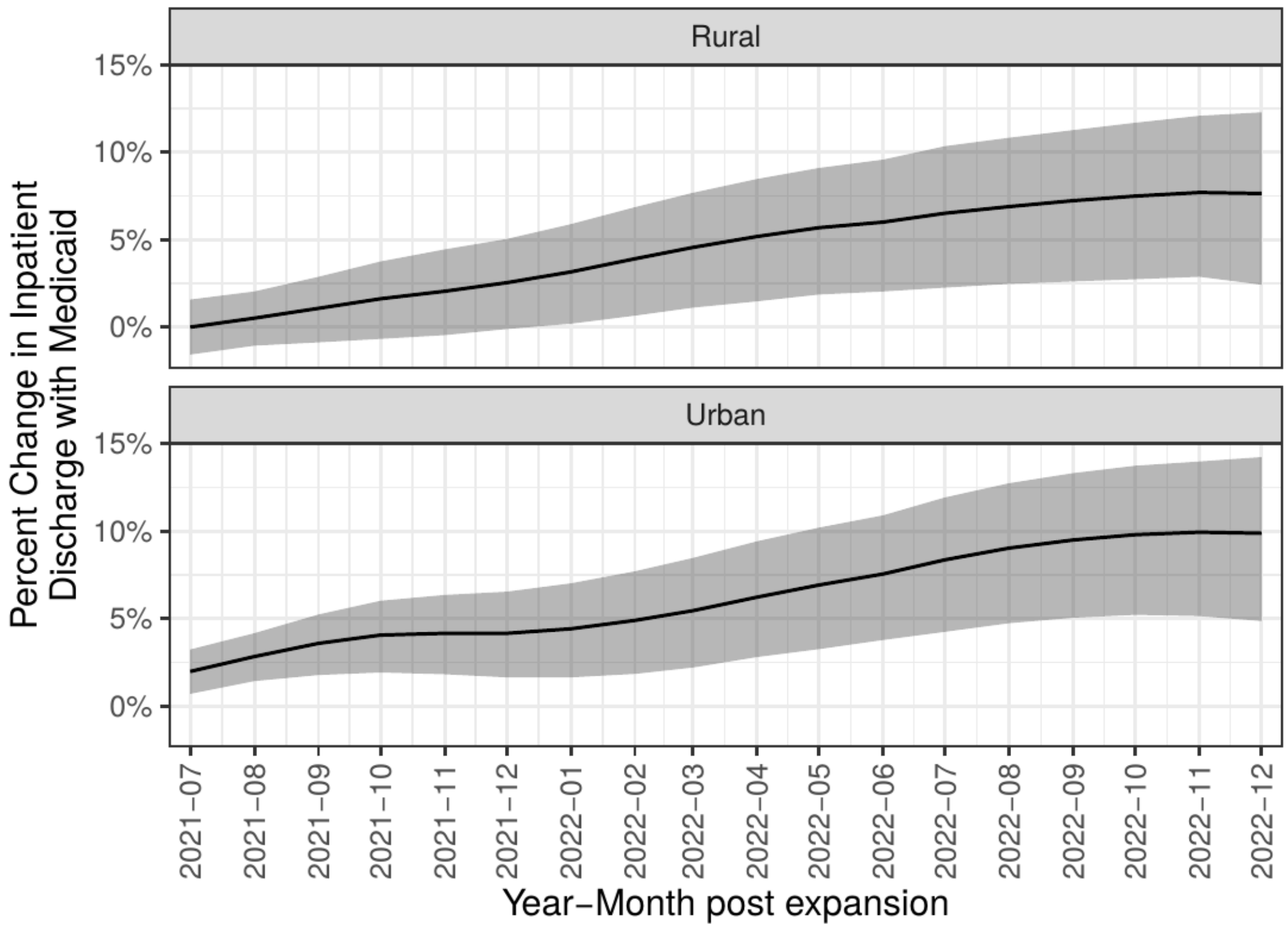

*Figure S3. Effect of Medicaid expansion on the share of inpatient discharges with Medicaid as primary payer, by rurality (95% credible intervals).*

# A8. Effect summaries

For draw $m$ of the posterior, the pointwise effect is $g^{(m)}(t,s) = \eta_{st}^{(m)} - \mu_{st}^{(m)} - o_{st}$ on the link scale (log incidence rate ratio or log odds ratio), and on the outcome scale the difference between expected values with and without the interruption (e.g. $p_{st}^{(m)} - p_{st}^{(m),\mathrm{cf}}$ for Medicaid). Subgroup and overall summaries average over a set of series $G$ and post-period times $\mathcal{T} \subseteq \{T_1, \dots, T\}$ with weights $w_{st}$ (sample or population weights), per draw:

$$\bar{g}_{G,\mathcal{T}}^{(m)} = \frac{\sum_{s\in G} \sum_{t\in\mathcal{T}} w_{st}\ g^{(m)}(t,s)}{\sum_{s\in G} \sum_{t\in\mathcal{T}} w_{st}}.$$

On the ratio scale, $\exp(\bar{g})$ is the weighted geometric mean of the pointwise IRRs or ORs. Poststratified SEER estimates use US male population weights by year, age group and race.

For stroke, the IRR over a window $\mathcal{T}$ (weeks 1-12, weeks 13-44 or the full post-period) is instead the ratio of expected counts with and without the interruption, summed over the window:

$$\mathrm{IRR}_{G,\mathcal{T}}^{(m)} = \frac{\sum_{s\in G}\sum_{t\in\mathcal{T}} \exp\{o_{st} + \mu_{st}^{(m)} + g^{(m)}(t,s)\}}{\sum_{s\in G}\sum_{t\in\mathcal{T}} \exp\{o_{st} + \mu_{st}^{(m)}\}},$$

which weights each week by its expected count. The same estimand is used in the count simulation (A10.10).

Subgroup contrasts are pairwise within each grouping factor: for levels A and B, "B vs A" is computed per draw as the difference between their summaries (a ratio on the IRR or OR scale; a difference in percentage points for Medicaid), with 95% credible intervals from posterior quantiles. For stroke, contrasts between rurality levels and between windows (weeks 1-12 vs 13-44) are ratios of the IRRs above.

# A9. Computation

All models were fitted in Stan via rstan 2.32.7; R 4.6.0 for the simulation image, SEER, and Medicaid example, and R version 4.4.1 for the Medicare Stroke hospitalization example. Simulation Stan sampler control settings were: simulation, 4 chains × 2,000 iterations (1,000 warm-up), adapt_delta 0.99, max treedepth 12; SEER, 4 × 2,000 (1,500 warm-up), adapt_delta 0.95, max treedepth 12; Medicaid, 4 × 5,000 (4,000 warm-up), adapt_delta 0.99, max treedepth 12; stroke, 4 × 4,000 (3,000 warm-up), adapt_delta 0.999, max treedepth 12. Convergence was assessed with rank-normalized $\hat{R}$, bulk ESS, divergent transitions, treedepth saturation and E-BFMI[7]. Code: https://doi.org/10.5281/zenodo.24178292.

## A9.1 Sampler diagnostics (applications)

*Table S2. Sampler diagnostics for the three application fits. "All" covers every sampled parameter (excluding posterior predictive and other generated quantities); effect parameters are the level shifts and spline coefficients, overall and by subgroup.*

| | SEER | Stroke | Medicaid |
|---|---|---|---|
| Post-warm-up draws | 4 × 500 | 4 × 1,000 | 4 × 1,000 |
| Divergent transitions | 0 | 0 | 0 |
| Draws at maximum treedepth | 0 | 0 | 0 |
| Minimum E-BFMI | 0.84 | 0.94 | 0.78 |
| Maximum $\hat{R}$: all / effect parameters | 1.009 / 1.003 | 1.007 / 1.005 | 1.006 / 1.001 |
| Minimum bulk ESS: all / effect parameters | 422 / 920 | 676 / 676 | 783 / 2,297 |

Treedepth saturation in the Medicaid fit reduces sampling efficiency but not validity; $\hat{R}$ and ESS were adequate for every parameter.

## A9.2 Posterior predictive checks and residual autocorrelation

We compute a randomized quantile residuals from the posterior predictive draws of the observed outcome, $r_{st} = \Phi^{-1}(u_{st})$ with $u_{st} = \Pr\left(y_{st}^{\text{rep}} < y_{st}\right) + U\Pr\left(y_{st}^{\text{rep}} = y_{st}\right), U \sim$

Uniform(0,1)[8], which follow (approximately) a standard normal distribution if the model is appropriate for the data. We use the randomized quantile residuals to generate QQ-plots as well as compute and plot autocorrelation of the residuals up to lag 8, and finally perform Ljung-Box tests[8]. We additionally report the share of observations inside the 50% and 95% posterior predictive intervals.

- **SEER** (Figure S4): 98% of pre-period and 99% of post-period observations fell inside the 95% predictive intervals (67% inside the 50% intervals); residual SDs were 0.40-1.11. The Ljung-Box p-value was below 0.05 in 1 of 12 series (White, 65-74; p = 0.04), and the lag-1 autocorrelation was outside $\pm 1.96/\sqrt{n}$ in 2 of 12, both negative.
- **Stroke** (Figure S5): 99% of pre-period and 100% of post-period observations inside the 95% intervals (71% inside the 50% intervals); residual SDs 0.48-0.77; no Ljung-Box p-value below 0.05 and no lag-1 autocorrelation outside the bounds.
- **Medicaid** (Figure S6): 98% of pre-period and 99% of post-period observations inside the 95% intervals (63% inside the 50% intervals); residual SDs 0.59-1.03. The Ljung-Box p-value was below 0.05 in 2 of 12 series (rural women 27-44, p = 0.01, urban women 19-26 p=0.03), and the lag-1 autocorrelation was outside the bounds in 3 of 12, all negative (rural women 26-44, -0.22; rural men 19-26, -0.29; rural men 45-64, -0.22).

Across the 27 series, 3 Ljung-Box tests had $p < 0.05$, against about 1.4 expected by chance, and every lag-1 autocorrelation outside the bounds was negative, which points to the random walks absorbing slightly too much short-term variation rather than leaving positive serial dependence. Residual SDs mostly below or very near 1 and 95% predictive coverage of 98-100% (Figure S7) indicate predictive intervals that are, if anything, wide.

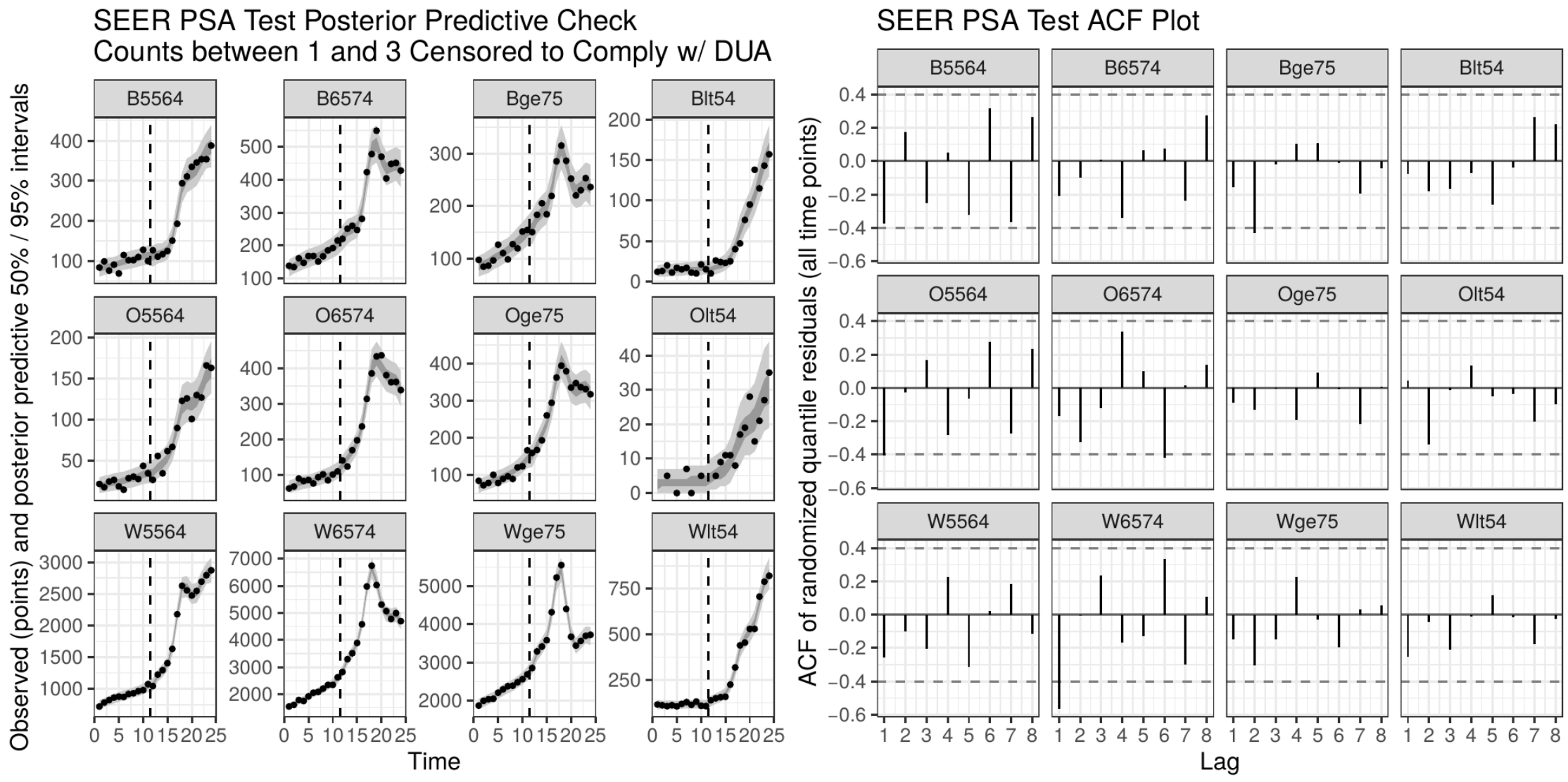

*Figure S4. SEER: observed counts with 50% and 95% posterior predictive intervals (left; the dashed line marks the interruption) and autocorrelation of randomized quantile residuals*

*(right; dashed lines $\pm 1.96/\sqrt{n}$). Small counts are censored in the plot to comply with the data use agreement.*

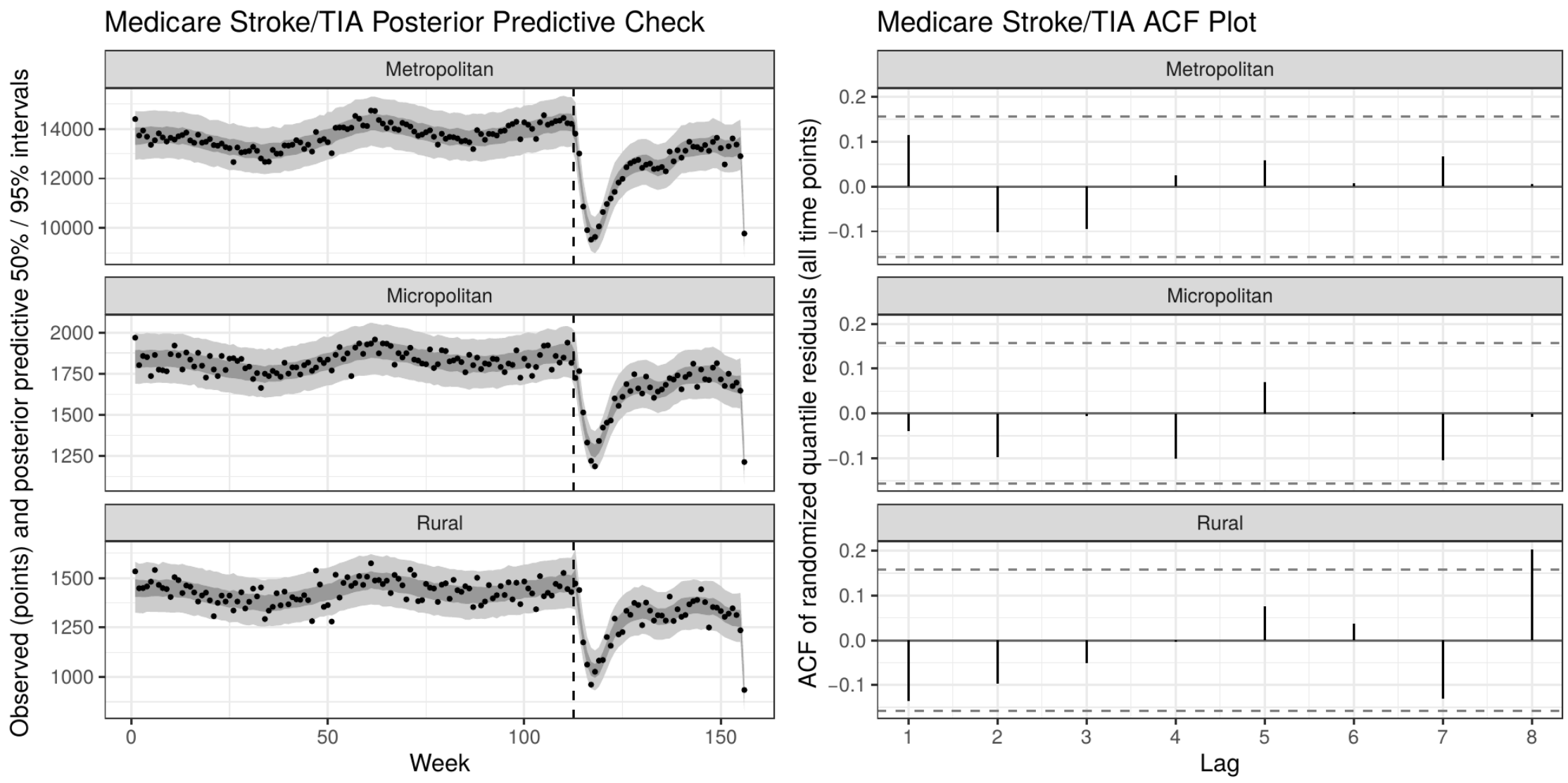

*Figure S5. Medicare stroke/TIA: posterior predictive check (left) and residual autocorrelation (right), as in Figure S4.*

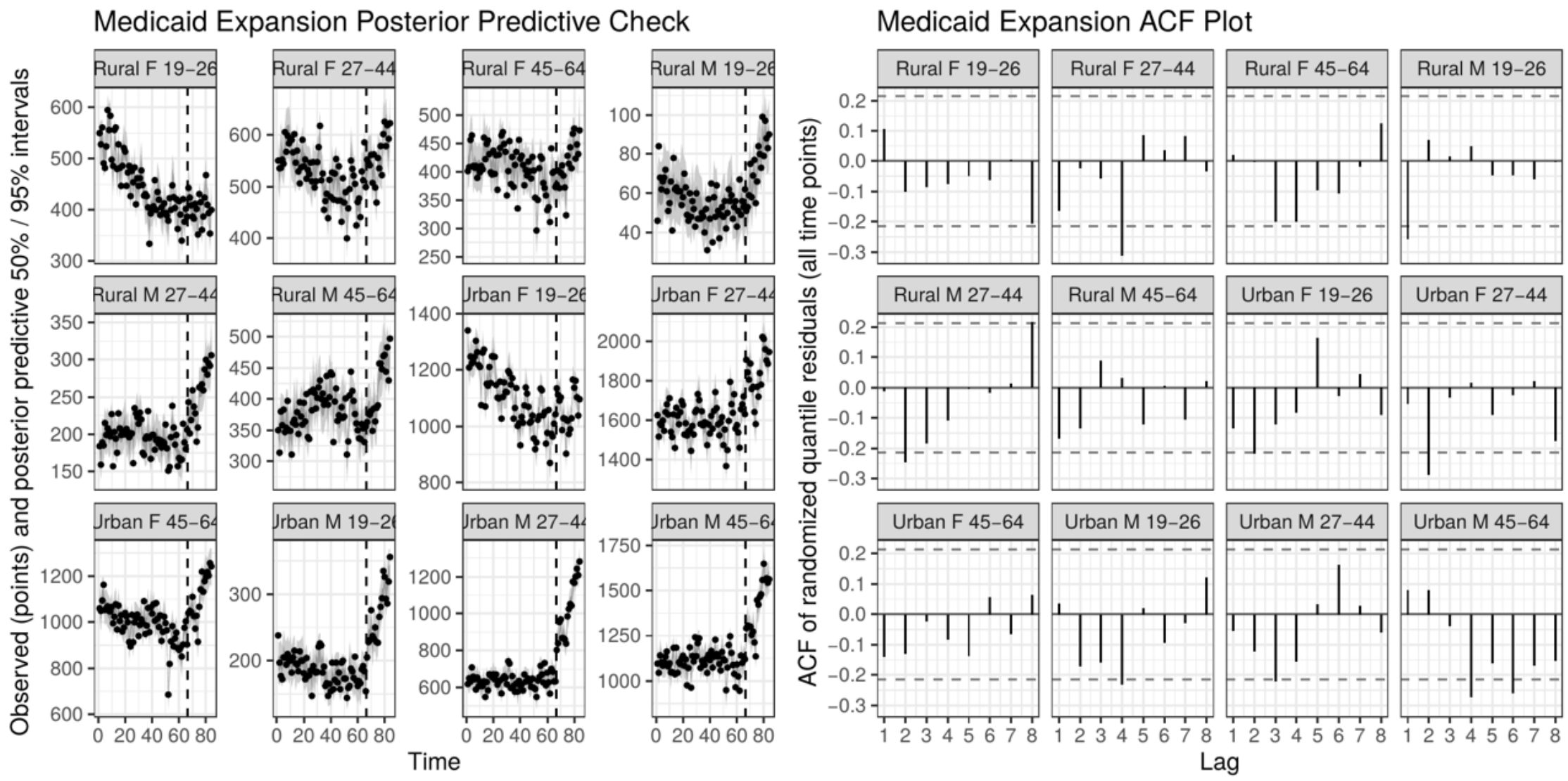

*Figure S6. Missouri Medicaid expansion: posterior predictive check (left) and residual autocorrelation (right), as in Figure S4.*

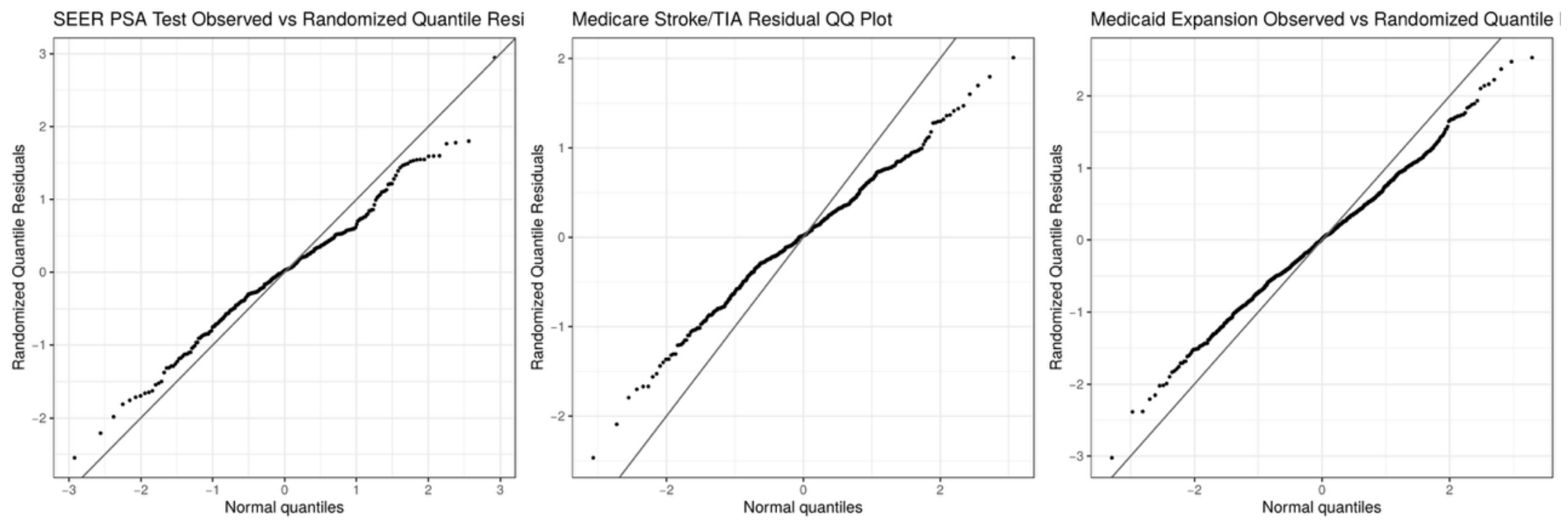

*Figure S7. Normal QQ plots of the randomized quantile residuals, all series pooled: SEER, stroke and Medicaid (left to right).*

# A10. Simulation study: design details and supplementary results

## A10.1 Data-generating process

Each dataset has two crossed factors, $F_1$ and $F_2$, with three levels each ($S = 9$ cells), $T = 60$ time points and the interruption at $T_1 = 41$, giving 20 post-period points ($\tau = 0, \dots, 19$). For cell $s$ in level $k_1$ of $F_1$ and level $k_2$ of $F_2$,

$$y_{st} = \gamma_t + \gamma_{1k_1t} + \gamma_{2k_2t} + g(t,s) + \varepsilon_{st}, \qquad \varepsilon_{st} \sim N(0,1).$$

**Background.** $\gamma_t$ is a random walk with innovation SD 0.30 and $\gamma_1 = 0$. For each factor, we generate three independent random walks with innovation SD 0.15 (starting at 0) and then center across levels at every $t$ within factor, so $\sum_k \gamma_{jkt} = 0$.

**Effect.** $F_1$ sets the scale, $s_k \in (1.0, 0.3, -0.6)$ with mean $\bar{s} = 0.233$. $F_2$ sets the shape $h_k(\tau)$: a shock ($h = 1$), a linear rise ($h = \tau/19$) and a logistic curve with its inflection at $\tau = 9.5$, rising from about 5% to 100% of its peak. $\bar{h}(\tau)$ is the average shape and $m = \text{SNR} \times 1$ (noise SD 1) the peak magnitude. For $t \geq T_1$ (zero before):

| Scenario | $g(t,s)$ | SNR |
|---|---|---|
| Additive | $m\left[\, s_{k_1}\bar{h}(\tau) + \bar{s}\{h_{k_2}(\tau) - \bar{h}(\tau)\}\right]$ | 0, 1, 2, 4 |
| Multiplicative | $m\, s_{k_1} h_{k_2}(\tau)$ | 1, 2, 4 |
| F2-homogeneous | $m\, s_{k_1}\bar{h}(\tau)$ | 1, 2, 4 |

The Null scenario sets SNR at 0 ($g = 0$ for all $s$ and $t$). We set every $F_2$ contrast to zero in the F2-homogeneous such that the true $F_1$ contrasts are equal across levels of $F_2$. We replicate this simulation experiment 300 times per scenario and SNR, and in reported analyses, no dataset was excluded, even if convergence diagnostics were poor; we do explore the effect on our simulation results when we require zero divergences and Rhat < 1.1 in all comparisons made in (A10.6), and we do not see any changes from the conclusions we come to in the paper.

**Methods.** The MLITS model (A6.1) is fit once per dataset. The single-level model (A6.2) and CausalImpact (A10.7) are fit to the cell-averaged series of each group at each level: 9 cells, 3 $F_1$ levels, 3 $F_2$ levels and the overall series (16 fits per dataset). MLITS and single-level: 4 chains × 2,000 iterations (1,000 warm-up), adapt_delta 0.99, max treedepth 12.

## A10.2 Estimands and metrics

All quantities reported in the simulation results in the paper are post-period averages over $\mathcal{T} = \{41, \dots, 60\}$ with equal weights, $\bar{g}_{G,\mathcal{T}}$, where $G$ = each individual cell, $G$ = each level of $F_1$, $G$ = each level of $F_2$, and $G$ = all nine cells overall. The true value is the equal-weight mean of the cells' true effects. We compute subgroup contrasts for levels of $F_1$ and $F_1$ per posterior draw, which is the same approach used for pairwise contrasts in the applications (A8).

We record the posterior mean and the 95% equal-tailed credible interval for each of the $\bar{g}_{G,\mathcal{T}}$ described above. All simulation results are presented with error bars representing {estimate of interest} ± 1.96 * Monte Carlo standard errors across datasets[9].

- Bias: mean of (estimate − truth).
- RMSE: $\sqrt{\text{mean}\{(\text{estimate} - \text{truth})^2\}}$.
- Coverage: proportion of 95% intervals containing the truth.
- Declared: proportion of 95% intervals excluding zero. With a zero truth this is a false-positive rate; with a nonzero truth, a detection rate. For contrasts we also report the familywise rate (any of the three contrasts declared in a dataset).
- Contrast estimate / truth: mean ratio for contrasts with |truth| $> 1$, a measure of attenuation.
- Pre-period coverage (MLITS only): coverage of the true counterfactual background (latent) and of observed $y$ (predictive) for $t < T_1$, by cell.

## A10.3 Sampler diagnostics

**MLITS.** For our model of interest, 93-96% of fits had no divergent transitions (mean 0.04-0.08 per fit, maximum 3), 96-99% had $\hat{R} < 1.01$ for every saved quantity (maximum 1.020), the minimum bulk and tail ESS of the effect quantities were 610 and 873, minimum E-BFMI was 0.65, and no fit reached the maximum treedepth across the 10 scenario × SNR settings.

**Single-level model.** Diagnostics were weaker for the single-level comparator than for MLITS. 72-74% of cell fits, 59-67% of $F_1/F_2$ fits (with data aggregated across $F_2/F_1$ respectively), and 33-41% of overall (aggregated across all series) fits were free of divergences (mean 1.1-1.6, 1.6-4.6 and 7.9-15.1 divergences per fit), and 88-91%, 74-81% and 45-53% had $\hat{R} < 1.01$. The aggregated series, with smaller noise relative to the random-walk scale, are the hardest to fit. While our main simulation study results includes these fits, we report the same comparisons in A10.6 with converged fits only and find no differences in our conclusions.

## A10.4 Bias by inferential level

Figure S8 shows the bias of the post-period average overall and for each level of $F_1$ and $F_2$; Figure S9 shows each cell. Biases manifest as attenuation toward zero effect in all cases. The overall effect is underestimated by both models to a similar degree (Additive, SNR 4: $-0.19$ MLITS, $-0.23$ single-level). For the 63 overall and subgroup-level comparisons at SNR > 0, MLITS has smaller absolute bias in 55 and a larger one in 8, all 8 within Monte Carlo error. At SNR 4 the single-level model's bias is about twice that of MLITS for the extreme $F_1$ levels (scale 1.0: $-0.85$ vs $-0.42$; scale $-0.6$: $+0.45$ vs $+0.07$, Additive and F2-homogeneous).

At the cell level, MLITS bias is about half the single-level bias in the Additive and F2-homogeneous scenarios (mean absolute bias at SNR 4: 0.23 vs 0.51), and the single-level model is never significantly less biased in any cell. In the Multiplicative scenario, where MLITS is misspecified, the single-level model is less biased (beyond Monte Carlo error) in the two extreme shock cells (scale 1.0 and $-0.6$ with the shock shape) at every SNR, where the additive approximation to scale × shape is worst (SNR 4: $-1.35$ vs $-0.37$ and $+1.46$ vs $+0.17$); MLITS remains less biased in 10 other Multiplicative cell comparisons.

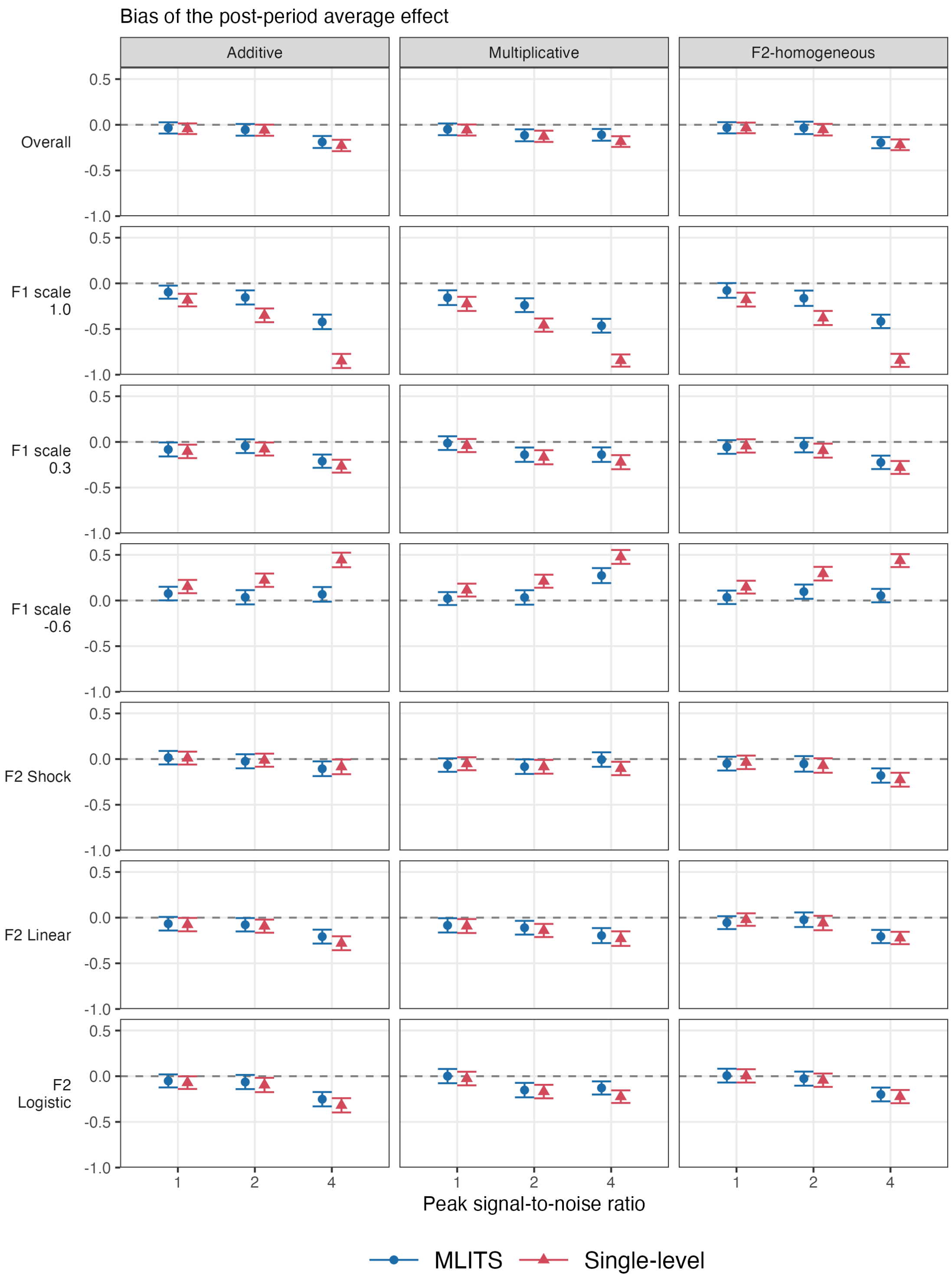

*Figure S8. Bias of the post-period average effect, overall and by level of $F_1$ and $F_2$, for MLITS and the single-level model. Bars: $\pm 1.96$ Monte Carlo SE; dashed line: zero.*

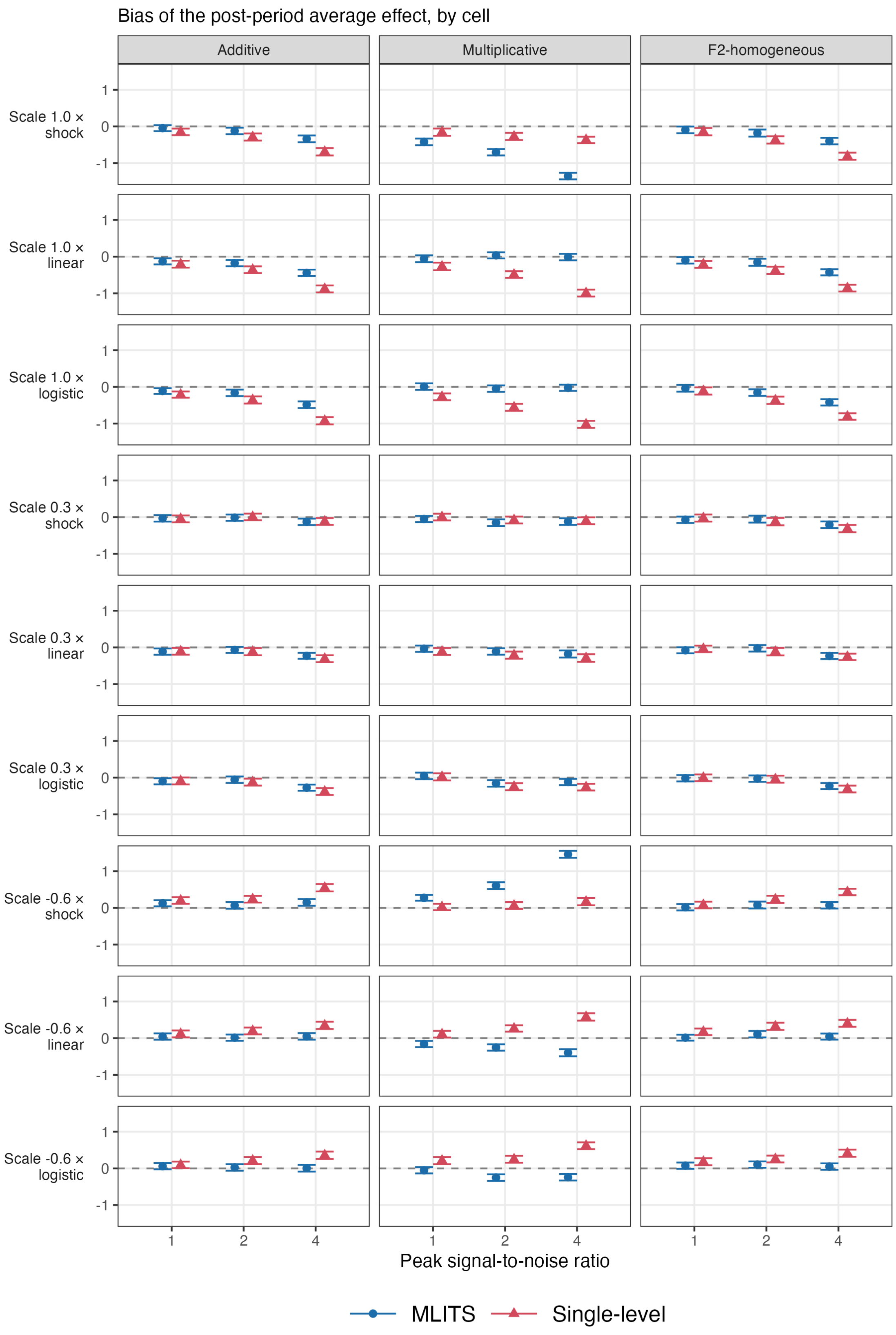

*Figure S9. Bias of the post-period average effect by cell ($F_1$ scale × $F_2$ shape).*

## A10.5 Coverage and bias over the post-period

Figure S10 shows the coverage and bias of the pointwise effect at each post-period time point, $\tau = 0, \ldots, 19$, for the overall effect, each level of $F_1$ and $F_2$ and each cell, for MLITS, the single-level model and CausalImpact (data-scaled prior). Pages come in pairs, coverage then bias: first the Null scenario, then each scenario and SNR. Each panel shows the estimate across 300 datasets with a $\pm 1.96$ Monte Carlo SE band.

- Under the Null scenario, pointwise coverage was 0.94-0.99 for MLITS and the single-level model and 0.85-0.94 for CausalImpact.
- With effects present, MLITS coverage rises over the post-period, from 0.93-0.97 at $\tau = 0$ for the overall effect and the $F_1$ and $F_2$ levels to 0.96-0.99 at $\tau = 19$ at every level. The exception is the Multiplicative scenario at SNR 4, where MLITS is misspecified and cell-level coverage at $\tau = 0$ is 0.73.
- Single-level coverage falls below nominal late in the post-period for the $F_1$ levels at SNR 4 (0.88 at $\tau = 19$). CausalImpact coverage is 0.84-0.97 throughout.
- Bias grows with $\tau$ and is toward zero. In the Additive scenario at SNR 4, MLITS bias at $\tau = 19$ is $-0.82$, $-0.38$ and $+0.20$ for the $F_1$ levels (scale 1.0, 0.3, $-0.6$), against $-1.57$, $-0.46$ and $+0.90$ for the single-level model.

A latent random walk can absorb part of an effect that builds gradually, which explains the attenuation late in the post-period for both models. This design is also a conservative setting for MLITS: the subgroup effects are widely dispersed around a small overall effect, including effects of opposite sign, so most of the signal sits in the subgroup deviations, the component partial pooling shrinks most. If subgroup effects were share a direction, more of the effect is carried by the overall component, which borrows strength from every series.

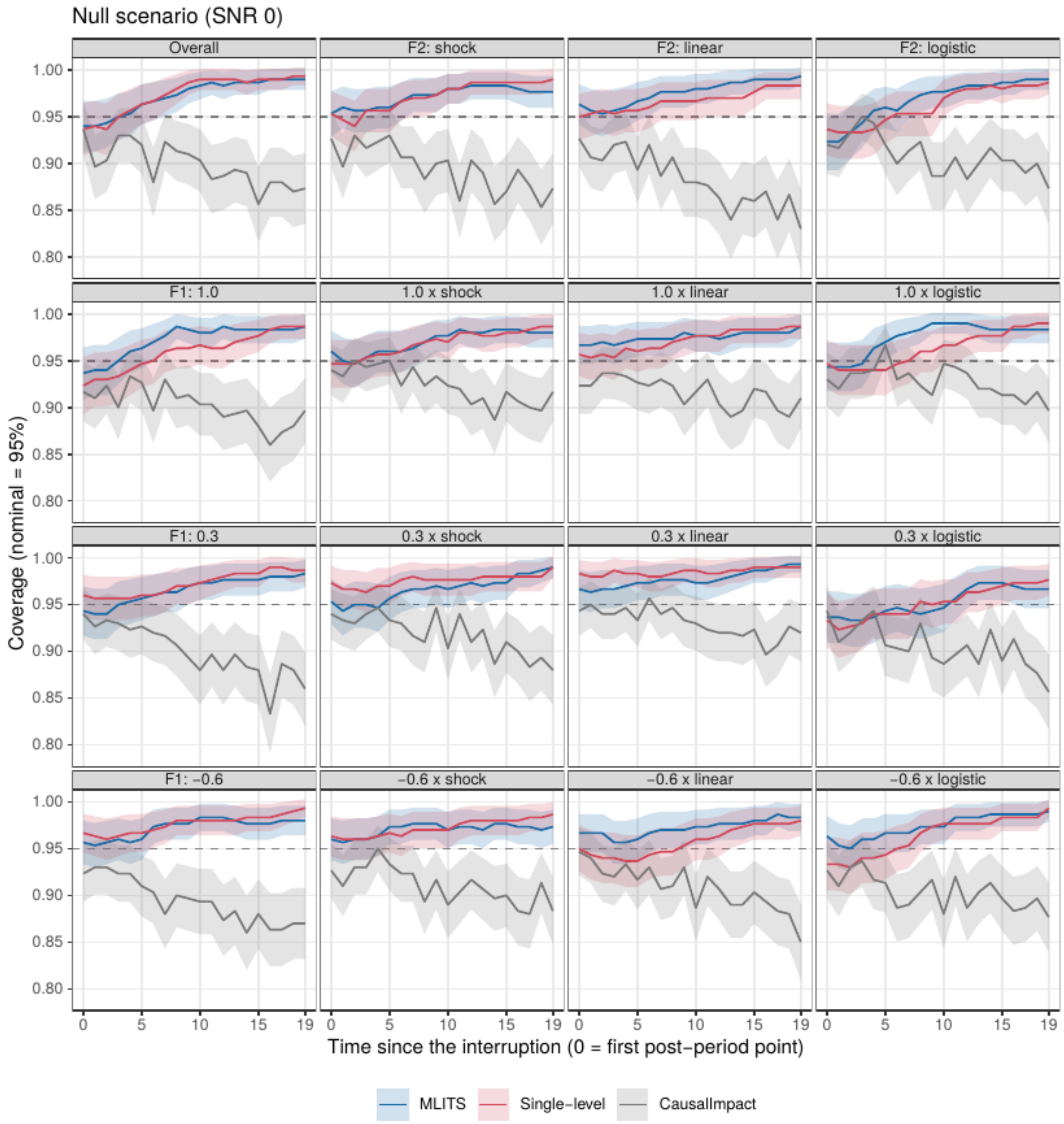

*Figure S10.1. Pointwise coverage by time since the interruption: Null scenario (SNR 0). Top left: overall; top row: $F_2$ levels; left column: $F_1$ levels; interior: cells ($F_1$ scale $\times$ $F_2$ shape).*

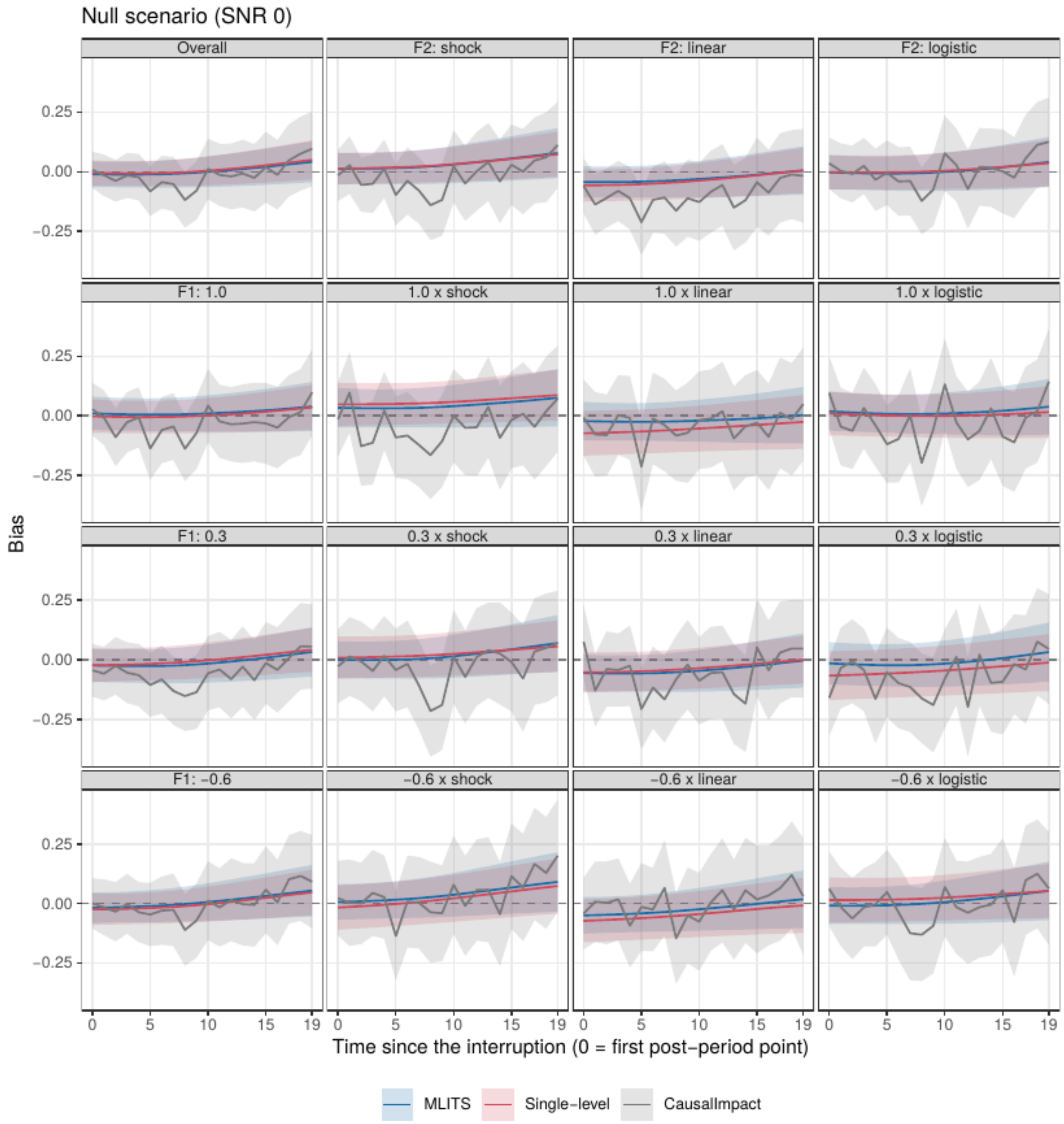

*Figure S10.2. Pointwise bias by time since the interruption: Null scenario (SNR 0). Top left: overall; top row: $F_2$ levels; left column: $F_1$ levels; interior: cells ($F_1$ scale $\times$ $F_2$ shape).*

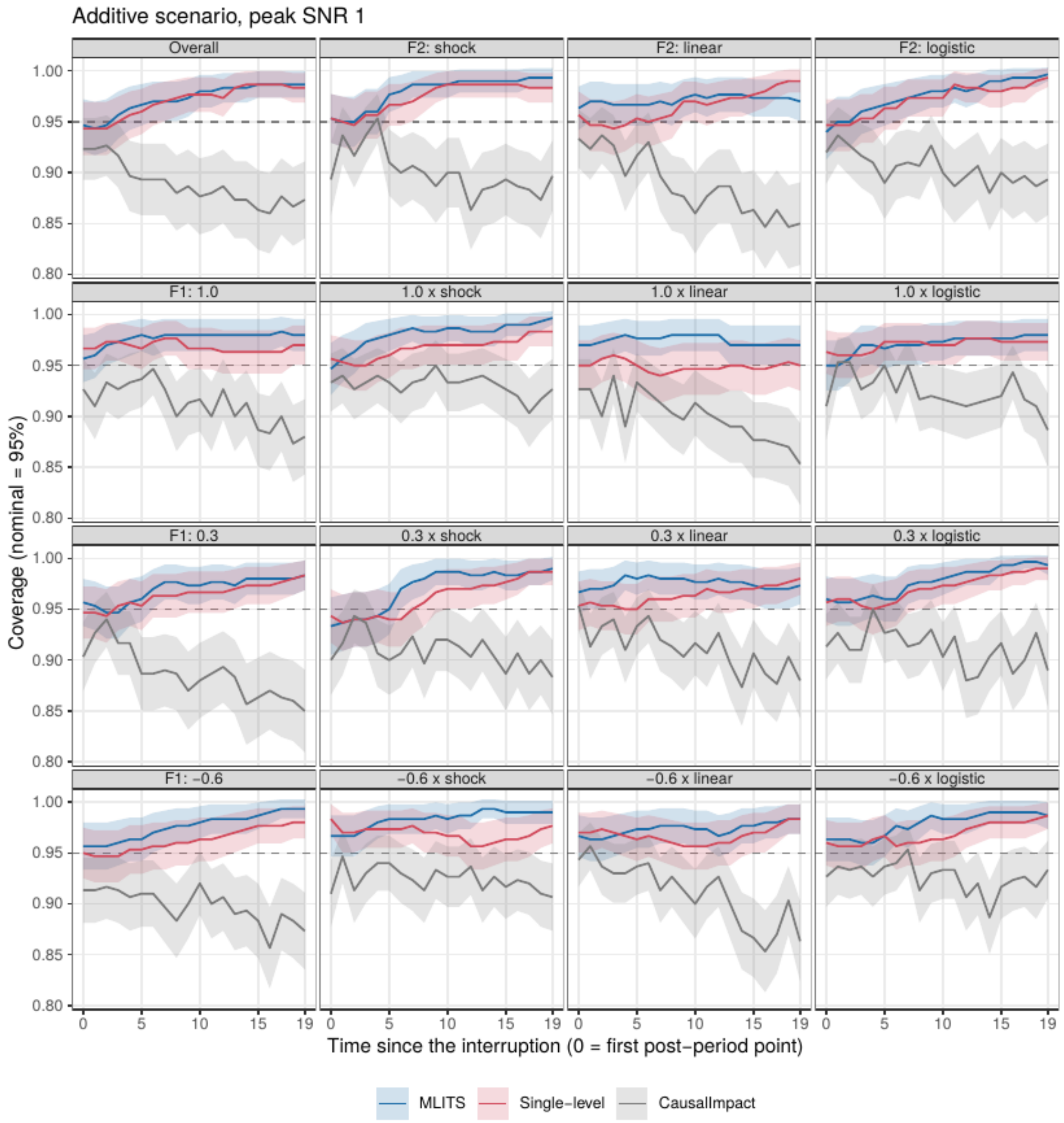

*Figure S10.3. Pointwise coverage by time since the interruption: Additive scenario, peak SNR 1. Top left: overall; top row: $F_2$ levels; left column: $F_1$ levels; interior: cells ($F_1$ scale × $F_2$ shape).*

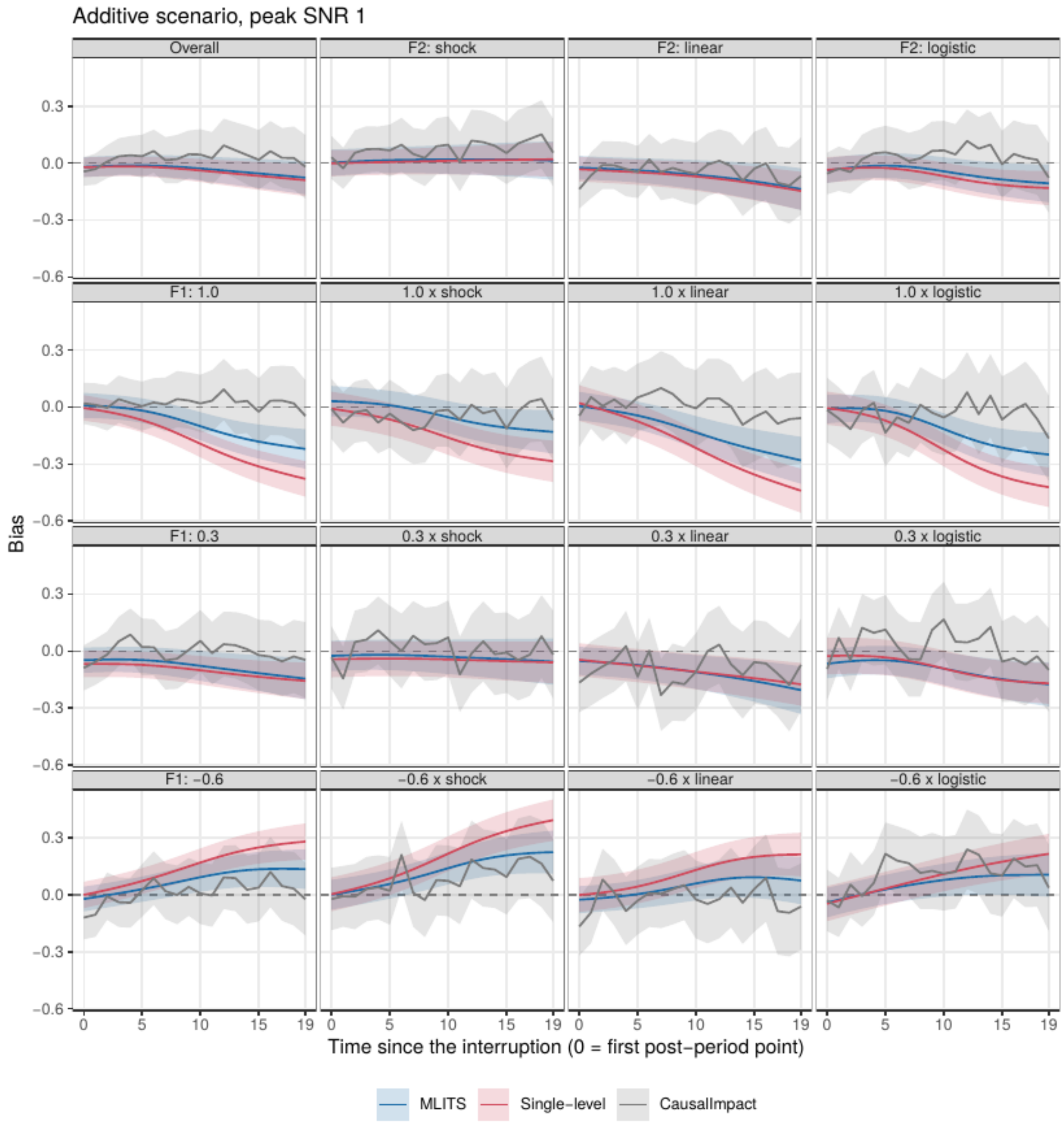

*Figure S10.4. Pointwise bias by time since the interruption: Additive scenario, peak SNR 1. Top left: overall; top row: $F_2$ levels; left column: $F_1$ levels; interior: cells ($F_1$ scale $\times$ $F_2$ shape).*

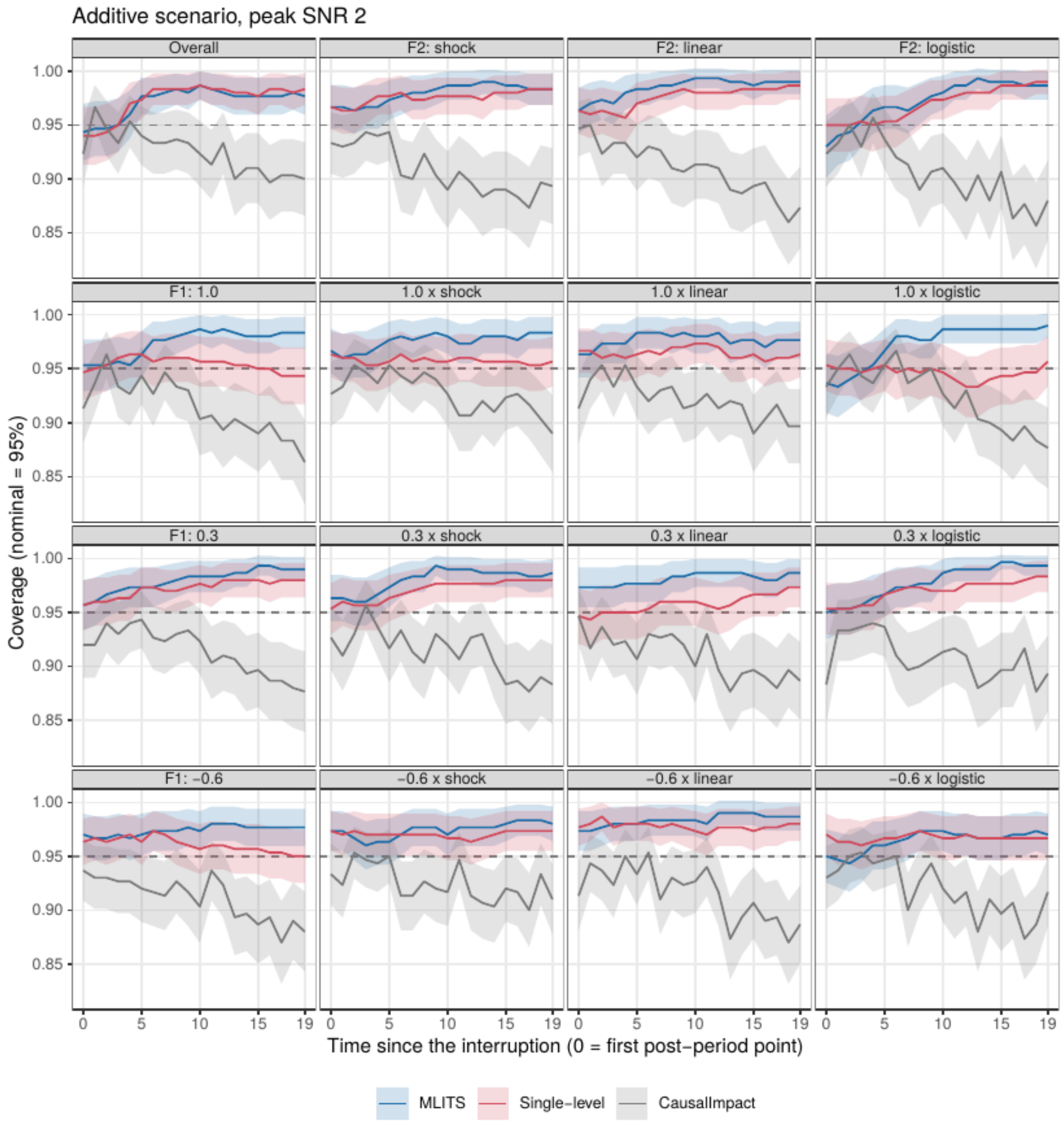

*Figure S10.5. Pointwise coverage by time since the interruption: Additive scenario, peak SNR 2. Top left: overall; top row: $F_2$ levels; left column: $F_1$ levels; interior: cells ($F_1$ scale $\times$ $F_2$ shape).*

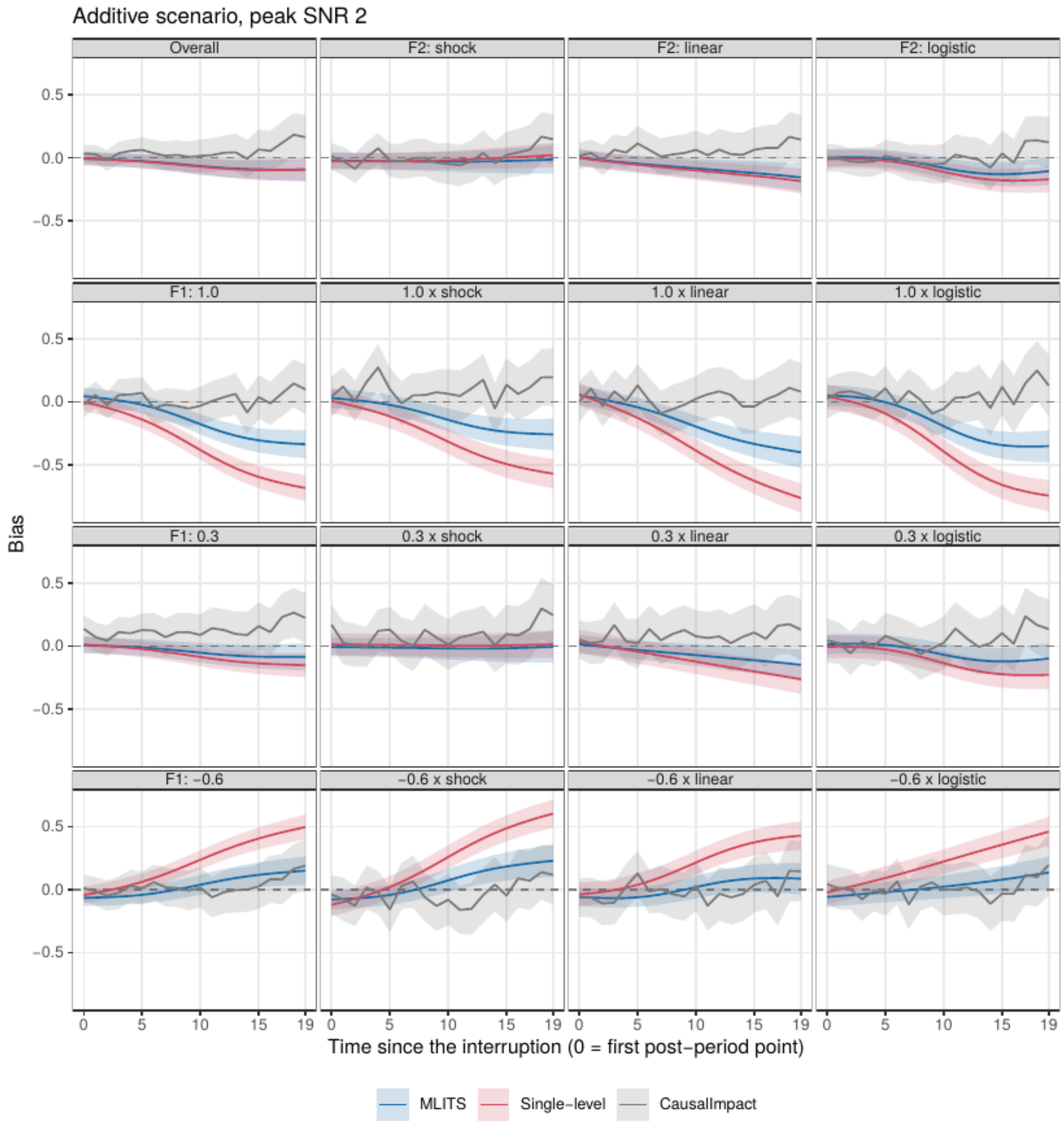

*Figure S10.6. Pointwise bias by time since the interruption: Additive scenario, peak SNR 2. Top left: overall; top row: $F_2$ levels; left column: $F_1$ levels; interior: cells ($F_1$ scale $\times$ $F_2$ shape).*

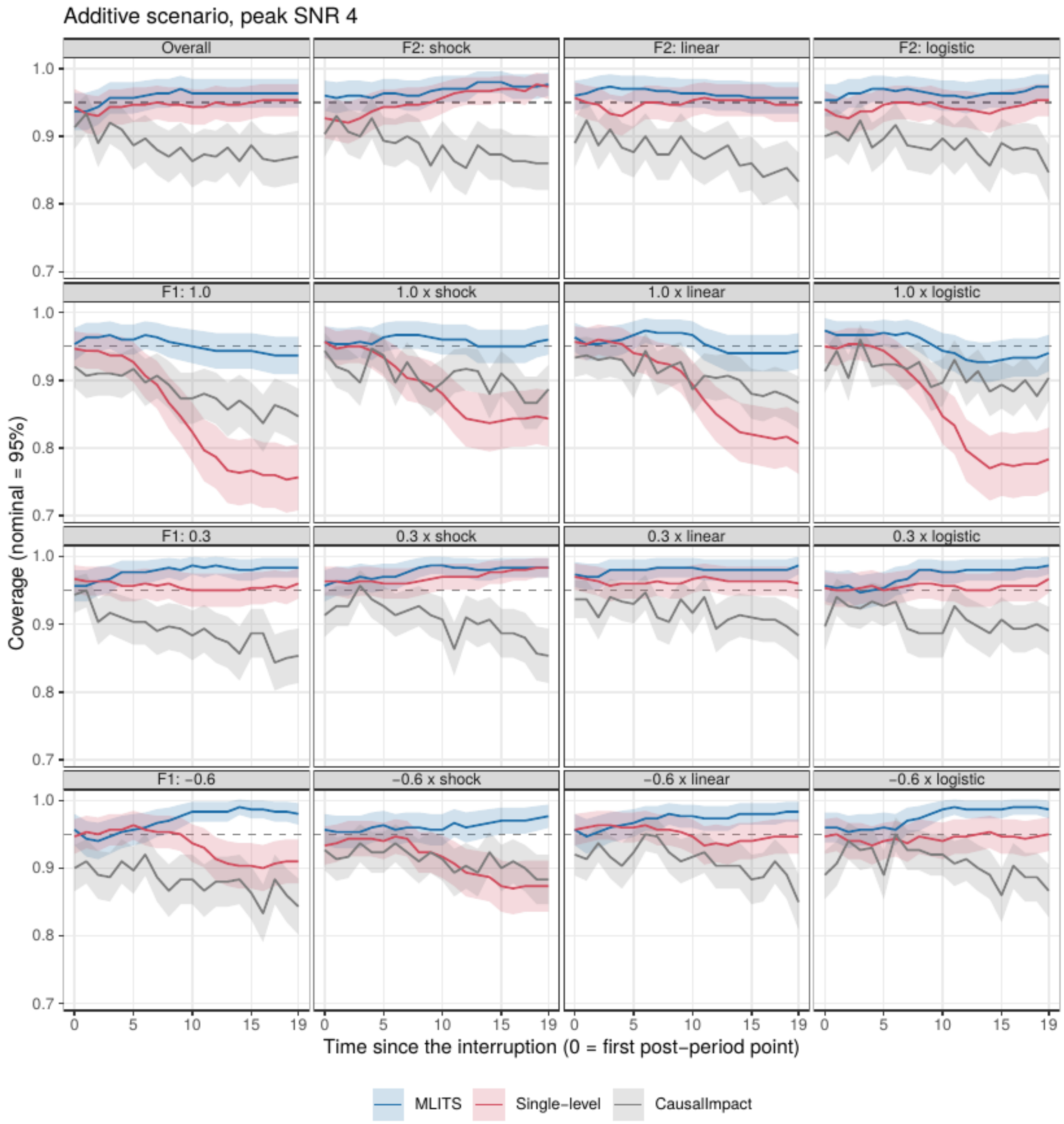

*Figure S10.7. Pointwise coverage by time since the interruption: Additive scenario, peak SNR 4. Top left: overall; top row: $F_2$ levels; left column: $F_1$ levels; interior: cells ($F_1$ scale $\times$ $F_2$ shape).*

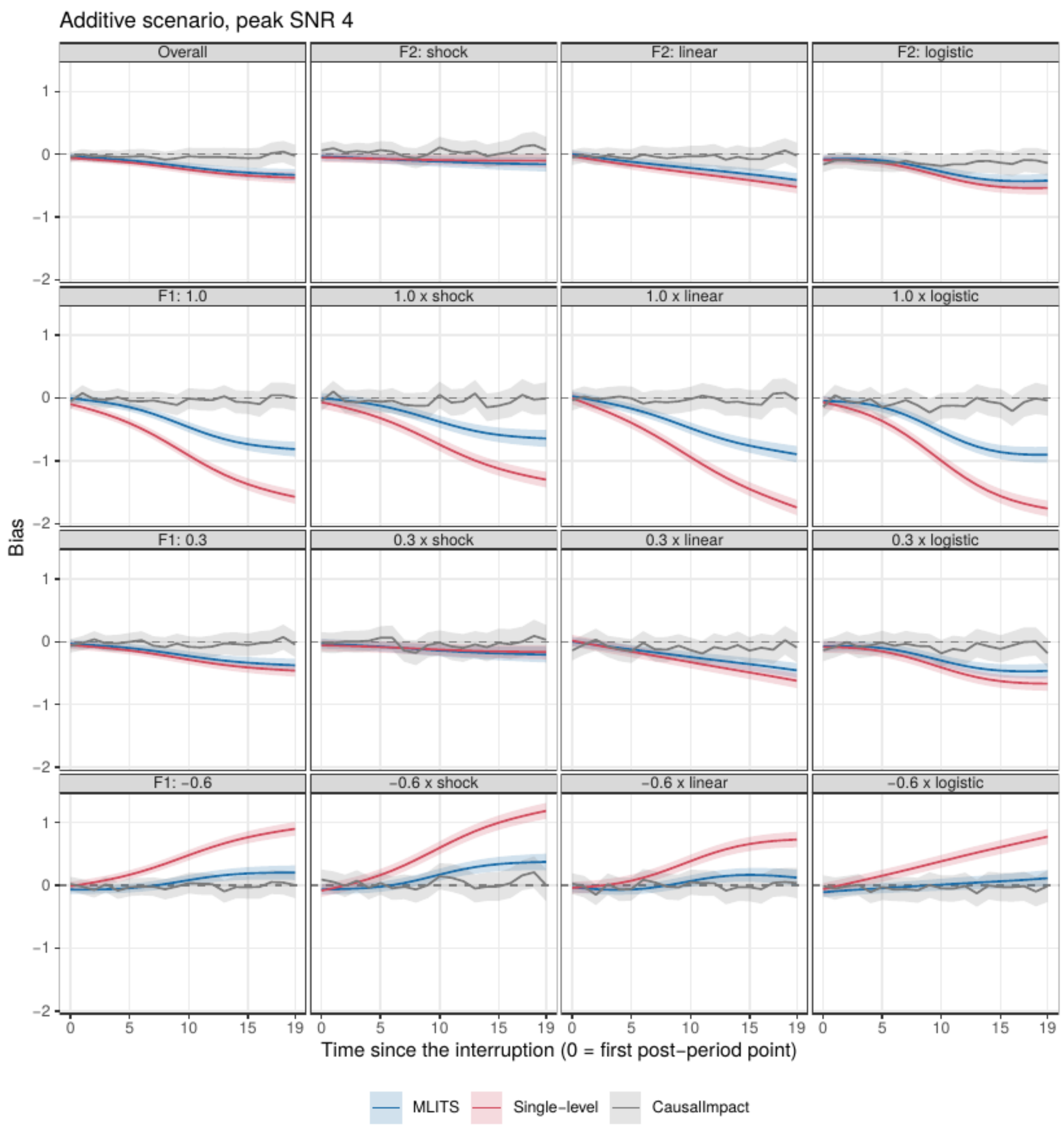

*Figure S10.8. Pointwise bias by time since the interruption: Additive scenario, peak SNR 4. Top left: overall; top row: $F_2$ levels; left column: $F_1$ levels; interior: cells ($F_1$ scale $\times$ $F_2$ shape).*

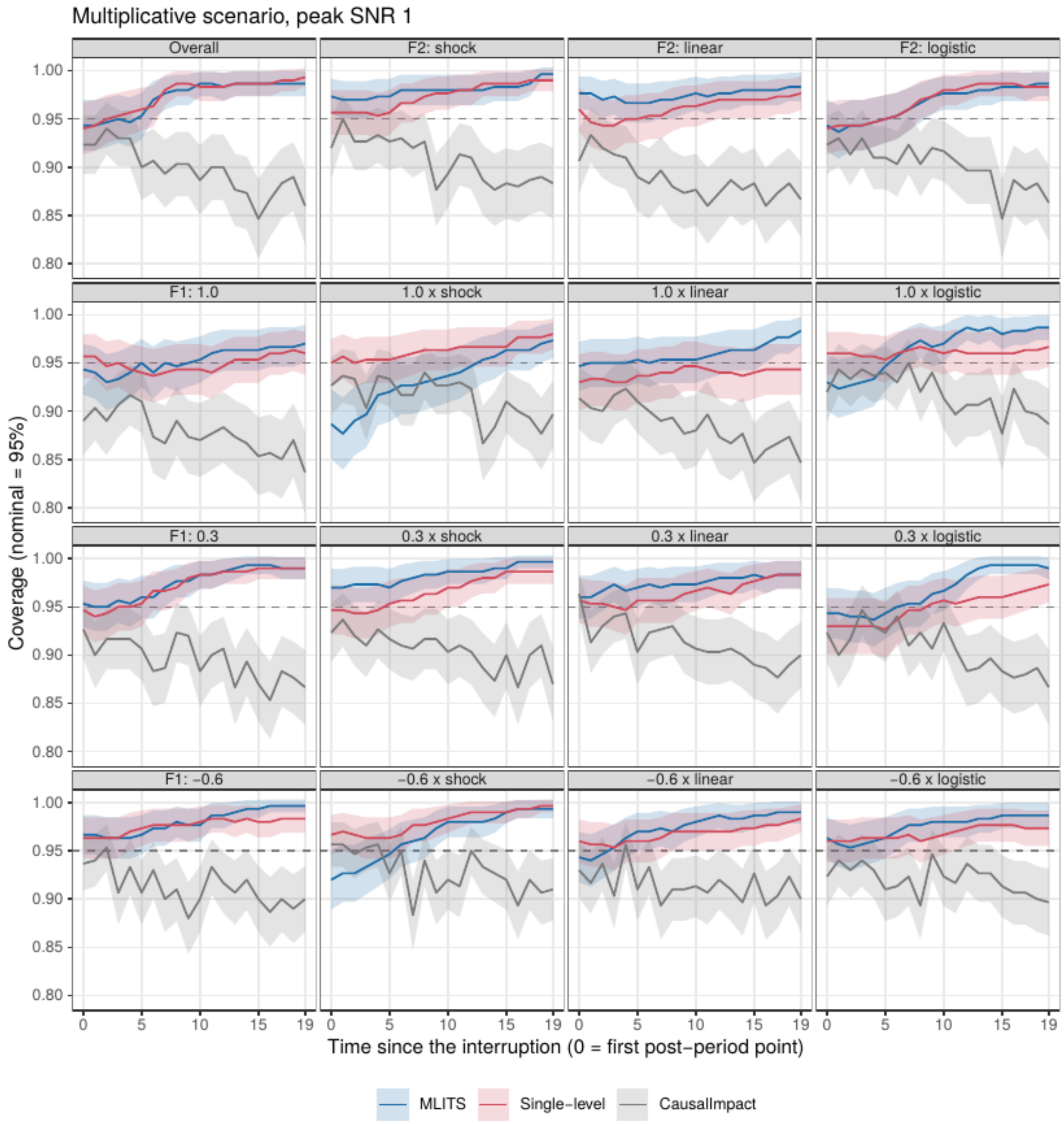

*Figure S10.9. Pointwise coverage by time since the interruption: Multiplicative scenario, peak SNR 1. Top left: overall; top row: $F_2$ levels; left column: $F_1$ levels; interior: cells ($F_1$ scale $\times$ $F_2$ shape).*

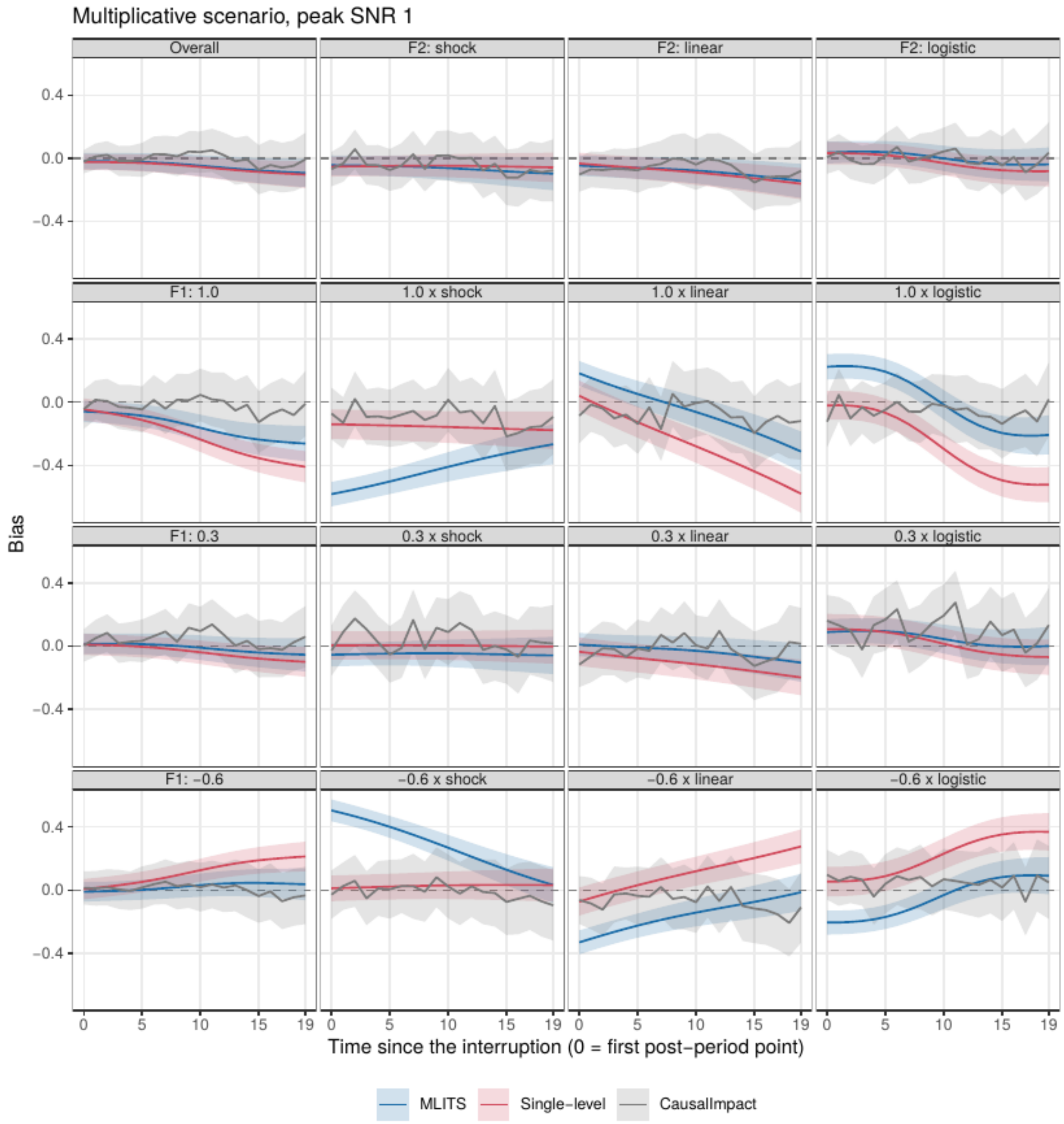

*Figure S10.10. Pointwise bias by time since the interruption: Multiplicative scenario, peak SNR 1. Top left: overall; top row: $F_2$ levels; left column: $F_1$ levels; interior: cells ($F_1$ scale $\times$ $F_2$ shape).*

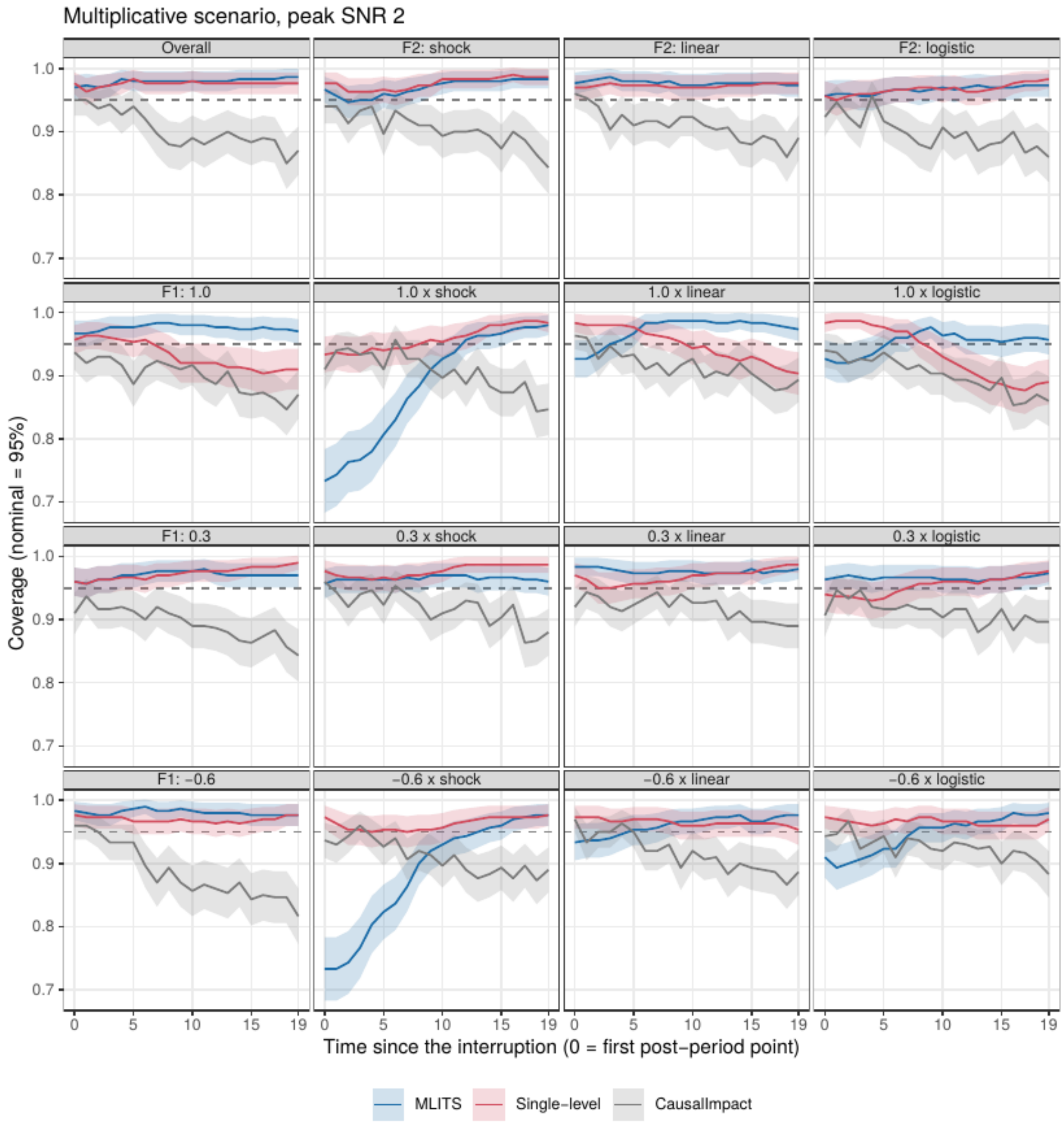

*Figure S10.11. Pointwise coverage by time since the interruption: Multiplicative scenario, peak SNR 2. Top left: overall; top row: $F_2$ levels; left column: $F_1$ levels; interior: cells ($F_1$ scale $\times$ $F_2$ shape).*

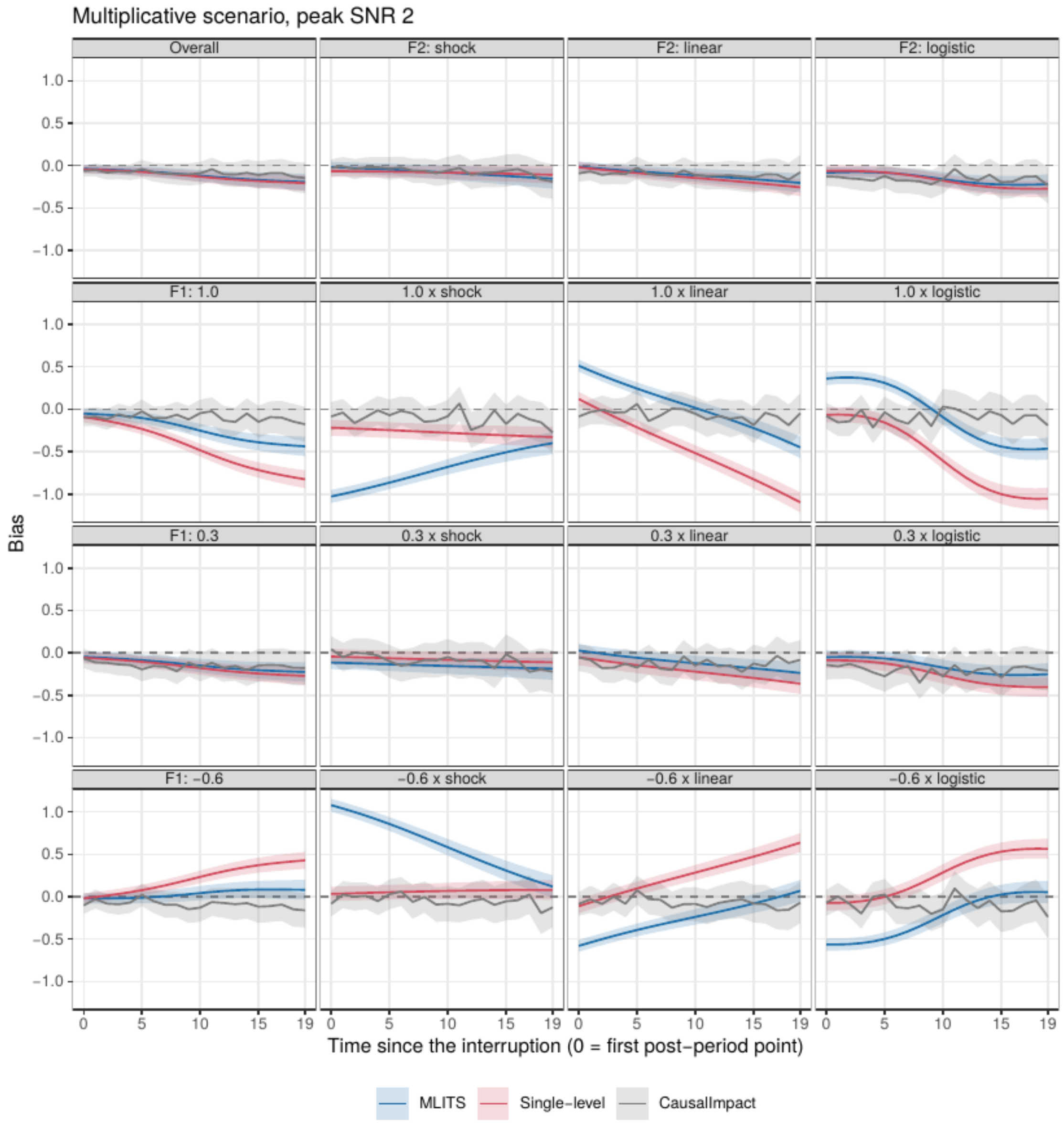

*Figure S10.12. Pointwise bias by time since the interruption: Multiplicative scenario, peak SNR 2. Top left: overall; top row: $F_2$ levels; left column: $F_1$ levels; interior: cells ($F_1$ scale × $F_2$ shape).*

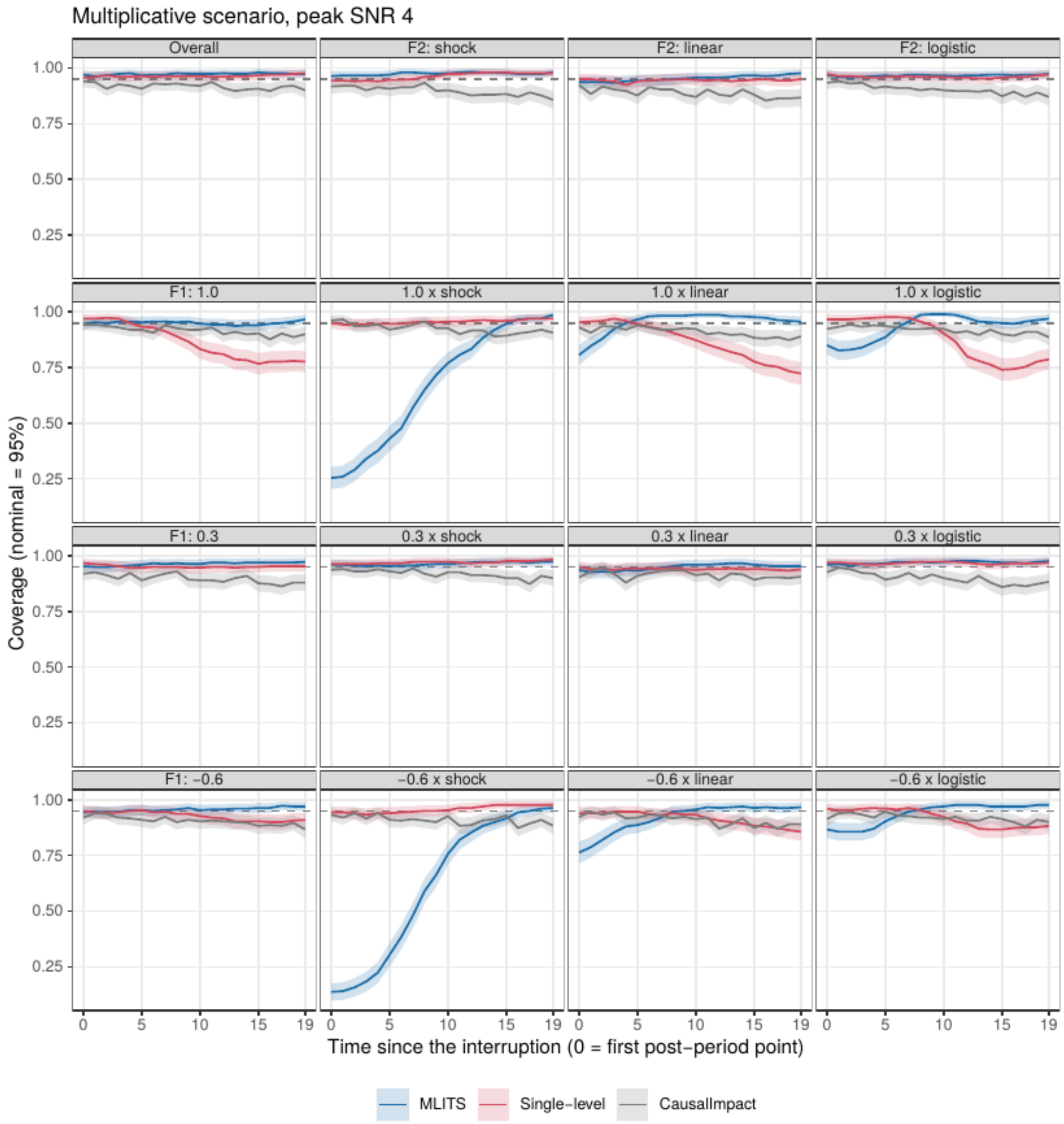

*Figure S10.13. Pointwise coverage by time since the interruption: Multiplicative scenario, peak SNR 4. Top left: overall; top row: $F_2$ levels; left column: $F_1$ levels; interior: cells ($F_1$ scale × $F_2$ shape).*

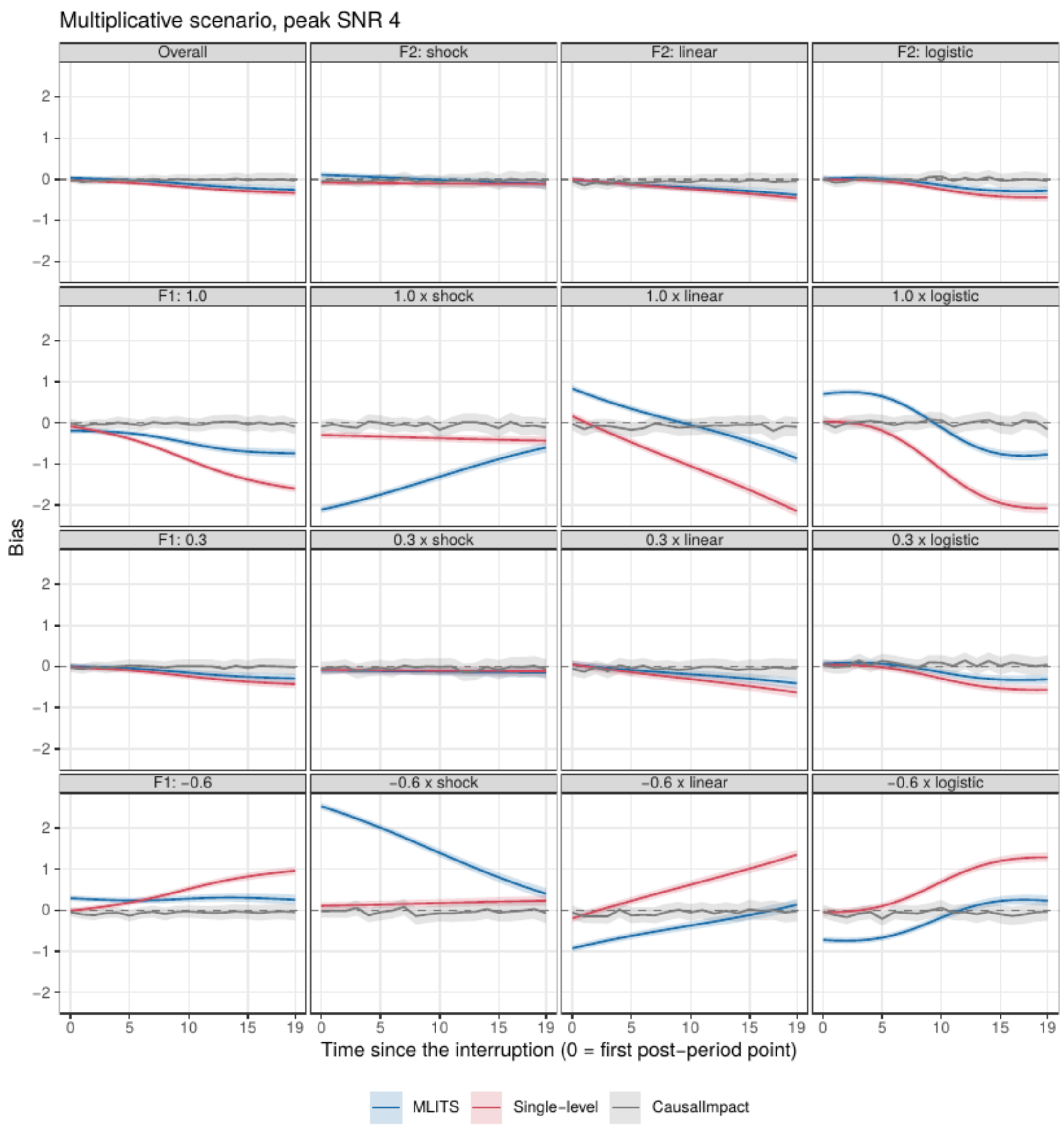

*Figure S10.14. Pointwise bias by time since the interruption: Multiplicative scenario, peak SNR 4. Top left: overall; top row: $F_2$ levels; left column: $F_1$ levels; interior: cells ($F_1$ scale × $F_2$ shape).*

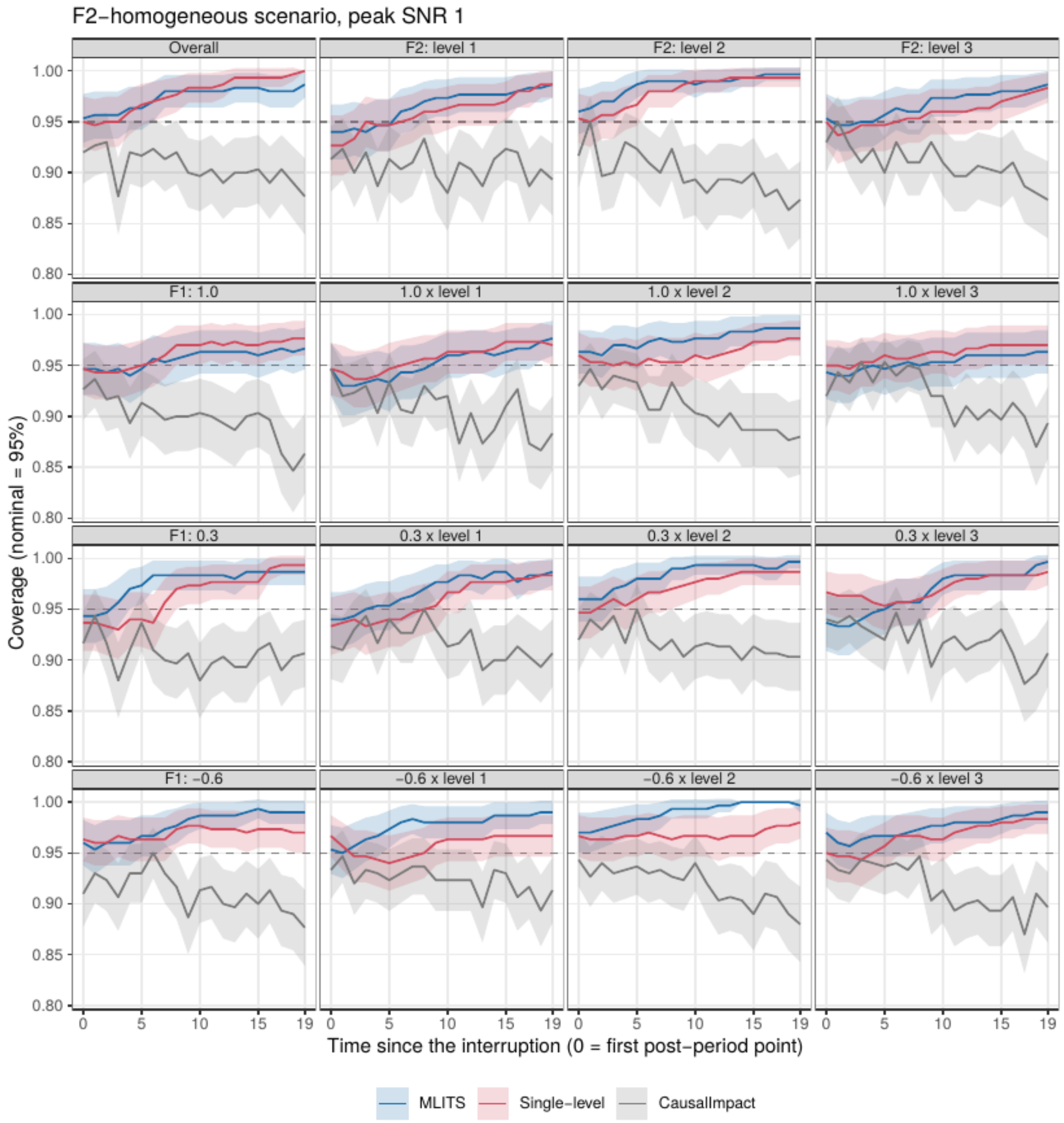

*Figure S10.15. Pointwise coverage by time since the interruption: F2-homogeneous scenario, peak SNR 1. Top left: overall; top row: $F_2$ levels; left column: $F_1$ levels; interior: cells ($F_1$ scale $\times$ $F_2$ shape).*

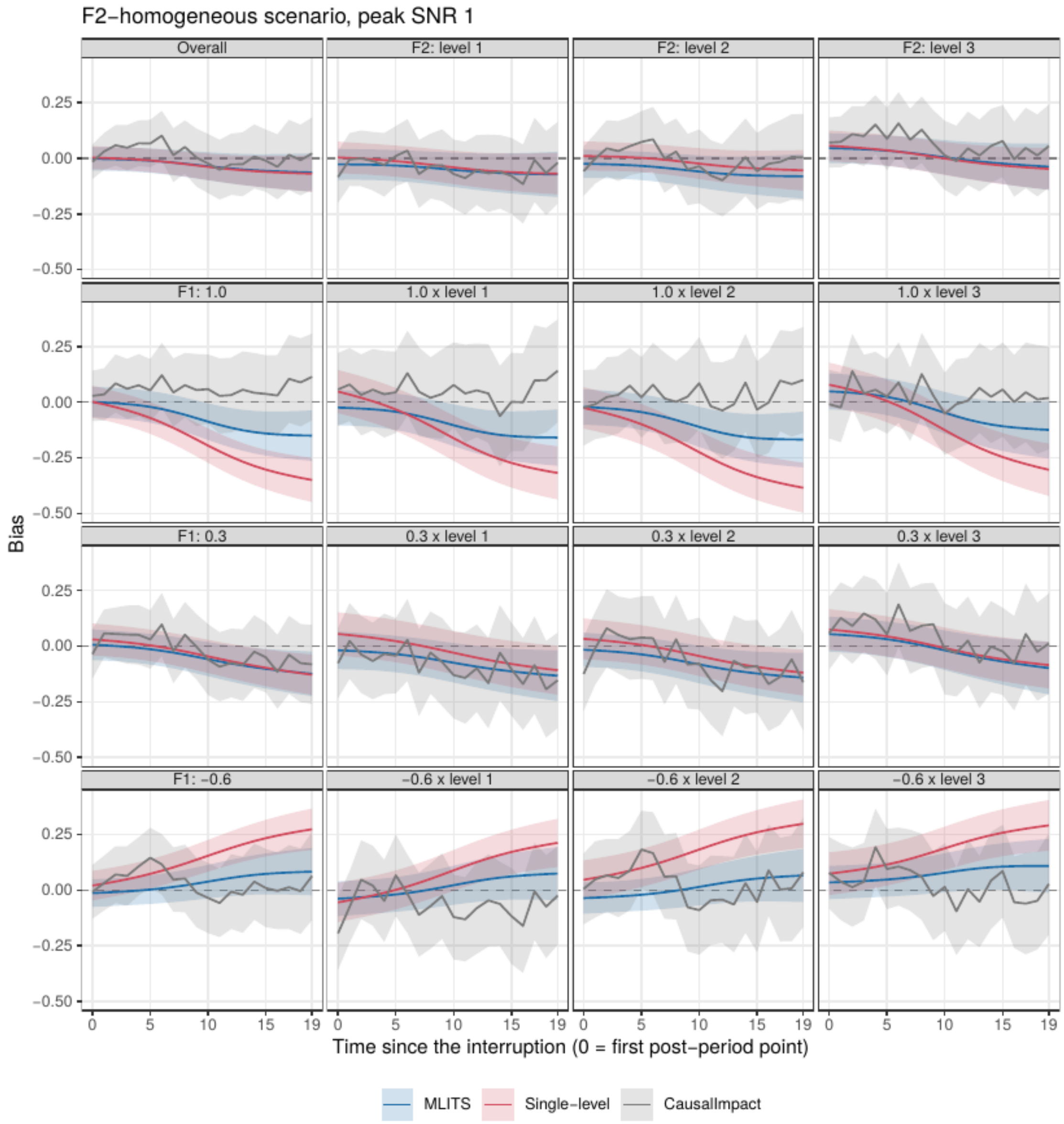

*Figure S10.16. Pointwise bias by time since the interruption: F2-homogeneous scenario, peak SNR 1. Top left: overall; top row: $F_2$ levels; left column: $F_1$ levels; interior: cells ($F_1$ scale $\times$ $F_2$ shape).*

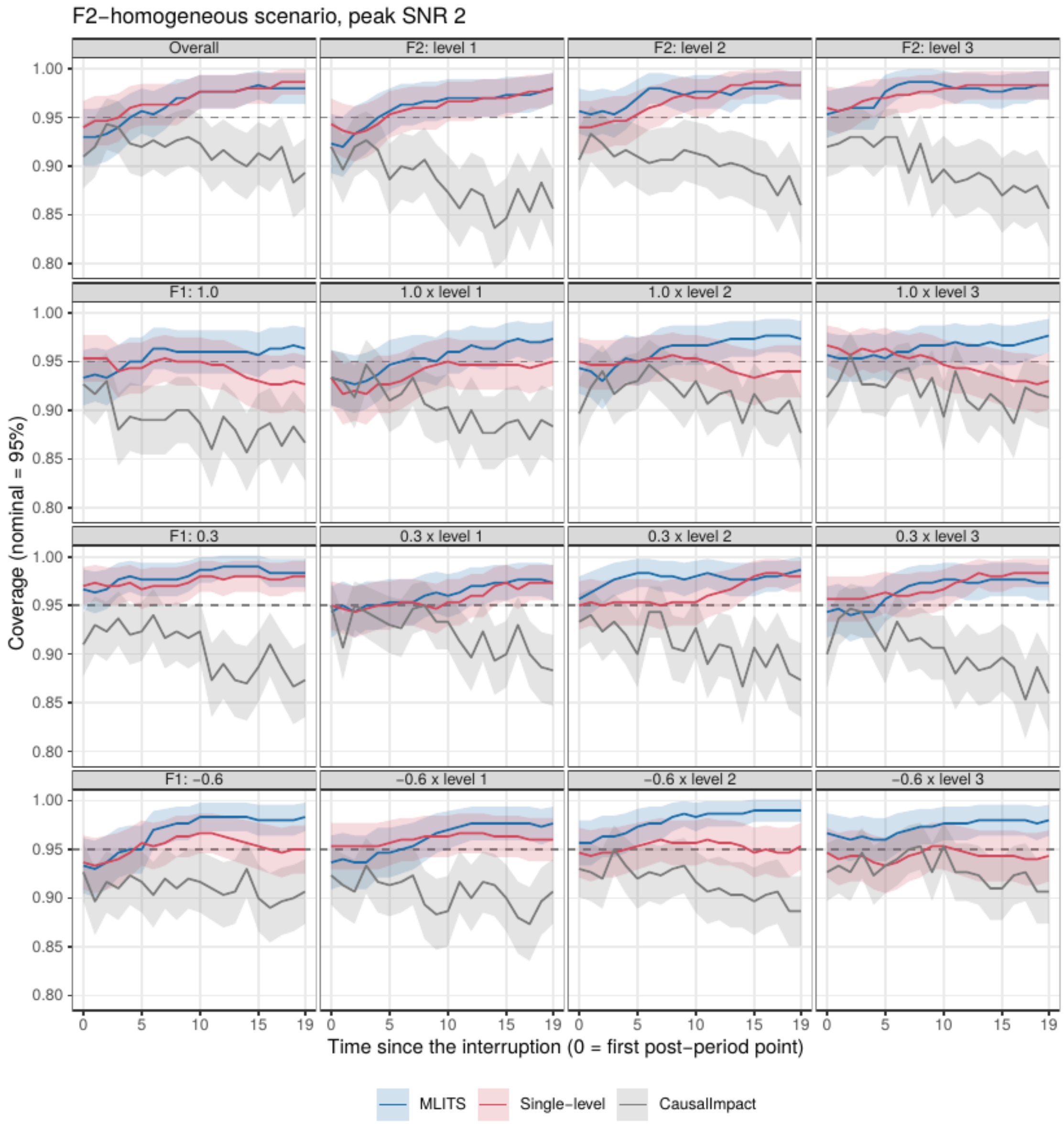

*Figure S10.17. Pointwise coverage by time since the interruption: F2-homogeneous scenario, peak SNR 2. Top left: overall; top row: $F_2$ levels; left column: $F_1$ levels; interior: cells ($F_1$ scale × $F_2$ shape).*

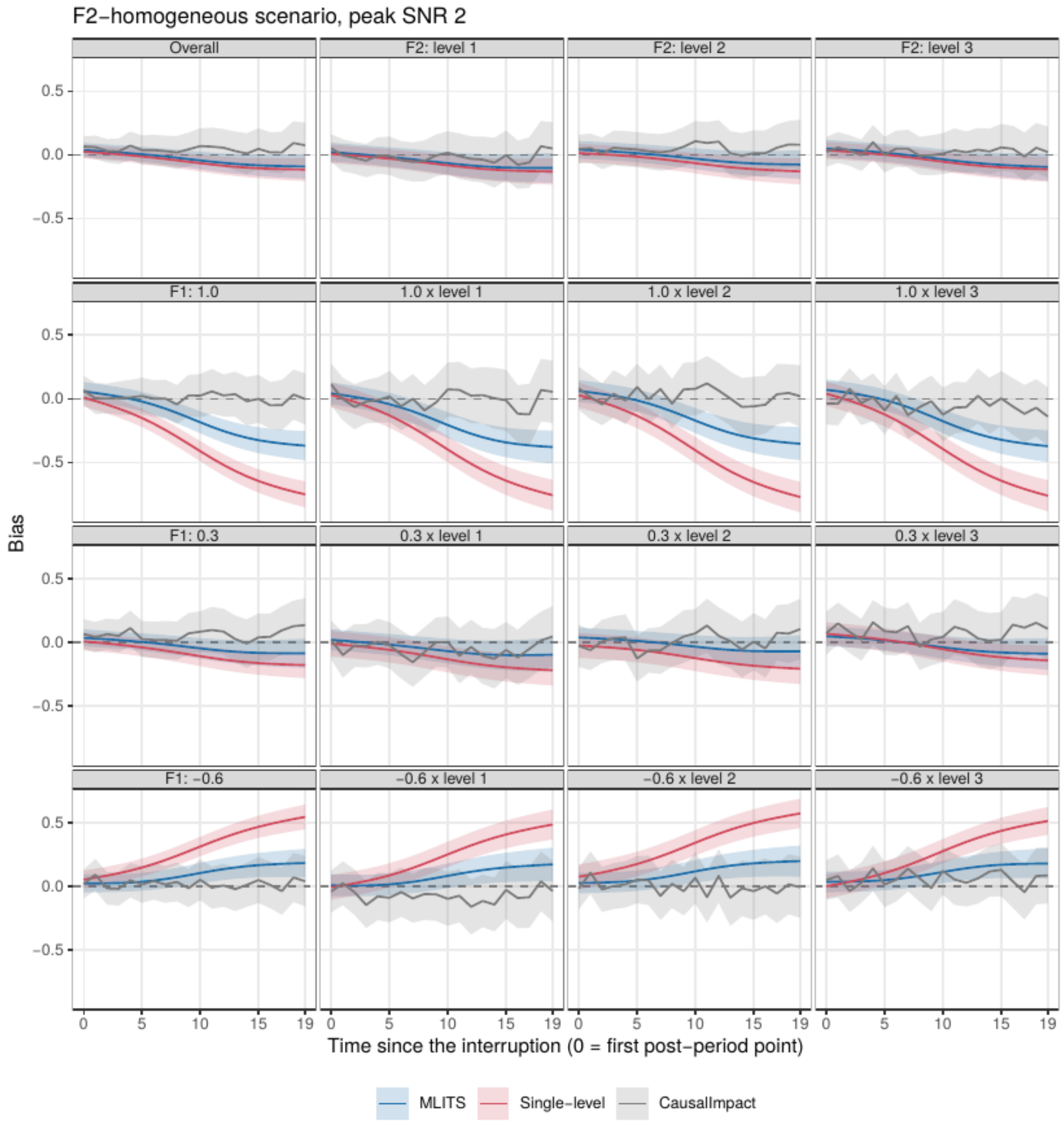

*Figure S10.18. Pointwise bias by time since the interruption: F2-homogeneous scenario, peak SNR 2. Top left: overall; top row: $F_2$ levels; left column: $F_1$ levels; interior: cells ($F_1$ scale × $F_2$ shape).*

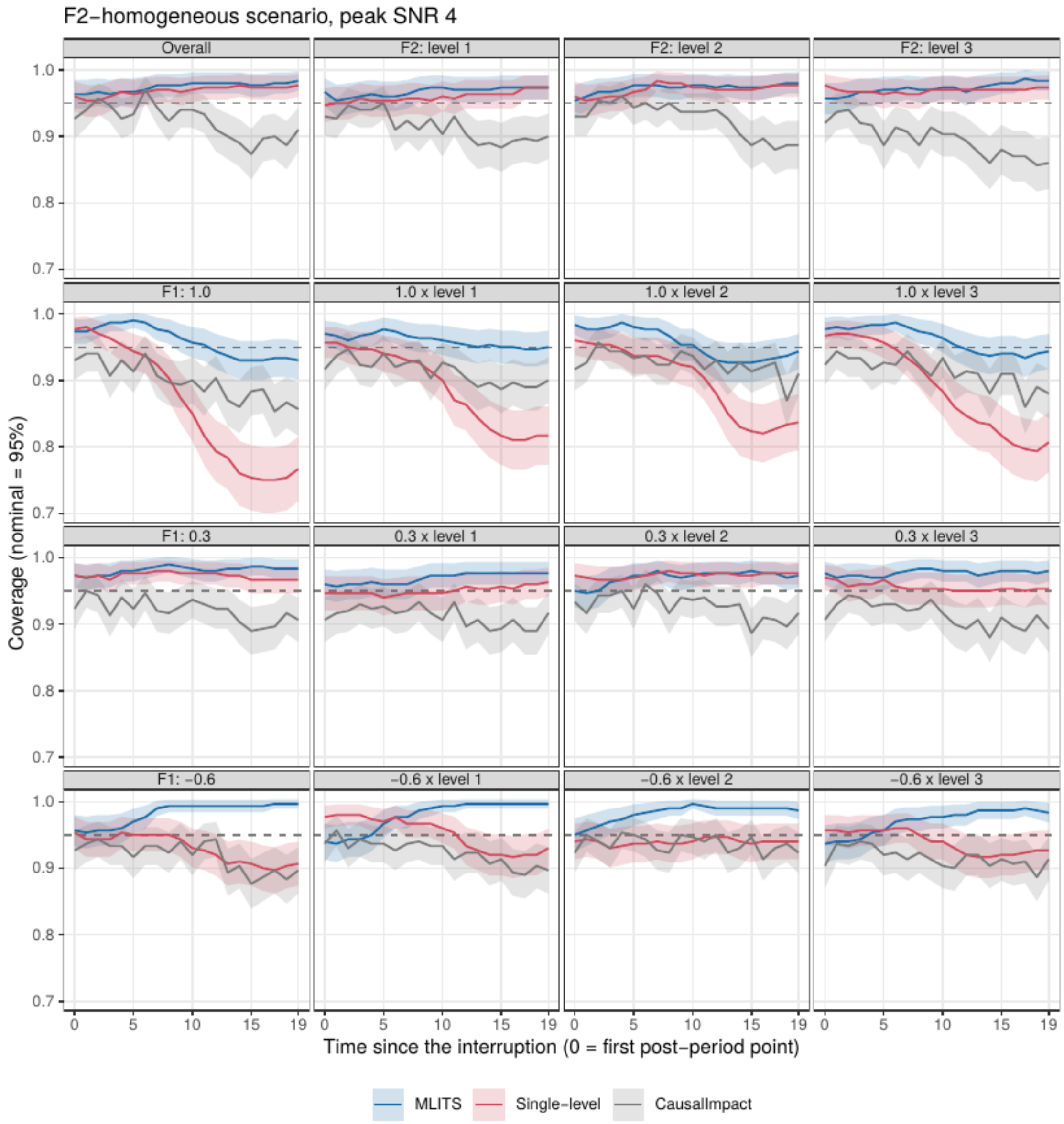

*Figure S10.19. Pointwise coverage by time since the interruption: F2-homogeneous scenario, peak SNR 4. Top left: overall; top row: $F_2$ levels; left column: $F_1$ levels; interior: cells ($F_1$ scale $\times$ $F_2$ shape).*

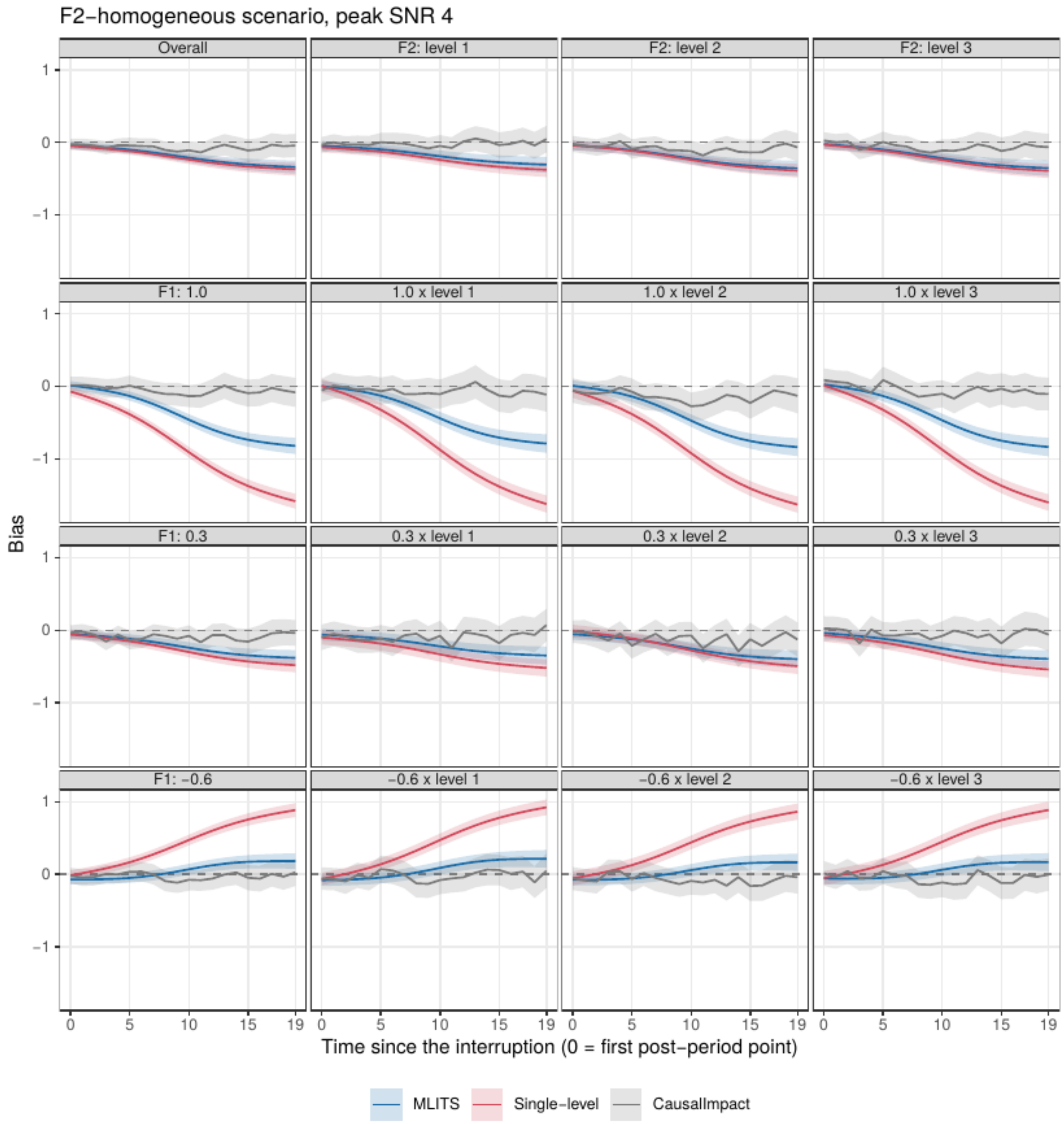

*Figure S10.20. Pointwise bias by time since the interruption: F2-homogeneous scenario, peak SNR 4. Top left: overall; top row: $F_2$ levels; left column: $F_1$ levels; interior: cells ($F_1$ scale $\times$ $F_2$ shape).*

## A10.6 Simulation results for fits that converged

A comparison was counted as failed if it involved any divergent transition or a maximum $\hat{R} \geq 1.1$. Failures were 4-7% of MLITS fits per scenario and SNR, and for the single-level model 26-28% of cell fits, 33-41% of $F_1/F_2$ fits and 59-67% of overall fits, almost all because of divergent transitions. Figure S11 repeats Figure 1 of the main text using, for each comparison, only the comparisons for which every fit coverged are reported below (95-300 datasets per comparison). No conclusion changes.

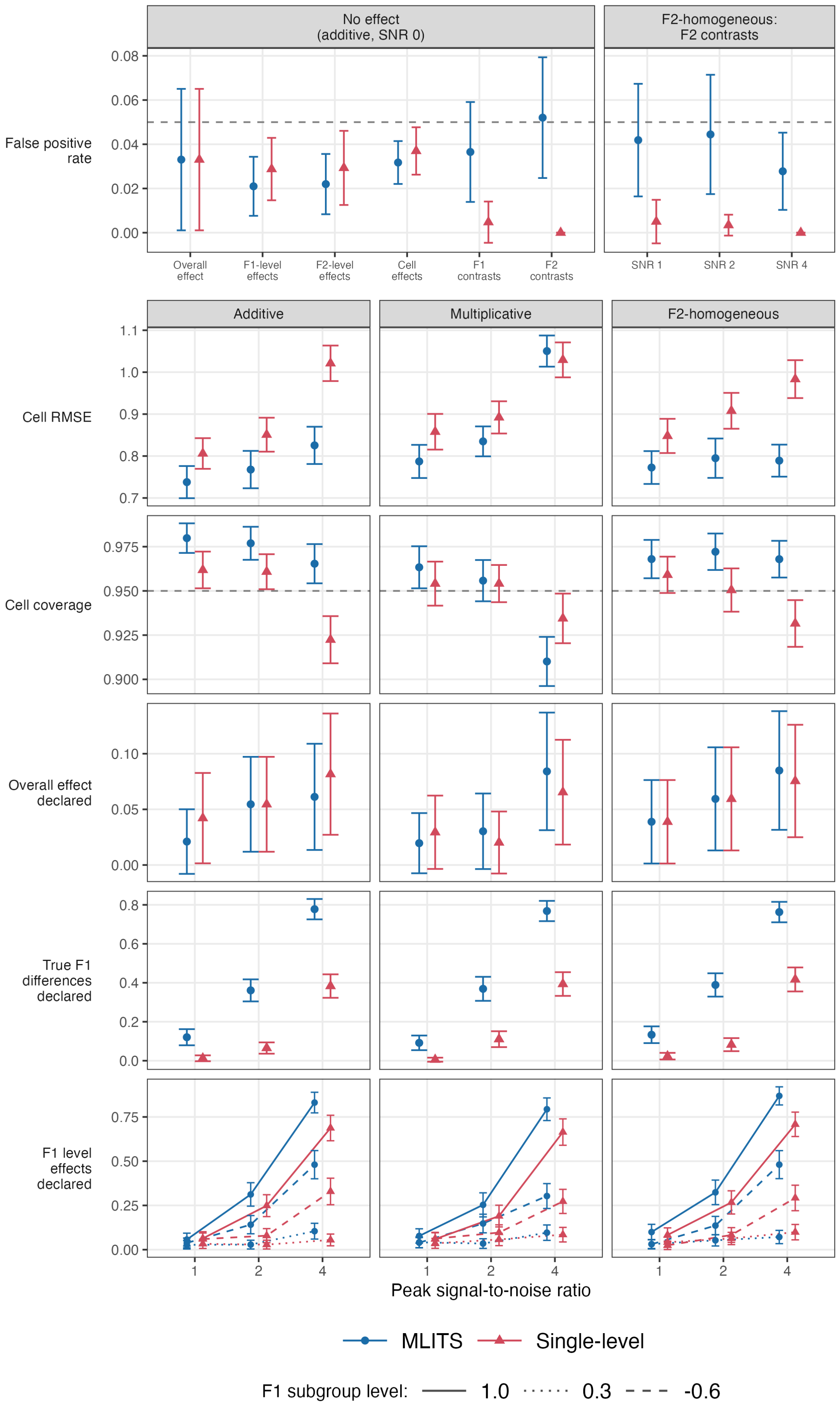

*Figure S11. Main-text Figure 1 recomputed from converged fits only (no divergent transitions and maximum $\hat{R} < 1.1$ for every fit entering a comparison).*

## A10.7 Comparison with CausalImpact (BSTS)

**Design.** We use the CausalImpact [10] package to fit Bayesian structural time series models [11] to the same 3,000 datasets, refit to the aggregated series at each level as for the single-level model (16 fits per dataset, 1,000 MCMC iterations). We used two settings for the local-level prior SD (*prior.level.sd*): a data-scaled value, estimated from each series' standardized pre-period data (moment estimate, truncated to [0.01, 1]), and the package default (0.01). The post-period average effect is CausalImpact's average absolute effect with its interval; contrasts between levels use a normal approximation from the two independent fits. Because CausalImpact's pointwise effect is observed minus predicted, it retains observation noise; the post-period average largely removes it.

**Results (Figure S12).**

- Data-scaled prior: cell RMSE 1.29-1.37 (MLITS 0.75-1.05), cell coverage 0.71-0.73 and overall coverage 0.79-0.86. Under the Null scenario, an overall effect was declared in 20% of datasets, false $F_1$ and $F_2$ differences in 7-8% of contrasts (familywise 15-18%). True $F_1$ differences at SNR 4 were detected in 50-52% of contrasts (MLITS 73-79%).
- Package default: intervals are far too narrow (cell coverage 0.32-0.36, overall 0.19-0.27); an overall effect was declared in 73% of null datasets and false subgroup differences in 42-45% of contrasts.
- The data-scaled level prior was at its 0.01 floor (the package default) for 38-40% of cell series, 25-30% of $F_1/F_2$ series and 12-17% of overall series; no CausalImpact fit failed.

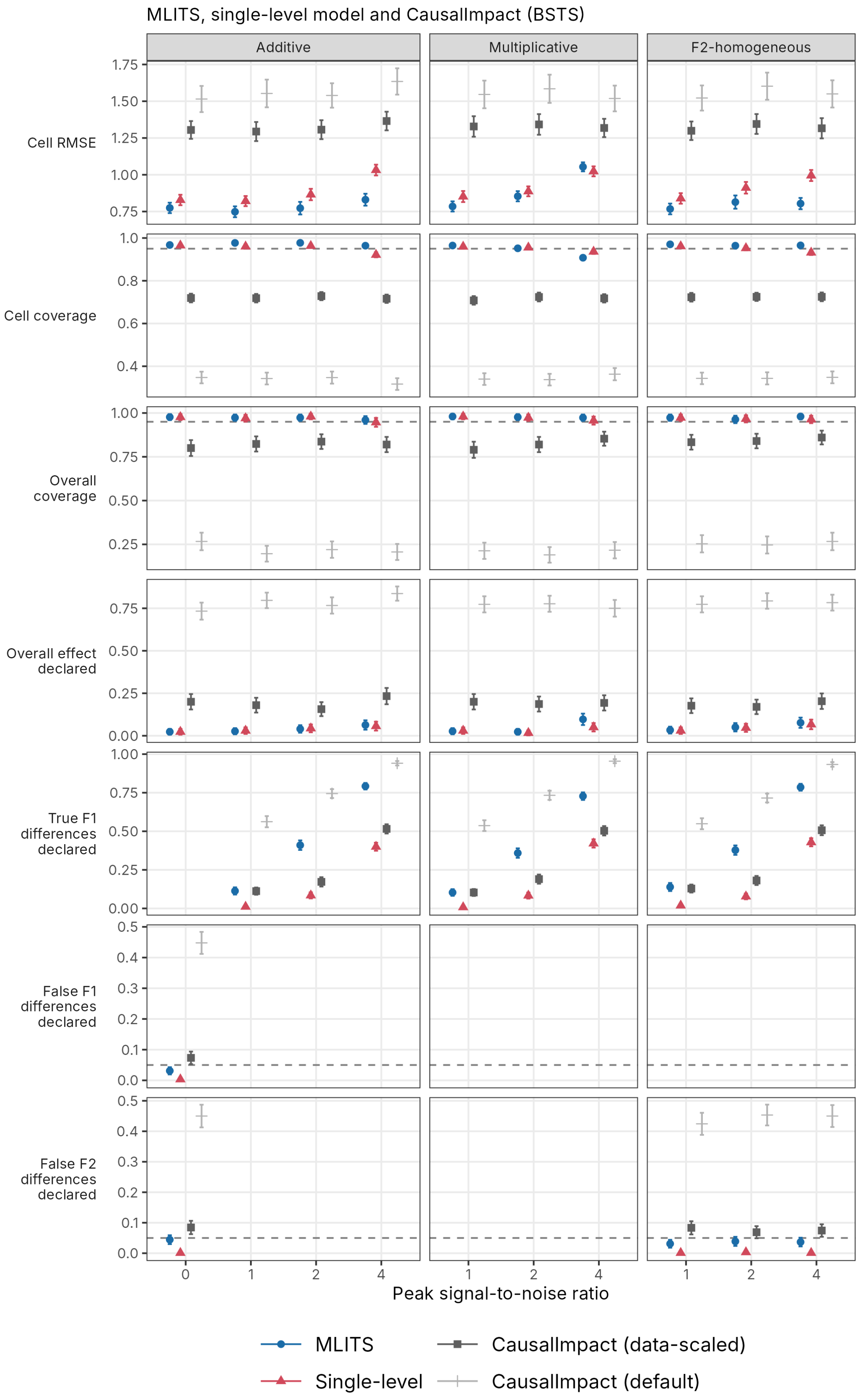

MLITS, single-level model and CausalImpact (BSTS)
Additive
Multiplicative
F2-homogeneous
Cell RMSE
Cell coverage
Overall coverage
Overall effect declared
True F1 differences declared
False F1 differences declared
False F2 differences declared
Peak signal-to-noise ratio
MLITS
Single-level
CausalImpact (data-scaled)
CausalImpact (default)

*Figure S12. MLITS, the single-level model and CausalImpact with a data-scaled and the default level prior. Rows as in Figure 1, plus overall coverage and false $F_1$ differences.*

## A10.8 Prior sensitivity

**Design.** The MLITS model was refit with one prior changed at a time, in the Additive scenario at SNR 0, 2 and 4 over 50 simulation replications for each prior change by SNR combination. The reference ("Base") uses fixed shift SDs (3 for subgroups, 2 overall); "Shift: data-scaled" is the prior used in the main run (A5), which is about 3 in these data. Variants: shift SD 1 or 10; subgroup random-walk SD $N^+(0,0.1)$ or $N^+(0,1)$ (base $N^+(0,0.25)$); overall random-walk SD $N^+(0,1.5)$ (base $N^+(0,0.5)$); smoothing $N^+(1,5)$ (base $N^+(5,10)$). With 50 datasets the Monte Carlo SE is about 0.03 for a coverage near 0.95 and about 0.07 for a rate near 0.5.

**Results (Figure S13, top three rows).**

- Most changes have little effect. The near-flat shift (SD 10), the data-scaled shift, the wider overall random-walk SD and the weaker smoothing prior stay close to Base: at SNR 4, $F_1$ contrasts declared 0.80-0.83, contrast coverage 0.97-0.99, cell coverage 0.96-0.98; under the null, false $F_1$ contrasts 3-4%. The data-scaled shift is essentially identical to Base.
- A tight shift prior (SD 1) shrinks the effect: at SNR 4, $F_1$ contrast estimate / truth 0.73 (Base 0.89), detection 0.65 (0.81) and contrast coverage 0.88. This is the reason for scaling the shift prior to the data.
- A tight subgroup random walk ($N^+(0,0.1)$) attributes subgroup background drift to the subgroup effects: under the null, false $F_1$ differences rise to 15% and false $F_2$ differences to 10%, with $F_1$ contrast coverage 0.85. Pre-period latent coverage falls to 0.92-0.93 (0.94-0.96 for the other variants), so the pre-period diagnostic (Methods) flags it. Sampling was also worse at SNR 4 (6.7 divergences per fit, maximum $\hat{R}$ 1.13).
- A wide subgroup random walk ($N^+(0,1)$) is conservative: detection at SNR 4 falls to 0.72, with no loss of coverage.

Across variants, 44-100% of fits were divergence-free (mean 0.0-1.3 divergences per fit, except the tight subgroup random walk at SNR 4), maximum $\hat{R}$ was 1.03 (1.13 for the same tight subgroup variant) and minimum E-BFMI 0.65.

## A10.9 Misspecified background: AR(1)

**Design.** In the same prior-sensitivity run, the background was also generated as a stationary AR(1) process (coefficient 0.5; innovation SDs 0.30 overall and 0.15 by level, centered as above) and fit with the random-walk MLITS model, in the Additive scenario at SNR 0 and 2 (50 datasets each). This was the background in the original submission.

**Results (Figure S13, bottom two rows).** Effect inference becomes conservative: under the null, cell and overall coverage are 1.00 and false subgroup differences 0% ($F_1$) and 0-1% ($F_2$); at SNR 2, contrast coverage is 0.98-1.00. Detection is retained: true $F_1$ differences are declared in 60% of contrasts at SNR 2 (43% with the random-walk background), with estimate / truth about 0.89. Pre-period latent coverage drops to about 0.91 while predictive coverage stays at 0.97, so the diagnostic again flags the misspecification. The over-coverage of the original submission's Table 1 is consistent with this: a random walk, whose variance grows with the horizon, fit to a mean-reverting background.

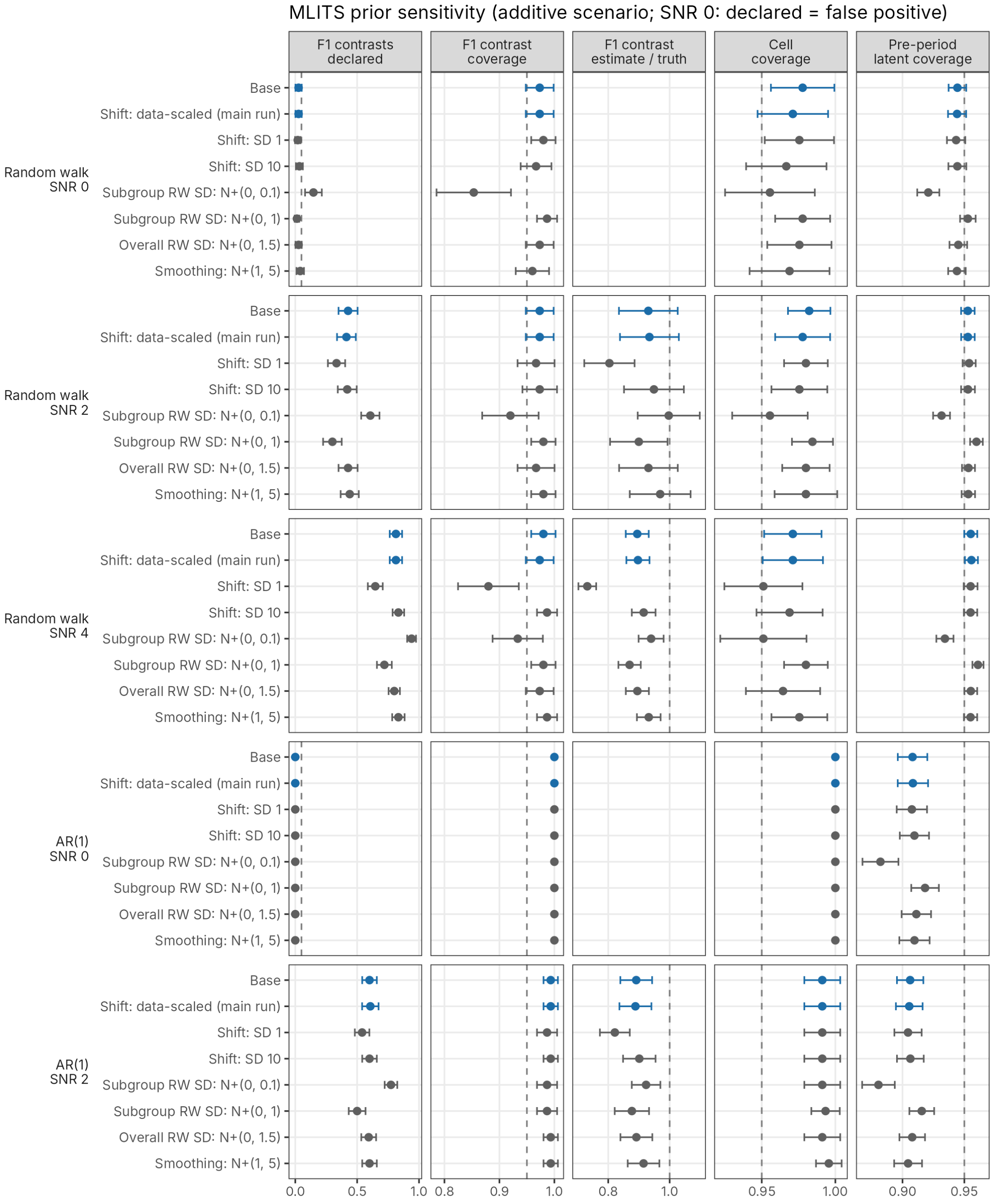

*Figure S13. MLITS prior sensitivity in the Additive scenario, with a random-walk (top three rows) and an AR(1) background (bottom two rows), 50 datasets per setting. Blue: Base and*

*the data-scaled shift prior used in the main run. At SNR 0, "F1 contrasts declared" is a false-positive rate (dashed line 0.05).*

## A10.10 Count outcomes: negative binomial

**Design.** We additionally show simulations for the MLITS model for count data. We simulate negative binomial MLITS model (A6.3); the single-level model and CausalImpact were not fit. Each cell's baseline mean is drawn from a lognormal with median 45 and log-SD 0.3; the dispersion is $\phi_s = 5 \times$ baseline (variance 1.2 times the mean at baseline). Everything latent is expressed in noise units $u = \sqrt{1.2/45} = 0.163$, the log-scale SD of a count with mean 45: random-walk SDs $0.30u$ and $0.15u$, and the effects of the Gaussian design multiplied by $u$. Peak SNR therefore keeps its meaning; at SNR 4 the largest cell effect is a log rate ratio of 0.65. $\log\lambda_{st}^{(0)} = \log(\text{baseline}_s) + u \times \text{background}_{st}$, $\log\lambda_{st} = \log\lambda_{st}^{(0)} + u\, g(t,s)$ and $y_{st} \sim \text{NB}(\lambda_{st}, \phi_s)$. We replicate this experiment 300 times across all SNRs and scenarios.

**Results (Figure S14; RMSE divided by $u$ for comparison with the Gaussian run).** Performance closely matches the Gaussian simulation. Cell RMSE is 0.75-0.83 noise units in the Additive and F2-homogeneous scenarios (Gaussian 0.75-0.83) and cell coverage 0.96-0.97. Under the Null scenario an overall effect was declared in 2.3% of datasets and false $F_1$ and $F_2$ differences in 2.7% and 5.2% of contrasts; false $F_2$ differences in the F2-homogeneous scenario were 4-6%. True $F_1$ differences were declared in 77-81% of contrasts at SNR 4. As in the Gaussian run, the Multiplicative scenario at SNR 4 has the highest cell RMSE (1.08) and the lowest cell coverage (0.89), and pre-period latent coverage drops there (0.92, against 0.94-0.95 elsewhere). The overall effect is declared more often than in the Gaussian run at SNR 4 (15-22% vs 6-10%). The dispersion is weakly identified at these counts (posterior median $\phi$ about 240-780 against a true median of about 225; 82-99% of upper interval limits above 1,000), which is expected so close to Poisson and does not affect the effect estimates.

**Diagnostics.** 63-84% of fits were divergence-free (mean 0.2-1.1 per fit; two settings had a single fit with more than 200), 88-95% had $\hat{R} < 1.01$ (maximum 1.06), minimum bulk and tail ESS of the effect quantities 645 and 404, minimum E-BFMI 0.70; median fit time about 2 minutes.

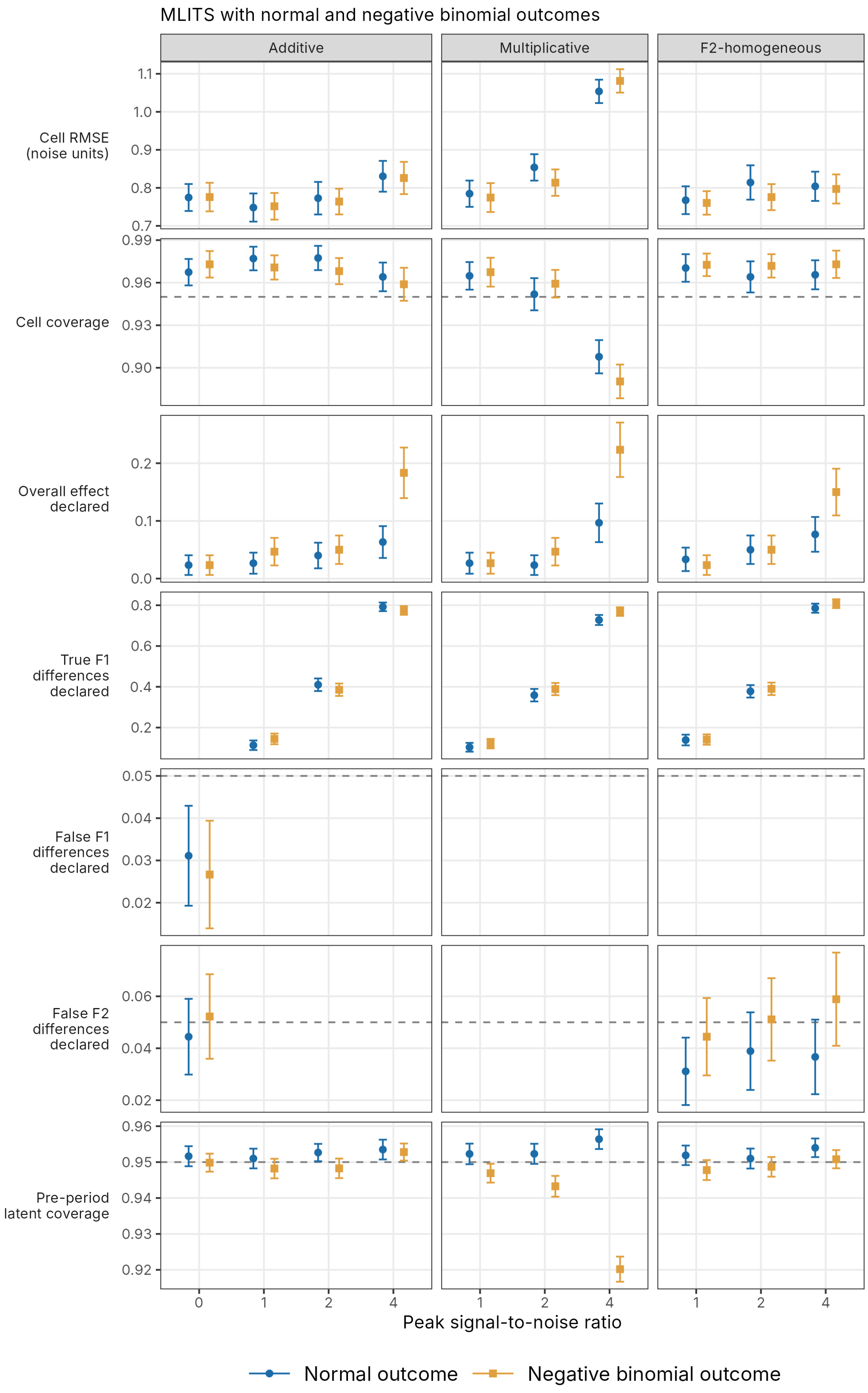
MLITS with normal and negative binomial outcomes
Additive
Multiplicative
F2-homogeneous
Cell RMSE (noise units)
Cell coverage
Overall effect declared
True F1 differences declared
False F1 differences declared
False F2 differences declared
Pre-period latent coverage
Peak signal-to-noise ratio
Normal outcome
Negative binomial outcome

*Figure S14. MLITS with normal (main simulation) and negative binomial outcomes. Count-outcome RMSE is in noise units ($u = 0.163$).*